%% file: mainpart.tex
\documentclass[12pt]{article}
\pdfoutput=1
\usepackage{graphicx,psfrag,epsf,color}
\usepackage[dvipsnames]{xcolor}
\usepackage{amsmath,amssymb,amsfonts, booktabs}
\usepackage{appendix}
\usepackage{multirow}
\usepackage{geometry}
\usepackage{float}
\usepackage{mathrsfs}  
\usepackage{authblk}
\usepackage{array}
\usepackage{cite}
\usepackage{slashed}
\usepackage{graphicx}
\usepackage{braket}
\usepackage{comment}
\usepackage{todonotes}
\usepackage{amsmath}
\usepackage{soul}

\usepackage{savesym}
\savesymbol{checkmark}
\usepackage{pifont}
\usepackage{tikz, dingbat, multirow}
\usetikzlibrary{positioning}
\usetikzlibrary{tikzmark,fit,shapes.geometric}
\usetikzlibrary{arrows,shapes}

\usepackage{siunitx}
\usepackage[section]{placeins}
\usepackage{xspace}

\usepackage{hyperref}
\usepackage[capitalise]{cleveref}

\graphicspath{ {Images/} }
\usepackage{rotating}
\usepackage{slashed, cancel}
\usepackage[caption=false]{subfig}
\usepackage[compat=1.1.0]{tikz-feynman} 

\usepackage{siunitx}

\usepackage{tabularx}
\newcolumntype{C}{>{\centering\arraybackslash}X}

\def\be{\begin{equation}}
\def\ee{\end{equation}}
\def\beq{\begin{eqnarray}}
\def\eeq{\end{eqnarray}}
\newcommand{\bea}{\begin{eqnarray}}
\newcommand{\eea}{\end{eqnarray}}
\newcommand{\beas}{\begin{eqnarray*}}
\newcommand{\eeas}{\end{eqnarray*}}

\newcommand{\qcut}{ q^2_\text{cut}}
\newcommand{\qcutineq}{{ q^2 \ge  q^2_\text{cut}}}
\newcommand{\qq}{{ q^2}}
\newcommand{\dd}{{\mathrm{d}}}

\definecolor{UBCblue}{rgb}{0.04706, 0.13725, 0.26667}
\definecolor{darkred}{rgb}{0.68, 0.09, 0.13}
\definecolor{darkgreen}{rgb}{0.31, 0.49, 0.16}
\definecolor{yaleblue}{rgb}{0.06, 0.3, 0.57}
\definecolor{grey}{rgb}{0.66, 0.66, 0.66}

\newcommand{\darkred}[1]{\textcolor{darkred}{#1}}
\newcommand{\darkgreen}[1]{\textcolor{darkgreen}{#1}}

\newcommand{\Vcb}{\ensuremath{V_{cb}}\xspace}
\newcommand{\absVcb}{\ensuremath{\left| \Vcb \right| }\xspace}

\newcommand{\mb}{\ensuremath{m_b}\xspace}
\newcommand{\mc}{\ensuremath{m_c}\xspace}

\newcommand{\muG}{\ensuremath{\mu_{G}^2}\xspace}
\newcommand{\mupi}{\ensuremath{\mu_\pi^2}\xspace}

\newcommand{\rhoD}{\ensuremath{\rho_{D}^3}\xspace}

\newcommand{\sB}{\ensuremath{s_{B}^4}\xspace}
\newcommand{\sqB}{\ensuremath{s_{qB}^4}\xspace}
\newcommand{\sE}{\ensuremath{s_{E}^4}\xspace}
\newcommand{\rE}{\ensuremath{r_{E}^4}\xspace}
\newcommand{\rG}{\ensuremath{r_{G}^4}\xspace}

\newcommand{\chisq}{\ensuremath{\chi^2}\xspace}
\newcommand{\chisqmin}{\ensuremath{\chisq_\mathrm{min}}\xspace}

\newcommand{\rhocut}{\ensuremath{\rho_\mathrm{cut}}\xspace}
\newcommand{\rhomom}{\ensuremath{\rho_\mathrm{mom}}\xspace}

\DeclareMathOperator{\sign}{sign}

\begin{document}
\allowdisplaybreaks

\begin{titlepage}

\begin{flushright}
{\small
Nikhef 2022-006\\
TTP22-031, P3H-22-052\\
\today \\
}
\end{flushright}

\vskip1cm
\begin{center}
{\Large \bf\boldmath First extraction of inclusive $V_{cb}$ from $q^2$ moments }
\end{center}

\vspace{0.5cm}
\begin{center}
{\sc Florian Bernlochner$^{a}$, Matteo Fael$^{b}$, Kevin Olschewsky$^c$, Eric Persson$^{a}$, Raynette van Tonder$^{d}$, K. Keri Vos$^{e,f}$, Maximilian Welsch$^{a}$} \\[3mm]
{\it $^a$Physikalisches Institut,\\
Nu\ss{}allee~12, 
University of Bonn,\\
53115 Bonn, Germany\\[0.3cm]

{\it $^b$ Institut f\"ur Theoretische Teilchenphysik,
  Karlsruhe Institute of Technology (KIT), \\
  76128 Karlsruhe, Germany
}\\[0.3cm]

{\it $^c$ Theoretische Physik 1,
  Naturwiss. techn. Fakult\"at, Universit\"at Siegen, \\
  57068 Siegen, Germany
}\\[0.3cm]

{\it $^d$Department of Physics,
McGill University,\\
Montréal, Québec, H3A 2T8, Canada}\\[0.3cm]

{\it $^e$Gravitational 
Waves and Fundamental Physics (GWFP),\\ 
Maastricht University, Duboisdomein 30,\\ 
NL-6229 GT Maastricht, the
Netherlands}\\[0.3cm]

{\it $^f$Nikhef, Science Park 105,\\ 
NL-1098 XG Amsterdam, the Netherlands}}
\end{center}

\vspace{0.6cm}
\begin{abstract}
\vskip0.2cm\noindent
We present the first determination of $V_{cb}$ from inclusive $B\to X_c \ell \bar\nu_\ell$ using moments of the dilepton invariant mass, $q^2$. These moments are reparametrization invariant quantities and depend on a reduced set of non-perturbative parameters.  
This reduced opens a new path to extract these parameters up to $1/m_b^4$ purely from data and thereby reducing the uncertainty on $V_{cb}$. In this paper, we present our first determination of $V_{cb}$ using this method. Combining the recent measurements of $q^2$ moments by Belle and Belle~II, our default fit gives $|V_{cb}| = (41.69\pm 0.63)\cdot 10^{-3}$. This results presents an important independent cross check of, and is consistent with, the previous state-of-the-art inclusive determinations using lepton energy and hadronic invariant mass moments. \end{abstract}
\end{titlepage}



\section{Introduction}
\input{intro.tex}

\section{\boldmath The new method to determine $V_{cb}$}
\input{detail.tex}

\section{The fit procedure}\label{sec:fit}

\input{fit.tex}

\section{Fit results}

\input{results.tex}

\section{Conclusion and outlook}\label{sec:conl}
We have presented the first determination of $V_{cb}$ from $q^2$ moments of the inclusive $B\to X_c \ell\bar\nu_\ell$ spectrum based on \cite{Fael:2018vsp}. These moments have the benefit that they depend on an RPI reduced set of HQE parameters, requiring only 8 non-perturbative parameters up to order $1/m_b^4$. This opens a new path to determine $V_{cb}$ from inclusive decays and to probe $1/m_b^4$ terms constraint solely from data. In this first determination, we are able to include two out of five $1/m_b^4$ parameters. In addition, we performed an in-depth analysis of the theoretical correlations for the moments predictions, with a default scenario where these parameters are left free in the fit.

Using the recently measured $q^2$ moments from both Belle and Belle~II, we find
\begin{equation}
\label{eq:vcbour2}
    |V_{cb}| = (41.69\pm 0.59|_{\rm fit} \pm 0.23|_{\rm h.o.})\cdot 10^{-3} = (41.69 \pm 0.63)\cdot 10^{-3} \ ,
\end{equation}
which has an incredible percent-level precision. Our value presents an independent cross-check of previous inclusive $V_{cb}$ determinations, using both new data and a new method. We find good agreement with the previously obtained inclusive $V_{cb}$ in \eqref{eq:vcb1} from \cite{Bordone:2021oof} which was obtained from lepton-energy and hadronic invariant mass moments. This provides strong evidence that inclusive $V_{cb}$ can be reliably obtained using the HQE and that the uncertainties are well under control. In light of the persisting $V_{cb}$ puzzle, this is an important finding.

We emphasize that any tension between the two determinations is fully due to a different input for the total branching ratio, which is a key input in the $V_{cb}$ determination. We therefore stress the need for new determinations of the $B\to X_c\ell\bar\nu_\ell$ branching ratio, preferably as a function of different $q^2$ thresholds. In addition, it would be interesting to obtain separate results for $B^0$ and $B^+$ decays, and also to carry out measurements of moments and partial branching fractions for $B\to X\ell\bar\nu_\ell$.

From the theoretical side, our analysis can be improved in the future by higher-order perturbative corrections as indicated in Table~\ref{tab:overview}. The found correlations between \rhoD, \rE and \rG will be studied in future works. In addition, it may be interesting to combine our analysis with that of \cite{Bordone:2021oof}, to include lepton moments and hadronic invariant mass moments. This would require using the full set of HQE parameters, but may have the benefit of higher statistics. Also observables like the forward-backward asymmetry \cite{Turczyk:2016kjf}, or the recently discussed differences of partial moments \cite{Herren:2022spb} could be included. These future improvement have the potential to push the inclusive $V_{cb}$ determination to below percent level uncertainty.

\subsubsection*{Acknowledgements}
We thank C.\ Schwanda and M.\ Rotondo for their input on the branching ratio. We thank P.\ Gambino for useful discussion about theory correlations.
We also like to thank \mbox{Th.\ Mannel} and M.\ Bordone for suggestions to improve the manuscript. This research was supported by the 
DFG Sonderforschungsbereich/Transregio 110 ``Symmetries and the Emergence of Structure in QCD''. 
The research of M.F.\ and K.O.\ was supported by the Deutsche Forschungsgemeinschaft (DFG, German Research Foundation) under grant 396021762 - TRR 257 ``Particle Physics Phenomenology after the Higgs Discover''.
\appendix
\label{sec:app:expr}
\input{app.tex}

\bibliographystyle{JHEP} 
\bibliography{refs.bib}
\end{document}

%% file: intro.tex
The Cabibbo-Kobayashi-Maskawa (CKM) matrix element $V_{cb}$ plays an important role in the Standard Model of particle physics (SM). Improving our knowledge and the uncertainty on $V_{cb}$ is crucial given the longstanding tension between $V_{cb}$ obtained from exclusive $B\to D^{(*)}\ell\bar\nu_\ell$ decays and the inclusive $B\to X_c \ell \bar\nu_\ell$ determination, known as the $V_{cb}$ puzzle (for recent works see e.g.~\cite{Gambino:2019sif, Bordone:2019vic, FermilabLattice:2021cdg, Martinelli:2021myh}). In addition, for several decays (like the rare $B_s\to \mu^+\mu^-$), the precision on $V_{cb}$ is already the dominant source of uncertainty on the SM predictions. 

The inclusive determination of $V_{cb}$ has reached an impressive percent level of accuracy due to both experimental and theoretical progress (see e.g.~\cite{Gambino:2020jvv} for a recent review). Using the Heavy Quark Expansion (HQE), the inclusive differential semileptonic rate can be written as an Operator Product Expansion (OPE) -- a power expansion in the inverse bottom quark mass, $1/m_b$. Consequently, at each order in $1/m_b$ non-perturbative matrix elements with perturbatively calculable coefficients are introduced. These HQE elements can be extracted from moments of the differential $B\to X_c \ell\bar\nu_\ell$ decay spectrum, typically lepton energy and/or hadronic invariant mass moments. Besides serving as an input in the total rate to extract $V_{cb}$, the extraction of the HQE elements themselves is important to check the convergence of the HQE. In addition, they form an input for e.g. the SM prediction for inclusive $B\to X_{s,d} \ell\ell$ decays~\cite{Huber:2020vup, Huber:2019iqf}.

The current most precise determination of $V_{cb}$ is \cite{Bordone:2021oof} (see also~\cite{Alberti:2014yda, Gambino:2013rza}):
\begin{equation}
     |V_{cb}| = (42.16\pm 0.30|_{\rm th}\pm 0.32|_{\rm exp}\pm 0.25|_{\Gamma}) \cdot 10^{-3} = (42.16 \pm 0.51)\cdot 10^{-3} \ ,
\label{eq:vcb1}
\end{equation}
which employs the kinetic scheme~\cite{uraltsev:heavy_flavour_sum_rules,uraltsev:kinetic_mass_scheme} for the $b$-quark mass. 
This state-of-the-art determination is obtained from centralized moments of the lepton energy and hadronic invariant mass decay spectra of the inclusive $B\to X_c \ell \bar{\nu}_\ell$. On the perturbative side, it includes the recently calculated $\alpha_s^3$ corrections to the total rate~\cite{Fael:2020tow} and on the non-perturbative side, power corrections up to $1/m_b^3$. At this level, only four hadronic parameters are present:  $\mu_\pi^2$ and $\mu_G^2$ at order $1/m_b^2$, and $\rho_{LS}^3$ and $\rho_D^3$ at $1/m_b^3$. The number of hadronic matrix elements increases rapidly at each subsequent order in $1/m_b$ with an additional 9 elements at $1/m_b^4$ and 18 at $1/m_b^5$~\cite{Dassinger:2006md,Mannel:2010wj,Kobach:2018pie}. 
This proliferation of HQE elements hinders their full extraction from data. On the other hand, using the lowest-lying state saturation (LLSA) assumption~\cite{Mannel:2010wj,Heinonen:2014dxa}, the size of the $1/m_b^4$ and $1/m_b^5$ parameters can be estimated. Using these estimates as Gaussian priors, the $1/m_b^4$ and $1/m_b^5$ parameters were included in an analysis of the inclusive semileptonic $b\to c$ data~\cite{Gambino:2016jkc, Bordone:2021oof}. The inclusion of the above-mentioned higher power corrections is found to reduce the value of $V_{cb}$ by about $0.25\%$~\cite{Gambino:2016jkc}. In addition, a recent analysis using the 1S-scheme found a value of $V_{cb}$ consistent with~\eqref{eq:vcb1}, but exhibiting a slightly larger uncertainty~\cite{Hayashi:2022sjy}.  

In this paper, we perform the first determination of $V_{cb}$ using a novel approach proposed in~\cite{Fael:2018vsp} based on moments of the dilepton moment $q^2$ spectrum. 
These moments have only recently been measured for the first time by both Belle~\cite{Belle:2021idw} and Belle II~\cite{The:2022cbm}, with both experiments already reaching a high precision. The benefit of using $q^2$ moments is that they are reparametrization invariant quantities, and therefore depend on a reduced set of HQE parameters as shown in \cite{Mannel:2018mqv}. Up to $1/m_b^4$, only 8 parameters are required, whereas at this order the traditional lepton energy and hadronic mass moments depend on the full set of 13 elements. This opens the road for a full data-driven $V_{cb}$ extraction including terms up to $1/m_b^4$. Combining both Belle and Belle II measurements of the $q^2$ moments, we find for our default fit scenario
\begin{equation}
 \label{eq:vcbour}
     |V_{cb}| = (41.69\pm 0.59|_{\rm fit} \pm 0.23|_{\rm h.o.})\cdot 10^{-3} = (41.69 \pm 0.63)\cdot 10^{-3} \ ,
\end{equation}
where the fit uncertainty is dominated by theoretical (input) uncertainties as well as the total branching ratio of $B\to X_c \ell \bar\nu_\ell$. In the above equation, h.o.\ represent a conservative additional uncertainty due to neglected power corrections, as discussed in detail in the remainder of this paper. Our value is in agreement with~\eqref{eq:vcb1} within uncertainty and has a similar uncertainty. The difference in central value between \eqref{eq:vcb1} and \eqref{eq:vcbour} can be attributed solely to a difference in the input for the branching ratio for $B\to X_c \ell\bar\nu_\ell$. Our result presents an important independent cross check of the inclusive $V_{cb}$ determination, showing a very consistent picture and an impressive uncertainty. We discuss how to improve on this precision in future works. 

This paper is outlined as follows: we first summarize the novel method to determine $V_{cb}$ from $q^2$ moments and discuss the available $\alpha_s$ corrections. In Sec.~\ref{sec:fit}, we discuss the fit procedure, inputs and the treatment of the theoretical uncertainties. Then we present and discuss our fit results. We present our conclusions and an outlook in Sec.~\ref{sec:conl}. In Appendix~\ref{app:sec:HQEpara}, we report the definitions of the matrix elements. In Appendix~\ref{app:addimat} and~\ref{sec:corrscan}, we present supplemental material for the combined Belle and Belle~II analysis, while appendices \ref{sec:belleonly} and \ref{sec:belle2only} contain fit results using only Belle and Belle~II data, respectively. 

%% file: detail.tex
\subsection{Preliminaries}
We consider the semileptonic decay of a $B$ meson $B \to X_c \ell \bar\nu_\ell $, with $\ell=e,\mu$, described by the effective Hamiltonian,
\begin{equation}
  \mathcal{H}_W = \frac{4 G_F}{\sqrt{2}} V_{cb}
  \left( \bar{c} \gamma^\mu P_L b \right)
  \left( \bar{\ell} \gamma_\mu P_L \nu \right) + \mathrm{h.c.} =
  \frac{4 G_F}{\sqrt{2}} V_{cb} J^\mu_q J_{\ell \mu} + \mathrm{h.c.},
  \label{eqn:HW}
\end{equation}
where $P_L$ is the left-handed projector, $J_q^\mu$ and $J_\ell^\mu$ are the hadronic and leptonic currents, respectively.

The total semileptonic rate $\Gamma_\mathrm{sl}$ and spectral moments $\langle M^k \rangle$ of inclusive decays are defined via phase-space integration of the differential rate multiplied by a certain weight function $w(v,p_\ell,p_\mu)$, raised to an integer power $n$:
\begin{equation}\label{eq:genmom}
  \langle M^n[w] \rangle = 
  \int \dd\Phi \,  w^n(v,p_\ell,p_\nu) \,
  W^{\mu\nu}L_{\mu\nu},
\end{equation}
where $v = p_B/m_B$ is the $B$ meson velocity. The momenta of the neutrino and the charged lepton are denoted by $p_\nu$ and $p_\ell$, respectively. Further, $p_B$ and $p_X$ denote the momenta of the $B$ meson and inclusive $X_c$ system. The leptonic and hadronic tensors, $W^{\mu\nu}$ and $L^{\mu\nu}$, are integrated over the phase space of the decay $\mathrm{d}\Phi$.
The prediction for $\Gamma_\mathrm{sl}$ is obtained with $w(v,p_\ell,p_\nu)=1$.
Moments of the leptonic invariant mass ($q^2$ moments) correspond to the weight function $w(v,p_\ell,p_\nu) = q^2$, where $q=p_\ell+p_\nu$ is the momentum of the lepton pair.
For the hadronic invariant mass moments, one sets $w(v,p_\ell,p_\nu) = (m_B v - q)^2$ with $p_X = p_B-q$, while moments of the charged lepton energy in the $B$ meson rest frame are obtained with $w(v,p_\ell,p_\nu) = v \cdot p_\ell$. 

The hadronic tensor $W^{\mu\nu}$ parametrizes all strong interactions relevant in the inclusive $B$ decay and can be related to the imaginary part of the forward scattering amplitude such that
\begin{align}
  W^{\mu\nu} &= 
  (2\pi)^4 \sum_{X_c}
  \delta^{4}(p_B-q-p_X)
  \bra{B} J_q^{\dagger \mu} \ket{X_c}
  \bra{X_c} J_q^{\nu} \ket{B} \, ,
  \notag \\ &
  =  2 \, \mathrm{Im} \, \bra{B} 
  i \int \dd^4 x \, e^{-i m_b S \cdot x} \,
  T \Big\{ \bar{b}_v(x) \gamma^\mu P_L c(x) \, \bar{c}(0) \gamma_\mu P_L b_v(0) \Big\}
  \ket{B} \, ,
  \label{eqn:Tdef}
\end{align}
where $S=v-q/m_b$ and $b_v(x) = \exp(im_b \, v\cdot x)\, b(x)$ is the re-phased $b$-quark field.

The calculation of the total semileptonic rate and the various moments proceeds via an Operator Product Expansion (OPE) of the time-ordered product in~\eqref{eqn:Tdef}, which yields the Heavy Quark Expansion (HQE) (for details see for instance~\cite{Blok:1993va,Manohar:1993qn,Manohar:2000dt}). 
It allows to predict the above-mentioned observables as a series in inverse powers of $m_b$, the bottom quark mass, and has the schematic form
\begin{equation}
  \langle M^n[w] \rangle = 
  \sum_{n=0}^\infty \frac{C^{(n)}_{\mu_1 \dots \mu_n}}{m_b^{n+3}} \otimes 
  \bra{B} \bar{b}_v (iD^{\mu_1} \ldots iD^{\mu_n})  b_v \ket{B} ,
  \label{eqn:moments}
\end{equation}
where the symbol $\otimes$ is shorthand notation for contraction of spinor indices not explicitly shown. 
On the r.h.s.\ of ~\eqref{eqn:moments} there are certain matrix elements of local 
operators taken between $B$ meson states. They parametrize the short-distance dynamics in the decay.
The Wilson coefficients $C^{(n)}_{\mu_1 \dots \mu_n}$ describe short-distance effects and are computed in perturbative QCD.

The total rate and the $q^2$ moments are invariant under a reparametrization (RP) transformation $\delta_{\rm RP} $ that shifts $ v_\mu \longrightarrow v_\mu + \delta v_\mu $.
On the contrary, hadronic invariant mass moments and charged-lepton energy moments are not invariant under reparametrization, since their associated weight functions $\delta_{\rm RP} w(v,p_\ell,p_\nu) \neq 0$.
In~\cite{Mannel:2018mqv,Fael:2018vsp} it was shown that the invariance under reparametrization (RPI) of total rate and $q^2$ moments, which implies $\delta_\mathrm{ RP} \Gamma_\mathrm{sl}=0$ and $\delta_{\rm RP} \langle (q^2)^n \rangle$, holds also for their OPE and connects subsequent orders in $1/m_b$ in ~\eqref{eqn:moments}. 
Certain relations between the coefficients $C$ at order $n$ and $n+1$ must be fulfilled in order to preserve RPI~\cite{Mannel:2018mqv}
\begin{equation}
  \delta_\mathrm{RP} C^{(n)}_{\mu_1 \dots \mu_n}(S) =
  m_b \, \delta v^\alpha 
  \Big[ 
  C^{(n+1)}_{\alpha \mu_1 \dots \mu_n}(S)
  +C^{(n+1)}_{\mu_1 \alpha \dots \mu_n}(S)
  + \dots +
  C^{(n+1)}_{\mu_1 \dots \mu_n \alpha}(S)
  \Big] \, .
  \label{eqn:RPIrelation}
\end{equation}
The operators $\bra{B} \bar{b}_v (iD_{\mu_1} \cdots iD_{\mu_n})  b_v  \ket{B}$ are usually rewritten in terms of a set of non-redundant (scalar) operators. 
Their matrix elements taken between $B$ meson states yield the HQE parameters. 
The relations in~\eqref{eqn:RPIrelation} allow to restrict the set of parameters for the total rate and the $q^2$ moments. They depend on certain fixed linear combinations of the matrix elements defined for the general case. As a result, only eight independent parameters at tree level are present at order $1/m_b^4$, defined in~\cite{Mannel:2018mqv} and given in Appendix~\ref{app:sec:HQEpara}. 

At the $B$ factories, the moments of the spectrum are generally measured with various threshold selections on the charged-lepton energy, with the lowest thresholds dictated by the detector lepton identification performance and the experimental understanding of low-energy electron and muon backgrounds. 
These measurements with different selections provide additional information on the HQE parameters.  
Predictions for the moments with threshold selections on the charged-lepton energy are obtained by limiting the phase-space integration, e.g. by using \ $w(v,p_\ell,p_\nu) \to w(v,p_\ell,p_\nu) \, \theta(v\cdot p_\ell -E_\mathrm{cut})$. 
However, introducing a similar phase-space constraint in the $q^2$ moments breaks their invariance under reparametrization. Therefore, Ref.~\cite{Fael:2018vsp} suggested to study $q^2$ moments with a threshold selection directly on $q^2$, which still preserves the RPI of the observable. 
Note that the constraint on the minimum value of $q^2$ bounds also the
charged-lepton energy, since $q^2 \le 4 E_\ell E_\nu$, in particular 
\begin{equation}
    E_\ell \ge 
    \frac{m_B^2+q^2_\mathrm{cut} -m_D^2-\lambda^{1/2}(m_B^2,q^2_\mathrm{cut},m_D^2)}{2m_D},
\end{equation}
where $\lambda^{1/2}(m_B^2,q^2_\mathrm{cut},m_D^2)$ is the K\"all\'en function.
The $q^2$ threshold selection cannot be chosen to be too large, since a value at high $q^2$ would significantly reduce the available phase space and render the decay no longer sufficiently inclusive, as pointed out in~\cite{Fael:2018vsp}.

\subsection{\boldmath $q^2$ moments in the HQE}
According to~\eqref{eq:genmom}, the moments of the $q^2$ spectrum with a minimum threshold selection are given by
\begin{equation}
    \mathcal{Q}_{n} (q^2_\mathrm{cut}) \equiv
    \frac{1}{\Gamma_0}
    \int \dd \Phi \, (q^2)^n \,
    \theta(q^2 - q^2_\mathrm{cut}) \,
    W^{\mu\nu} L_{\mu\nu}
    =
    \frac{1}{\Gamma_0}
    \int_{\qcut}^{m_b^2(1-\sqrt{\rho})^2} \dd \qq \, (\qq)^n \, \frac{\dd \Gamma}{\dd \qq},
\end{equation}
where $\Gamma_0 = G_F^2 m_b^5 |V_{cb}|^2 A_\mathrm{ew} /(192 \pi^3)$ and $\rho=m_c^2/m_b^2$. 
The factor $A_{\rm ew} = (1+ \frac{\alpha}{\pi} \log(M_Z/m_b))^2\simeq 1.01435$ stems from short-distance radiative corrections at the electroweak scale~\cite{Sirlin:1977sv}. 
We define the normalized moments as follows:
\begin{equation}
  \left\langle (q^2)^n \right\rangle_\qcutineq \equiv
    \left. \int_{\qcut}^{m_b^2(1-\sqrt{\rho})^2} \dd\qq \, (\qq)^n \, \frac{\dd\Gamma}{\dd\qq}
    \right/
     \int_{\qcut}^{m_b^2(1-\sqrt{\rho})^2} \dd\qq \, \frac{\dd\Gamma}{\dd\qq}  =
     \frac{\mathcal{Q}_n ( q^2_\mathrm{cut})}
     { \mathcal{Q}_0(q^2_\mathrm{cut})} \, 
     \label{eqn:q2momdef}.
\end{equation}
Central moments are given by
\begin{align}\label{eq:cenmom}
  q_1(q^2_\mathrm{cut}) \equiv \left\langle  q^2 \right\rangle_\qcutineq &\mbox{ for } n=1, \notag \\[5pt]
 q_n(q^2_{\textrm{cut}})\equiv \left\langle (  q^2 - \left\langle q^2 \right\rangle )^n \right\rangle_\qcutineq & \mbox{ for } n>1,
\end{align}
which are related to the moments in~\eqref{eqn:q2momdef} via the binomial formula:
\begin{equation}
  \left\langle (  q^2 - a )^n \right\rangle = 
  \sum_{i=0}^n 
  \binom{n}{i}
  \left\langle ( q^2)^i \right\rangle (-a)^{n-i}.
\end{equation}
In the HQE, the moments are expressed as a double expansion in $\alpha_s$ and $1/m_b$:
\begin{align}\label{eq:Qnexp}
    \mathcal{Q}_n(q^2_\mathrm{cut}) =
    (m_b^2)^n
    \Bigg\{
    &\mu_3 
    \Bigg[
    X_0^{(n)} 
    + \left( \frac{\alpha_s}{\pi} \right) X_1^{(n)}
    + \dots
    \Bigg] 
    \notag \\ &
    +\frac{\mu_G^2}{m_b^2}
    \Bigg[
    g_0^{(n)}
    + \left( \frac{\alpha_s}{\pi} \right) g_1^{(n)} + \dots
    \Bigg]
    +\frac{ \rho_D^3}{m_b^3}
    \Bigg[
    d_0^{(n)}
    + \left( \frac{\alpha_s}{\pi} \right) d_1^{(n)} + \dots
    \Bigg]
    \notag \\    & 
    +\frac{r_E^4}{m_b^4} l_{r_E}^{(n)}
    +\frac{r_G^4}{m_b^4} l_{r_G}^{(n)}
    +\frac{s_B^4}{m_b^4} l_{s_B}^{(n)} 
    +\frac{s_E^4}{m_b^4} l_{s_E}^{(n)}
    +\frac{s_{qB}^4}{m_b^4} l_{s_{qB}}^{(n)}
    \Bigg\} \, ,
\end{align}
where the strong coupling constant $\alpha_s \equiv \alpha_s^{(4)}(\mu_s)$ is taken at the renormalization scale $\mu_s$. 
The overall factor of $(m_b^2)^n$ is introduced to ensure that the various functions appearing in ~\eqref{eq:Qnexp} are dimensionless.

The tree level expressions up to $O(1/m_b^4)$ of $\mathcal{Q}_n(q^2_\mathrm{cut})$, with $n=0,\dots,4$ are computed and listed in Ref.~\cite{Fael:2018vsp}. 
The $q^2$ spectrum in the free-quark approximation is calculated at next-to-leading (NLO) order several times~\cite{Jezabek:1988,Aquila:2005hq,Trott:2004xc}, while next-to-next-to-leading order (NNLO) corrections are currently not known. 
The calculations in Refs.~\cite{Melnikov:2008qs,Biswas:2009rb,Gambino:2011cq} focus only on hadronic invariant mass and charged-lepton energy moments.\footnote{The author of~\cite{Melnikov:2008qs} could not retrieve the original Monte Carlo code.} 
Recently, analytic results for the $q^2$ moments without threshold selection cuts have been presented in~\cite{Fael:2022frj}.
NLO corrections to the power suppressed terms of order $1/m_b^2$ are computed in~\cite{Becher:2007tk,Alberti:2012dn,Alberti:2013kxa}, while the corrections at $\mathcal{O}(1/m_b^3)$ are presented only recently~\cite{Mannel:2021zzr}.
The explicit definitions of the HQE parameters are given in Appendix~\ref{app:sec:HQEpara}. 

The total rate $\Gamma_\mathrm{sl}$ within the HQE has a structure similar to
~\eqref{eq:Qnexp}. To leading order in the HQE, perturbative QCD corrections are computed to NLO~\cite{Nir:1989rm}, NNLO~\cite{Melnikov:2008qs,Pak:2008qt,Pak:2008cp,Dowling:2008mc} and recently at next-to-next-to-next-to-leading (N$^3$LO) order~\cite{Fael:2020tow} (results for the Abelian color factors are also confirmed in Ref.~\cite{Czakon:2021ybq}).
At orders $1/m_b^2$ and $1/m_b^3$ the NLO corrections are computed in~\cite{Mannel:2014xza,Mannel:2021zzr}.

\begin{table}
\centering
\begin{tabular}{c|ccccl}
  $\Gamma $ & tree  & $\alpha_s$ & $\alpha_s^2$ & $\alpha_s^3$  \\
  \cline{1-5}
  \multirow{ 2}{*}{Partonic} &
  \multirow{ 2}{*}{\darkgreen{\ding{51}}} & 
  \multirow{ 2}{*}{\darkgreen{\ding{51}}} & 
  \multirow{ 2}{*}{\darkgreen{\ding{51}}} & 
  \multirow{ 2}{*}{ \darkgreen{\ding{51}}} & \\
  \multirow{ 2}{*}{$\mu_G^2$} &
  \multirow{ 2}{*}{\darkgreen{\ding{51}}} & 
  \multirow{ 2}{*}{\darkgreen{\ding{51}}} & 
  \multirow{ 2}{*}{} & & \\
  \multirow{ 2}{*}{$\rho_D^3$} &
  \multirow{ 2}{*}{\darkgreen{\ding{51}}} & 
  \multirow{ 2}{*}{\darkgreen{\ding{51}}} & 
  \multirow{ 2}{*}{} & & \\
  \multirow{ 2}{*}{$1/m_b^4$} &
  \multirow{ 2}{*}{\darkgreen{\ding{51}}} & 
  \multirow{ 2}{*}{} & 
  \multirow{ 2}{*}{} & & \\
  \multirow{ 2}{*}{$m_b^\mathrm{kin}/\overline{m}_c$} &
  \multirow{ 2}{*}{} & 
  \multirow{ 2}{*}{\darkgreen{\ding{51}}} & 
  \multirow{ 2}{*}{\darkgreen{\ding{51}}} & 
  \multirow{ 2}{*}{\darkgreen{\ding{51}}} &  \\
    & & & & &    
\end{tabular} 
\begin{tabular}{c|ccccl}
  $\left\langle (q^2)^n \right\rangle$ & tree  & $\alpha_s$ & $\alpha_s^2$ & $\alpha_s^3$  \\
  \cline{1-5}
  \multirow{ 2}{*}{Partonic} &
  \multirow{ 2}{*}{\darkgreen{\ding{51}}} & 
  \multirow{ 2}{*}{\darkgreen{\ding{51}}} & 
  \multirow{ 2}{*}{} & 
  \multirow{ 2}{*}{} & \\
  \multirow{ 2}{*}{$\mu_G^2$} &
  \multirow{ 2}{*}{\darkgreen{\ding{51}}} & 
  \multirow{ 2}{*}{\darkred{\ding{51}}} & 
  \multirow{ 2}{*}{} & & \\
  \multirow{ 2}{*}{$\rho_D^3$} &
  \multirow{ 2}{*}{\darkgreen{\ding{51}}} & 
  \multirow{ 2}{*}{\darkred{\ding{51}}} & 
  \multirow{ 2}{*}{} & & \\
  \multirow{ 2}{*}{$1/m_b^4$} &
  \multirow{ 2}{*}{\darkgreen{\ding{51}}} & 
  \multirow{ 2}{*}{} & 
  \multirow{ 2}{*}{} & & \\
  \multirow{ 2}{*}{} &
  \multirow{ 2}{*}{} & 
  \multirow{ 2}{*}{} & 
  \multirow{ 2}{*}{} & 
  \multirow{ 2}{*}{} &  \\
    & & & & &    
\end{tabular} 
\caption{Schematic overview of the perturbative corrections available for the partial rate $\Gamma$ and the $q^2$ moments. Green checkmarks denote corrections that are known and built into our code. Red checkmarks indicate corrections that are known, but currently not included in our package. For references and further information we refer to the text. }
\label{tab:overview}
\end{table}

\subsection{\boldmath NLO corrections to the $q^2$ moments}
NLO corrections to the differential $q^2$ rate were first derived in Ref.~\cite{Jezabek:1988} utilizing the on-shell (or pole) renormalization scheme for the
charm and bottom mass.
We refer to those as genuine NLO corrections (they correspond to $X_1^{(n)}$ in~\eqref{eq:Qnexp}) and they are given by 
\begin{equation}
\label{eq:genuine_corrections_moments}
 X^{(n)}_1 = \frac{1}{(m_b^2)^n \Gamma_0} \int_{q^2_\mathrm{cut}}^{m_b^2(1-\sqrt{\rho})^2}
 \dd \qq \, ( \qq)^n \, \frac{\dd\Gamma^{(1)}}{\dd \qq}\,,
\end{equation} 
where the explicit expression for $\dd\Gamma^{(1)}/\dd q^2$ can be obtained from~\cite{Jezabek:1988} (see also~\cite{Mannel:2021zzr}).
Re-expanding the $q^2$ moments in \eqref{eqn:q2momdef} in $\alpha_s$, gives the explicit dependence on $O(\alpha_s)$ terms for the normalized $q^2$ moments: 
\begin{align}
  \braket{(q^2)^n}\Big\vert_{\alpha_s} =
\frac{\alpha_s}{\pi}\frac{1}{\left(X_0^{(0)}\right)^2}\left(X_0^{(0)}X_1^{(n)} - X_0^{(n)} X_1^{(0)}\right)\,,
\end{align}
and equivalently for the centralized moments. 

The expressions in the on-shell scheme are affected by a renormalon ambiguity leading to a badly behaved perturbative series~\cite{uraltsev:convergence_pole_mass, beneke:renormalons_hqet}.
To avoid this, the heavy quark masses must be converted from the on-shell scheme to a short-distance scheme. In this work, we adopt the kinetic scheme~\cite{uraltsev:heavy_flavour_sum_rules, uraltsev:kinetic_mass_scheme} for the bottom quark mass and the $\overline{\mathrm{MS}}$ scheme for the charm mass. 

The definition of the kinetic mass is based on the relation between 
the masses of a heavy meson and the corresponding heavy quark. 
Using perturbation theory, then gives a relation between the on-shell mass ($m_b^\mathrm{OS}$) and the kinetic mass ($m_b^\mathrm{kin}(\mu)$): 
\begin{equation*}
    m_b^\mathrm{OS} = 
    m_b^\mathrm{kin}(\mu)
    +[\overline{\Lambda}(\mu)]_\mathrm{pert}
    +\frac{[\mu_\pi^2(\mu)]_\mathrm{pert}}{2m_b^\mathrm{kin}(\mu)}
    +\mathcal{O}\left(\frac{1}{m_b^2}\right).
\end{equation*}
The scale $\mu$ entering the definition of $m_b^\mathrm{kin}(\mu)$ and the perturbative version of the HQET parameters plays the role of a Wilsonian cutoff with 
$\Lambda_\mathrm{QCD} \ll \mu \ll m_b$. Currently the mass relation is known at NNLO~\cite{Czarnecki:1997sz} and N$^3$LO~\cite{Fael:2020iea,Fael:2020njb} and hence can be consistently applied to our results. In our analysis, we set $\mu=1$~GeV.

The perturbative versions of the HQET parameters are computed by making
use of the small velocity sum rules~\cite{uraltsev:heavy_flavour_sum_rules}.
In the kinetic scheme, the HQE parameters must be also redefined by subtracting the perturbative corrections:
\begin{align}
 \mu_\pi^2(0) & = \mu_\pi^2(\mu) - [ \mu_\pi^2(\mu)]_{\rm pert} \, \label{kin1} \, ,\\
 \mu_G^2(0) & = \mu_G^2(\mu) - [ \mu_G^2(\mu)]_{\rm pert} \, \label{kin2} \, ,\\
 \rho_D^3 (0) & = \rho_D^3(\mu) - [ \rho_D^3(\mu)]_{\rm pert} \, \label{kin3} \, ,
 \end{align}
 where the HQE parameter at $\mu=0$ denotes the pole scheme. In our analysis, we thus extract the values of the kinetic scheme HQE parameters. 

The conversion of the bottom (charm) mass to kinetic ($\overline{\mathrm{MS}}$) scheme is performed after re-expanding the expressions for the centralized $q^2$ moments in the on-shell scheme up to $O(\alpha_s)$. In that way, the mass scheme re-definitions yield additional $\alpha_s$ corrections in addition to the genuine $\alpha_s$ corrections previously described.

\begin{figure}
    \centering
    \includegraphics[trim=100 0 100 0,clip, width=1\textwidth]{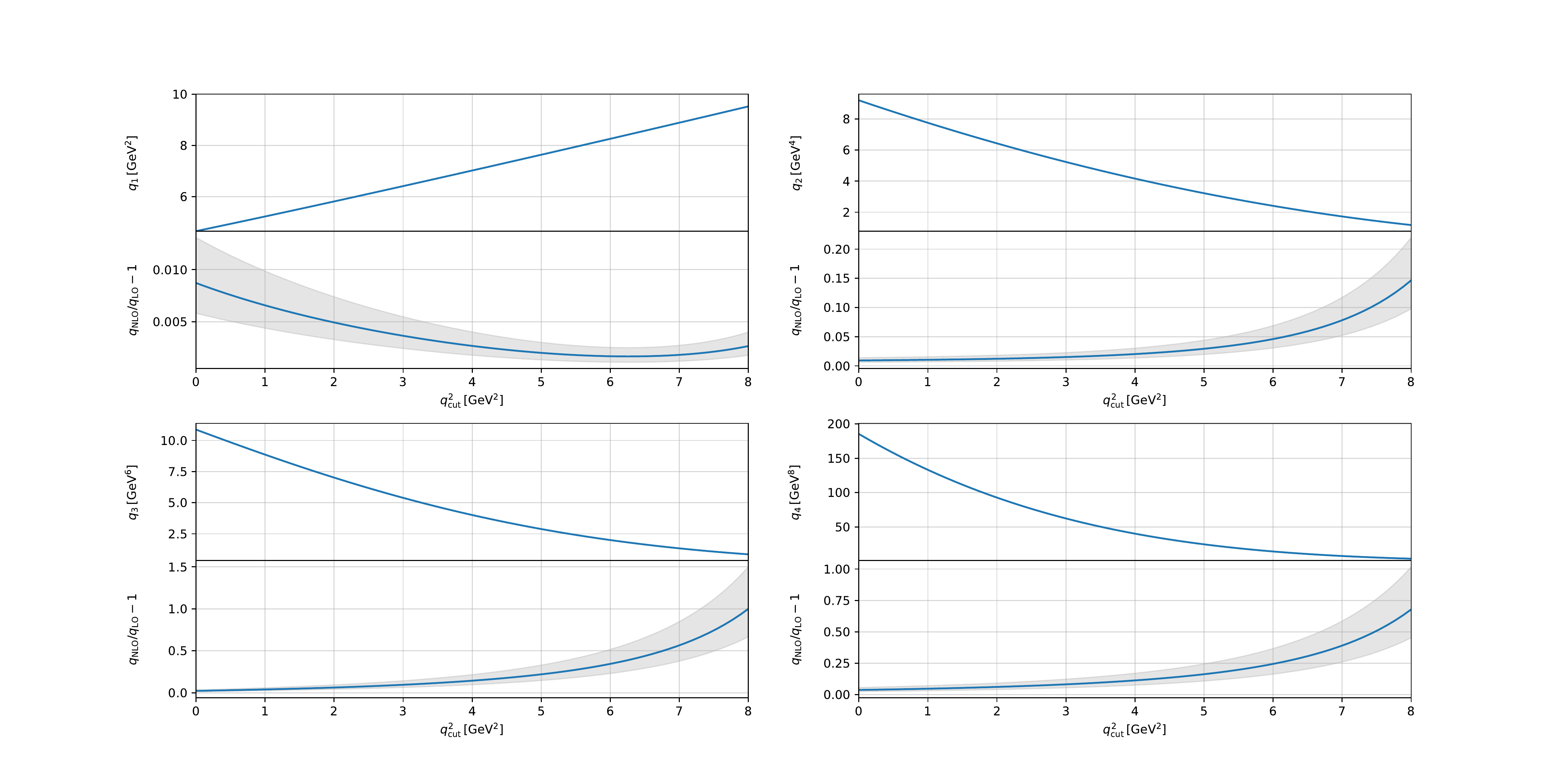}
    \caption{The first to fourth centralized $q^{2}$ moments (upper panels) and the relative size of the $\alpha_s$ corrections with respect to the LO (lower panels). The grey area in the ratios represents the error due to variation of the renormalization scale $\mu_s$.}
    \label{fig:moments_MS_as}
\end{figure}

Figure~(\ref{fig:moments_MS_as}) shows the theoretical predictions for the partonic part of the first four centralized moments, including both genuine and scheme change $\alpha_s$ corrections. 
In the lower panel of each plot we show the relative size of the $\alpha_s$ corrections compared to the LO prediction.

Note that for consistency, we neglected all terms $\sim \alpha_s \times \mathrm{HQET}$ parameters (i.e. corrections of $\mathcal{O}(\alpha_s/m_b^2)$), since we have not included the genuine $\alpha_s$ corrections to these parameters for the $q^2$ moments. 

The shaded area in the ratios represents the uncertainty obtained by varying the scale of $\mu_s$ in the range $m_b^\mathrm{kin}/2 < \mu_s < 2m_b^\mathrm{kin}$. We observe that for larger cuts in $q^2$ the NLO corrections become more important for the second to fourth moments. 

In \cite{software_package}, we provide an open-source software package which includes all genuine $\alpha_s$ corrections as well as all corrections associated with the scheme conversion. The integrals appearing in \eqref{eq:genuine_corrections_moments} are solved numerically. 
We implement the charm mass $m_c$ in the $\overline{\mathrm{MS}}$ scheme, allowing for the current, most precise value to be used as input.
For the bottom mass $m_b$ we fix the choice of the mass scheme to be kinetic. We finally point out that it is possible to extend this package to even higher order corrections, once they become available. In Table~\ref{tab:overview}, we present an overview of the perturbative corrections available and those included in the current version. Very recently, also the $\alpha_s$ corrections to $\rho_D^3$ for the moments have been calculated \cite{Mannel:2021zzr}. These will be implemented in an updated version of the code.

\subsection{\boldmath Dependence on $\mu_3$ and $\mu_\pi^2$}\label{sec:depmu3}
It is worth emphasizing that~\eqref{eq:Qnexp} does not explicitly dependent on the HQE parameter $\mu_\pi^2$. Due to RPI, the Wilson coefficient of $\mu_\pi^2/(2m_b^2)$ is always equal to the one
of the free quark decay (partonic result). The parameter always appears in the combination
\begin{equation}
\mu_3 = 1 +\frac{\mu_G^2 - \mu_\pi^2 }{2m_b^2}   \ ,
\end{equation}
where $\mu_3$ is defined by
 \begin{equation}
     \bra{B} \bar{b}_v b_v \ket{B} \equiv 2 m_B \mu_3 \ .
 \end{equation}
Due to the normalization of the $q^2$ moments in~\eqref{eqn:q2momdef}, we observe that the dependence on $\mu_\pi^2$ becomes effectively of order $1/m_b^4$ once the ratio in~\eqref{eqn:q2momdef} is re-expanded in $1/m_b$. 
This can be seen explicitly, as the normalized moments take the form
\begin{align}
  \frac{1}{(m_b^2)^2}\frac{  \mathcal{Q}_n(\hat q^2_\mathrm{cut}) }{  \mathcal{Q}_0(\hat q^2_\mathrm{cut}) }
  &=
   \frac{\mu_3    X_0^{(n)}     +\frac{\mu_G^2}{m_b^2}   g_0^{(n)}    +\mathcal{O}(\frac{1}{m_b^3}) } {\mu_3    X_0^{(0)}     +\frac{\mu_G^2}{m_b^2}   g_0^{(0)}    +\mathcal{O}(\frac{1}{m_b^3}) }  \, ,
   \notag \\ &
   = \frac{ X_0^{(n)}}{ X_0^{(0)}} \left[
   1
   +\frac{\mu_G^2}{m_b^2}
   \left(
   \frac{g_0^{(n)}}{X_0^{(n)}} 
   - \frac{g_0^{(0)}}{X_0^{(0)}} 
   \right)
   -\frac{(\mu_G^2)^2}{m_b^4}
   \left(\frac{1}{2}+\frac{g_0^{(0)}}{X_0^{(0)}} \right)   
   \left(
   \frac{g_0^{(n)}}{X_0^{(n)}} 
   - \frac{g_0^{(0)}}{X_0^{(0)}} 
   \right)
   \right. \notag \\&
   \quad + \left.
   \frac{\mu_\pi^2 \mu_G^2}{2m_b^4}
   \left(
   \frac{g_0^{(n)}}{X_0^{(n)}} 
   - \frac{g_0^{(0)}}{X_0^{(0)}} 
   \right)
   +\text{terms of } 
   \mathcal{O}\left(\frac{1}{m_b^3}\right)
   \right] .
 \end{align}
Consequently, the moments are rather insensitive to $\mu_\pi^2$, while $\mu_3$ is an important input for the total rate and thus the $V_{cb}$ determination. At the moment, we circumvent this problem by using an external constraint on $\mu_\pi^2$, as discussed in more detail in Sec.~\ref{sec:const}. For this reason our code contains the parameter $\mu_\pi^2$ instead of $\mu_3$. 

\subsection{\boldmath Predictions for central $q^2$ moments up to $1/m_b^4$ and extraction of $V_{cb}$}\label{sec:m4considerations}

The key benefit of using $q^2$ moments with different $q^2$ threshold selections is that these moments depend on a reduced set of HQE parameters. Up to order $1/m_b^4$, only eight parameters enter: $\mu_\pi^2$, $\mu_G^2$, $\rho_D^3$ and five $1/m_b^4$ non-perturbative quantities defined in Appendix~\ref{app:sec:HQEpara} denoted by $r_{E}^4, r_{G}^4, s_{E}^4 ,s_{B}^4$ and $s_{qB}^4$. 

\begin{figure}[tb]
    \centering
    \includegraphics[width=0.4\textwidth]{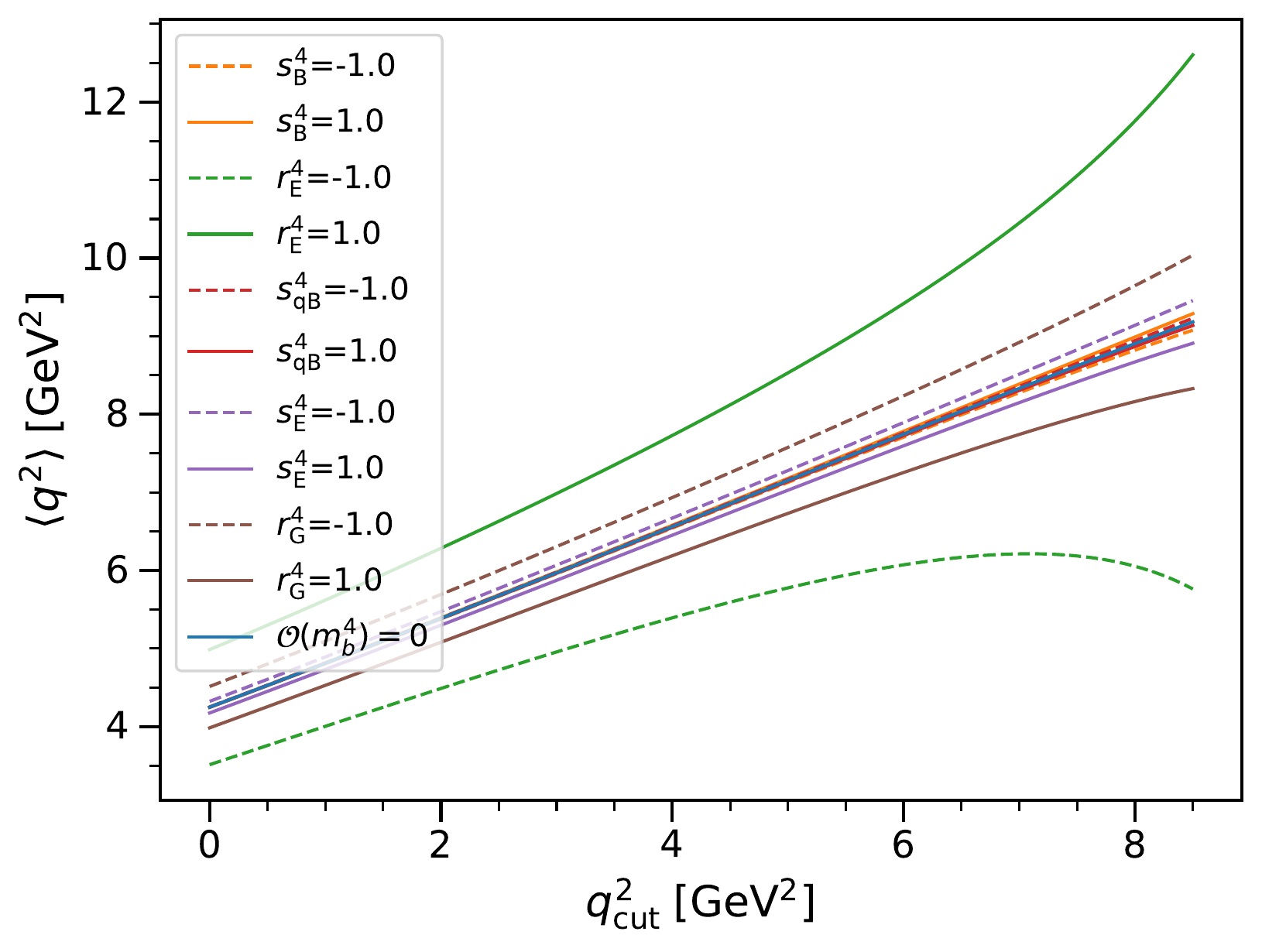}
    \includegraphics[width=0.4\textwidth]{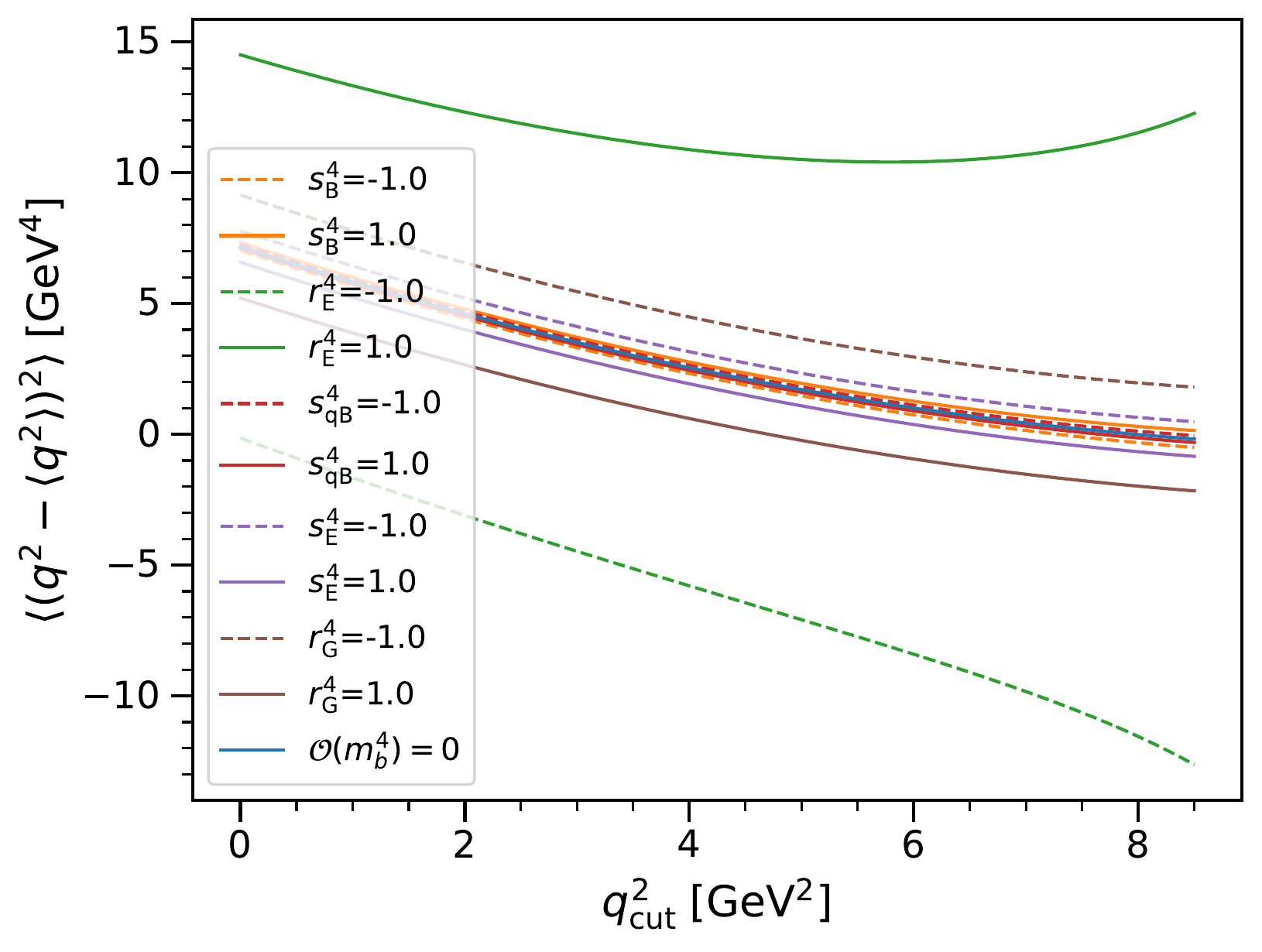}\\
    \includegraphics[width=0.4\textwidth]{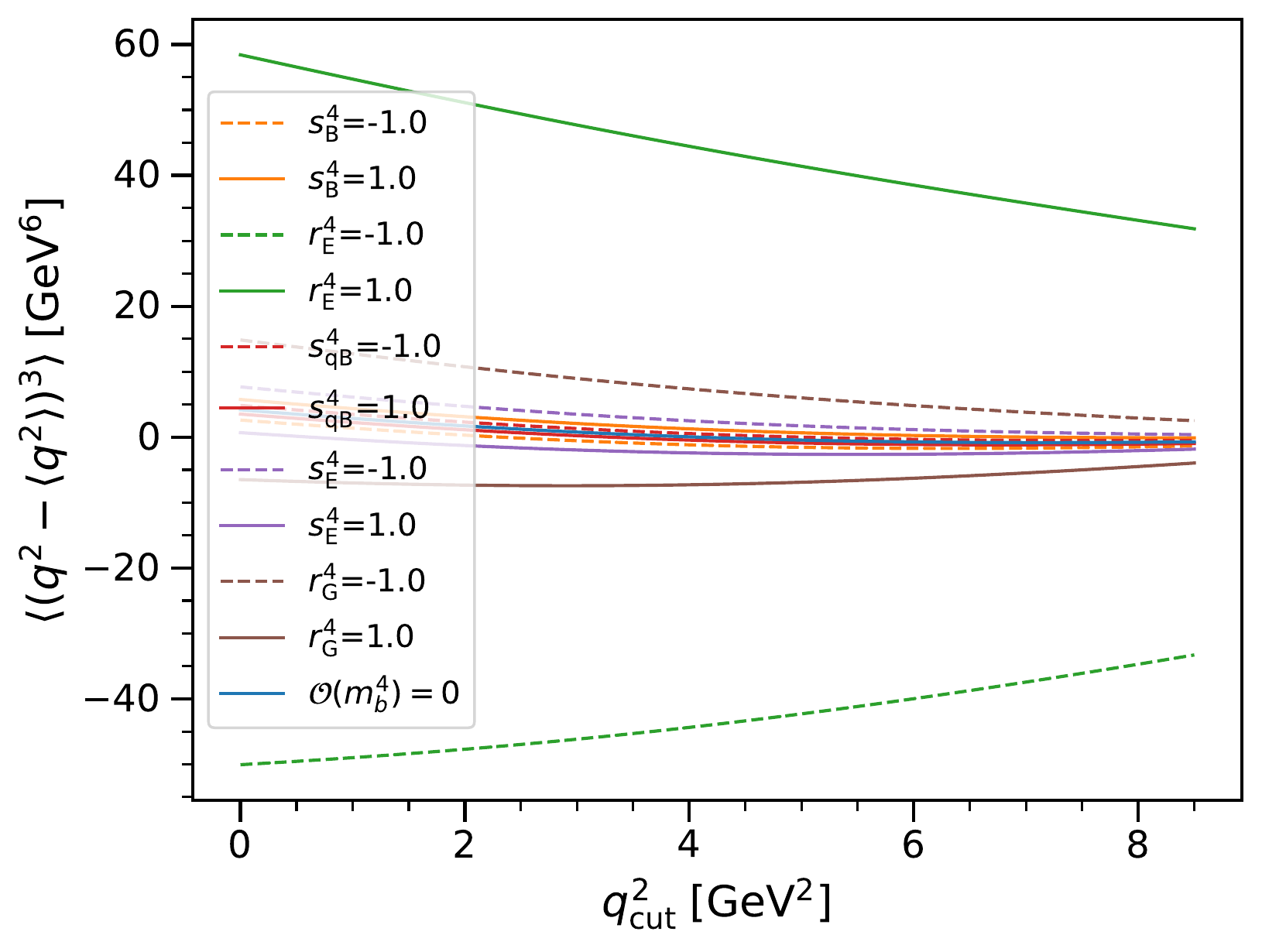}
    \includegraphics[width=0.4\textwidth]{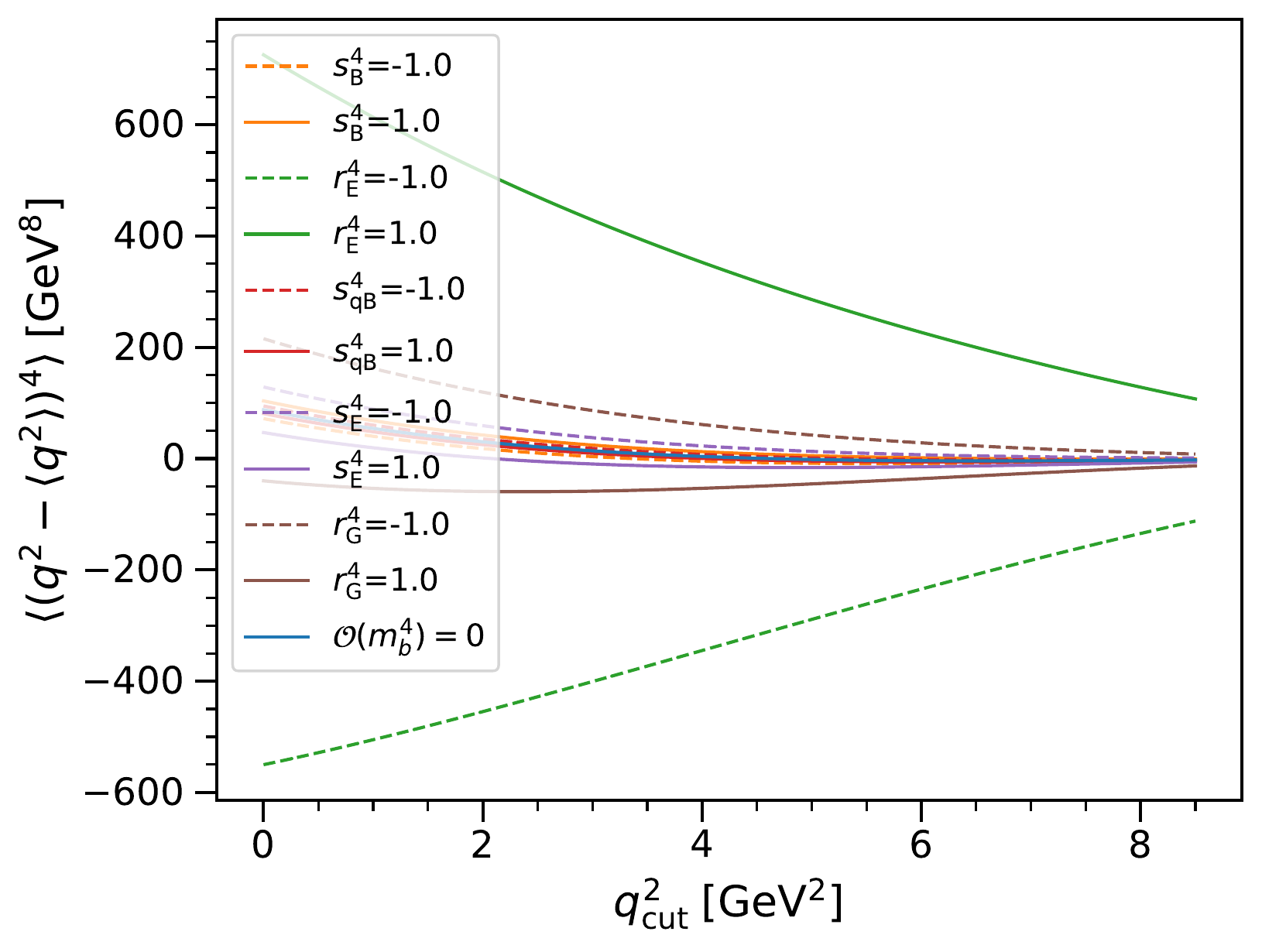}
    \caption{Sensitivity of the central $q^2$ moments for individual contributions of $1/m_b^4$ HQE terms obtained by varying their values between $\pm 1$ GeV$^4$. }
    \label{fig:variation}
    \end{figure}

\begin{figure}[tb]
    \centering
    \includegraphics[width=0.4\textwidth]{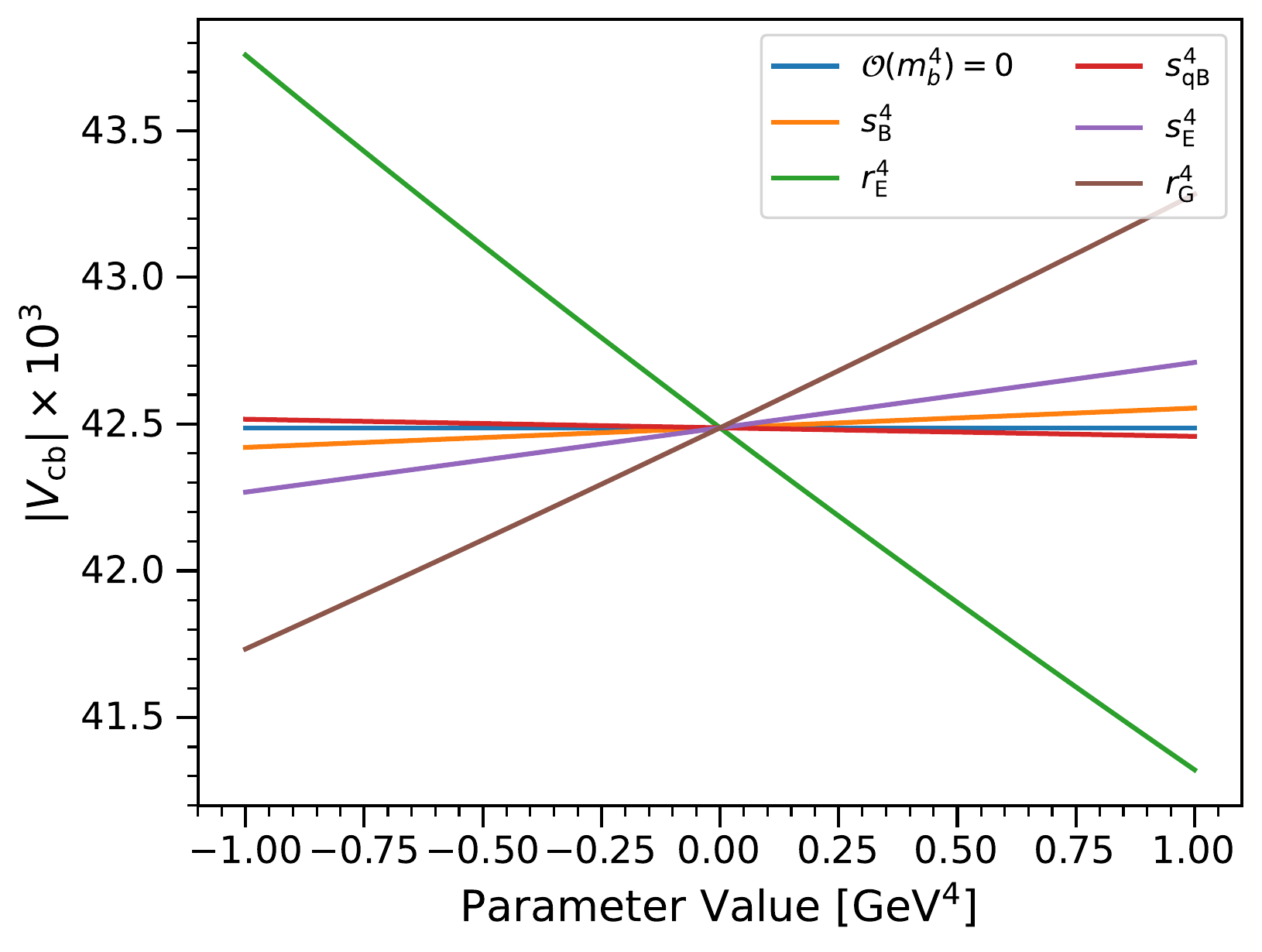}
    \caption{Illustration of the effect on $\left| V_{cb} \right|$ by varying the $1/m_b^4$ HQE terms independently between $\pm 1.5$ GeV$^4$. }
    \label{fig:vcb_variation_hqe_terms}
\end{figure}

Figure~\ref{fig:variation} shows the predicted central $q^2$ moments as a function of the $q^2$ threshold by individually varying the $1/m_b^4$ parameters between $\pm 1 \, \mathrm{GeV^4}$, while setting the other $1/m_b^4$ parameters to zero. From these variations we observe that the moments have the large sensitivity for $r_E^4$, followed by $r_G^4$. The other parameters only lead to small changes in the moments. We also note that different orders of moments exhibit different differential dependencies as a function of the $q^2$ threshold. In other words, the simultaneous analysis of different orders should allow us -- similar to the existing inclusive \Vcb hadronic and leptonic moment fits -- to separate parameter contributions from each other. Figure~\ref{fig:vcb_variation_hqe_terms} illustrates the impact of different choices for the sub-leading terms on a hypothetical \Vcb value from a measured $B \to X_c \ell \bar \nu_\ell$ branching fraction for variations of  $\pm 1.5 \, \mathrm{GeV^4}$. The largest shift in \Vcb is again from $r_E^4$, followed by $r_G^4$.  The other parameters exhibit only a small dependence on the overall rate. In the following, we thus include $r_E^4$ and $r_G^4$, along with $\rho_D^3, \mu_G^2$ and $\mu_\pi^2$, as we expect very little sensitivity to the other $1/m_b^4$ parameters. In turn, the precise value of the other parameters will only have a small impact on the description of the moments and due to their sub-leading contributions to the total rate on \Vcb. Since the measured experimental information from Belle and Belle~II do not provide partial branching fractions of $B \to X_c \ell \bar \nu_\ell$ with various $q^2$ thresholds, the extraction of \Vcb will rely on the total $B \to X_c \ell \bar \nu_\ell $ branching fraction as an input. The precise setup of the fit is further described in Sec.~\ref{sec:fit}.

%% file: fit.tex
\subsection{Fit Setup} 
The HQE parameters and $|V_{cb}|$ are determined in a simultaneous $\chi^2$ fit that takes into account both the experimental and theoretical correlations. The full $\chi^2$ function has the form
\begin{align} \label{eq:chi2}
    \chi^2(|V_{cb}|, \boldsymbol{\theta}) = &  \frac{ \left( \mathcal{B} - \Gamma(|V_{cb}|, \boldsymbol{\theta}) \, \tau_B / \hbar \right)^2 }{\sigma_{\mathcal{B}}^2 + \sigma_\Gamma^2} \nonumber \\ 
    &  + \left( \boldsymbol q (\boldsymbol \theta) - \boldsymbol q_{\mathrm{meas}}   \right) C^{-1} \left( \boldsymbol{q} (\boldsymbol{\theta}) - \boldsymbol{q_{\mathrm{meas}} }  \right)^T \nonumber \\
    & + \sum_{i=1}^{4} \frac{ \left(\theta_i - \theta_i^{\mathrm{cons}} \right)^2}{ \sigma_{\theta_i}^2 } \, ,
\end{align}
where $\mathcal{B}$ is the experimental branching ratio for $B\to X_c \ell \bar\nu_\ell$ and $\Gamma$ the theoretical expression for the rate. Here $\boldsymbol{q}$ denotes the vector of theoretical predictions for the $q_n$ central moments with different $q^2$ thresholds, whereas $\boldsymbol{q}_{\rm meas}$ denote the corresponding measured moments. We use the average of the charged and neutral $B$ meson lifetimes $\tau_B = \left(1.579 \pm 0.004\right) \, \text{ps}$~\cite{pdg:2020}  and $ \theta_i = \{m_b^{\rm kin}, \overline{m}_c, \mu_G^2, \mu_\pi^2 \}$ denote the constrained parameters in the fit, further discussed in Sec.~\ref{sec:const}. The covariance matrix is constructed from the sum of the statistical, systematical and theoretical covariance: $C = C_{\mathrm{stat}} + C_{\mathrm{syst}} + C_{\mathrm{theo}}$.

The statistical and systematical covariance matrices are from the Belle and Belle II measurements. In the fit systematic uncertainties are correlated between both sets of moment measurements, which we further discuss in Sec.~\ref{sec:bandb2}.

The theory covariance captures the uncertainty from missing higher-order corrections in $\alpha_s$ and the HQE. The prescription used to estimate the theory uncertainties is discussed in Sec.~\ref{sec:theun}. A priori the theory correlations between different thresholds and orders of moments are unknown. This lack of knowledge itself constitutes an uncertainty in the determination of \Vcb. We address this problem by introducing two parameters, $\rhocut$ and $\rhomom$, which parametrize the degree of correlation between different thresholds and orders of moments. 
We parametrize the correlation between two moments $q_n$ of the same order $n$ but with different thresholds $q^2_A$ and $q^2_B$ as
\begin{align}
 \rho_{n}[q_n( q^2_A), q_n( q^2_B)] = \rhocut^x \quad \mathrm{with} \quad x  =\frac{\left| q^2_A- q^2_B \right| }{\SI{0.5}{GeV^2}} .
\end{align}
This results in a decorrelation of moments with larger separation in terms of the $q^2$ thresholds. Note that a similar functional dependence  was used in~\cite{Gambino:2013rza} to parametrize the degree of decorrelation between moments with different lepton energy cuts. 
For moments of different orders $n$ and $m$ we construct the degree of correlation as
\begin{align}
    \rho_{nm}[q_m( q^2_A), q_n( q^2_B)] =  \sign(\rhomom) \cdot \left| \rhomom \right|^{\left|m-n\right|} \cdot \rho_{n}(q_n( q^2_A), q_n( q^2_B)).
\end{align}

   We do not choose fixed values for the correlation parameters $\rhocut$ and $\rhomom$, but allow them to float as nuisance parameters in the fit. This has the benefit that a large number of hypothesized values for both parameters are profiled, when the uncertainties on the HQE parameters and \Vcb are determined. To constrain both parameters to a meaningful range, we add two independent penalty terms into~\eqref{eq:chi2}. We parametrize these functions using a double Fermi-Dirac function:
\begin{align}
    f_{\mathrm{DFD}}(\rho, a, b) = \frac{1}{ 2(1 + e^{w(\rho - b)}) (1 + e^{-w(\rho - a)}) } \quad (b>a) \, ,
\end{align}
with $w = 50$. The parameters $a$ and $b$ denote the minimal and maximal values for the correlation of $\rho$. For $\rhocut$ we use $a=0$ and $b=1$ to avoid anti-correlations between moments at different thresholds. For $\rhomom$ we allow for anti-correlations and fix $a=-0.45$. This lower bound is motivated by the emergence of negative eigenvalues of the theory covariance matrix for $a < -0.45$, resulting in a non-singular covariance matrix. Finally, we add these two penalty terms to~\eqref{eq:chi2} by replacing
\begin{align}
    \chi^2\to \chi^2 + \chi^2_{\mathrm{DF}} =  \chi^2 - 2 \ln f_{\mathrm{DFD}}(\rhocut, 0, 1) - 2 \ln f_{\mathrm{DFD}}(\rhomom, -0.45, 1)  \, .
\end{align}

\subsection{External constraints to the fit}\label{sec:const} 

We constrain four parameters in our fit with external input. To avoid the complication, that only a linear combination of $m_b$ and $m_c$ can be constrained in moment fits (cf. discussion in~\cite{Gambino:2013rza}), we introduce constraints for both parameters. We work in the kinetic mass scheme for $m_b$ and using $\overline{m}_b(\overline{m}_b) = 4.198 (12)$ GeV~\cite{pdg:2020} and $\overline{m}_c(3$~GeV$)=0.988(7) $ GeV~\cite{Aoki:2019cca}, we find \cite{Fael:2020njb,Fael:2020iea}
\begin{equation}\label{eq:mbkinin}
    m_b^{\rm kin}(1\; \rm{GeV}) = 4.565 \pm 0.015 \pm 0.013\; \rm{GeV} = 4.565 \pm 0.020\; \rm{GeV} \ .
\end{equation}
The first error is the theoretical uncertainty of the scheme conversion~\cite{Fael:2020tow} and the second stems from the $\overline{m}_b(\overline{m}_b)$ error.

We use the charm mass $m_c$ in the $\overline{\mathrm{MS}}$ scheme, whose value has been determined precisely in lattice QCD~\cite{Aoki:2019cca} and QCD sum rules~\cite{Chetyrkin:2009fv,Chetyrkin:2017lif} computations. Using \texttt{RunDec} with 4-loop accuracy, we obtain
\begin{equation}\label{eq:mc2input}
    \overline{m}_c(2\; \rm{GeV}) = 1.093 \pm 0.008\; \rm{GeV}  \ .
\end{equation}
The constraints for both masses are introduced into~\eqref{eq:chi2} assuming Gaussian uncertainties.

We introduce a constraint on $\mu_G^2$ exploiting the relationship between the measured mass difference between the vector meson $B^*$ and the $B$ meson and this parameter
\begin{equation}\label{eq:massdif}
    m_{B^*}^2-m_B^2= \frac{4}{3} C_\mathrm{mag}(\mu_s) \mu_G^2(\mu_s) + \mathcal{O}(\alpha_s \mu_G^2, 1/m_b^3)\ ,
\end{equation}
with $C_\mathrm{mag}=1$. We define $\mu_G^2\equiv \mu_G^2(\mu_s=m_b^\mathrm{kin})$.

We recall that our HQE parameters (cf. Eq.~\eqref{eq:MEs}) are defined between full QCD states that know about the $b$ quark mass. The mass difference is, on the other hand, defined in the infinite mass (HQE limit) of the states. The difference between these two setups of the OPE do introduce higher order terms in the relation~\eqref{eq:massdif}. Ignoring these unknown higher order (and non-local) contributions, we use the mass difference to obtain a constraint and find (see cf. \cite{Gambino:2012rd}) 
\begin{equation}\label{eq:mugin}
    \mu_G^2  = \left(0.36 \pm 0.07 \right) \rm{GeV}^2 \ .
\end{equation}
This constraint is also implemented as a Gaussian penalty term into~\eqref{eq:chi2}. 

As explained in Sec.~\ref{sec:depmu3}, the fit is not very sensitive to $\mu_\pi^2$. To constrain this blind direction, we introduce a conservative constraint of 
\begin{equation}
   \mu_\pi^2 =  (0.43 \pm 0.24)\; \rm{GeV}^2\ . 
\end{equation}
This constraint is obtained by using the value computed in~\cite{Gambino:2016jkc} and inflating the uncertainty by a factor of four. Note that this value is in good agreement with the recent determination from the \Vcb analysis~\cite{Bordone:2021oof}, which determined $\mu_\pi^2 =  0.477 \pm 0.056 \;\rm{GeV}^2$. In future determinations, alternative direct extractions of the RPI combination $\mu_3$ by other non-perturbative methods may be possible (see discussion in~\cite{Mannel:2018mqv}), but this is beyond the scope of this paper. Future progress in inclusive lattice QCD calculations may also be able to provide constraints for this and other HQE parameter, which then in turn may be included into a future fit (see e.g.~\cite{Gambino:2020crt, Gambino:2017vkx,Gambino:2022dvu}).

\subsection{Theoretical uncertainties}\label{sec:theun}

To estimate the theoretical uncertainties on the prediction for the $q^2$ moments, we carry out variations of $\rho_D^3$, $\mu_G^2$, and the scale of $\alpha_s$. These variations are compared to a set of nominal predictions for the moments. We use $m_b^{\rm kin}$ and $\overline{m}_c$ as quoted in \eqref{eq:mbkinin} and~\eqref{eq:mc2input}. For the strong coupling constant we use
\begin{equation}
\alpha_s(\mu_s=m_b^{\rm kin}) = 0.2184 \, .
\end{equation}
This value was obtained using \texttt{RunDec}~\cite{Steinhauser:rundec_original, Steinhauser:rundec_update} with $n_f = 4$ active flavours and five loop accuracy and the initial value of $\alpha_s^{(5)}(M_Z) = 0.1179 (9)$~\cite{pdg:2020}. 

We consider three kinds of theory uncertainties on the $q^2$ moment predictions and the total semileptonic rate:
\begin{itemize}
    \item  Missing higher-order corrections in $\alpha_s$. These are estimated by varying the scale of $\alpha_s(\mu_s)$ between $m_b^\mathrm{{kin}}/2<\mu_s <2m_b^\mathrm{kin}$. 
    \item Missing higher-order $1/m_b$ corrections. The uncertainties are estimated by varying $\rho_D^3$ by $30\%$.
    \item Missing $\alpha_s/m_b^{2,3}$ corrections. These are estimated by varying $\mu_G^2$ by $20\%$.  
\end{itemize}
The latter two variations mirror similar procedures implemented in \cite{Alberti:2014yda, Gambino:2013rza}. To construct the theory uncertainties, we use $\mu_G^2$ in \eqref{eq:mugin} and $\rho_D^3 = 0.127$ GeV$^3$ as found in Appendix~A of \cite{Fael:2018vsp}. 

These variations give a conservative estimate of the uncertainty. We stress that in the future, with for example partonic $\alpha_s^2$ and $\alpha_s/m_b^2$ corrections included, these variations may be reduced. Finally, we do not include uncertainties coming from the input values of $m_b$ nor $m_c$, but treat both as fixed parameters when estimating the theory errors. As these inputs are very precisely known, they would not contribute to the theory uncertainty.


\subsection{Experimental input}


\subsubsection{\boldmath The $B \to X_c \ell \bar \nu_\ell$ branching fraction}
\label{sec:branchin_fraction}

Knowledge of the branching ratio of $B \to X_c \ell \bar \nu_\ell$ is instrumental to determine $|V_{cb}|$. The current measurements of this branching ratio either quote partial branching fractions as a function of a lepton energy cut, or provide the full $B
 \to X \ell \bar{\nu}_\ell$ rate. In the latter case, the total rate is also measured with a cut on the lepton energy and then extrapolated to the full phase space. Unlike in the fits of~\cite{Gambino:2013rza,Bordone:2021oof}, we cannot directly make use of the partial branching fraction results as a lepton energy cut spoils RPI, whose validity is required to reduce the number of HQE parameters in the $q^2$ moments and the total rate. We therefore need to use the total branching fraction. As only a small number of results are available that quote the total branching fraction, we extrapolate also partial branching fraction results with a cut on the lepton energy $\mathcal{B}_{\rm cut}$ to the full phase space. This is done with a correction factor $\Delta(E_{\rm cut})$, such that
\begin{equation}
    \mathcal{B}(B \to X_c \ell \bar \nu_\ell) = \Delta(E_{\rm cut}) \mathcal{B}_{\rm cut}(B \to X_c \ell \bar \nu_\ell) \ .
\end{equation}
The factor $\Delta(E_{\rm cut})$ can be calculated reliably in the local OPE. Using the partial branching rate at leading order in the HQE and including perturbative corrections up to NLO, we find
\begin{equation}\label{eq:cor}
    \Delta(0.6\; {\rm GeV}) = 1.047 \pm 0.004   \ , \quad\quad \Delta(0.4\; {\rm GeV}) = 1.014 \pm 0.001 \ .
\end{equation}
The quoted uncertainty presents an estimate obtained by including the power corrections up to $1/m_b^3$ and evaluated with the values obtained in \cite{Bordone:2021oof}. Additional uncertainties due to the variation of the renormalization scale $\mu_s$ are negligible. We note that, as expected, this correction factor is rather insensitive to power corrections due to the small extrapolation region. 

In Table~\ref{tab:branc}, we list the available measurements for $B \to X \ell \bar \nu_\ell$ and $B \to X_c \ell \bar \nu_\ell$  branching fractions and quote the extrapolations to the full phase space. We correct the branching fractions of $B \to X \ell \bar \nu_\ell$, by subtracting the $|V_{ub}|^2$ suppressed $B \to X_u \ell \bar \nu_\ell$ contribution via 
\begin{align}
   \mathcal{B}(B \to X_c \ell \bar \nu_\ell) = \mathcal{B}(B \to X \ell \bar \nu_\ell) - \Delta\mathcal{B}(B \to X_u \ell \bar \nu_\ell) / \epsilon_{\Delta \mathcal{B}} \, .
\end{align}
We use $\Delta\mathcal{B}(B \to X_u \ell \bar \nu_\ell) = (0.159 \pm 0.017)\%$ from~\cite{Belle:2021eni}, which has a lepton-energy cut of $E_\ell > 1 \, \mathrm{GeV}$. To correct for this cut, we use a factor of $\epsilon_{\Delta \mathcal{B}} = 0.858 \pm 0.008$ \footnote{We thank Lu Cao, author of \cite{Belle:2021eni}, for providing this correction.}. This results in a total $B \to X_u \ell \bar \nu_\ell$ branching fraction of $\mathcal{B}(B \to X_u \ell \bar \nu_\ell) = (0.185 \pm 0.020)\%$.

Averaging the indicated measurements listed in Table~\ref{tab:branc}, we obtain
\begin{align}\label{eq:btoxc}
    \mathcal{B}(B \to X_c \ell \bar \nu_\ell) = (10.48 \pm 0.13)\cdot 10^{-2} \, ,
\end{align}
which we use as our default branching ratio. 

As the central value of the branching ratio will dominate the central value of $V_{cb}$, it is worth commenting on it. Our default value differs by $\approx 1.4\, \sigma$ from the value obtained in \cite{Bordone:2021oof} which is $(10.66\pm0.15) \cdot 10^{-2}$. This difference is caused most notably by the inclusion of the Babar measurement~\cite{BaBar:2016rxh} into our average, while \cite{Bordone:2021oof} only include the Belle~\cite{Belle:2006kgy} and BaBar~\cite{BaBar:2009zpz} results. In their analysis, the total branching ratio of $B\to X_c \ell \bar\nu_\ell$ is a prediction of the fit, determined from the used partial branching ratio measurements at different $E_\ell$ cuts and the analyzed moments. This approach allows for a self-consistent extrapolation. To allow for an easier comparison with their results, we also determine \Vcb using an average based on the same branching fractions~\cite{Belle:2006kgy,BaBar:2009zpz}:
\begin{equation}\label{eq:btoxc2}
        \mathcal{B}(B \to X_c \ell \bar \nu_\ell) = (10.63 \pm 0.19)\cdot 10^{-2} \ .
\end{equation}
This value is in excellent agreement with the value obtained by Ref.~\cite{Bordone:2021oof}. 

\begin{table}[t!]
\vspace{1ex}
\begin{tabular}{lccc}
\hline
  & $\mathcal{B}(B \to X \ell \bar \nu_\ell)$ (\%)  & $\mathcal{B}(B \to X_c \ell \bar \nu_\ell)$  (\%) & In Average \\
 \hline
 Belle~\cite{Belle:2006kgy} $E_\ell > 0.6 \, \mathrm{GeV}$ &  - & $10.54 \pm 0.31$ & \checkmark \\
 Belle~\cite{Belle:2006kgy} $E_\ell > 0.4 \, \mathrm{GeV}$ & -  & $10.58 \pm 0.32$&  \\
 CLEO~\cite{CLEO:2004stg} incl.          &  $10.91 \pm 0.26$ & $10.72 \pm 0.26$ \\
 CLEO~\cite{CLEO:2004stg} $E_\ell > 0.6$  & $10.69 \pm 0.25$ & $10.50 \pm 0.25$ & \checkmark  \\
 BaBar~\cite{BaBar:2016rxh} incl.         &  $10.34 \pm 0.26$ & $10.15 \pm 0.26$ & \checkmark  \\
  BaBar SL~\cite{BaBar:2009zpz} $E_\ell > 0.6\, \mathrm{GeV}$ & - & $10.68 \pm 0.24$ &  \checkmark  \\ 
 \hline
 Our Average & - & $10.48 \pm 0.13$ \\
 \hline
 Average Belle~\cite{Belle:2006kgy} \&  BaBar~\cite{BaBar:2009zpz} & -  & $10.63 \pm 0.19$ \\ 
 ($E_\ell > 0.6\, \mathrm{GeV}$) & & \\
 \hline
\end{tabular}
\caption{
	Available measurements of the inclusive $B \to X \ell \bar \nu_\ell$ and $B \to X_c \ell \bar \nu_\ell$  branching fractions, extrapolated to the full region using the correction factors in \eqref{eq:cor}. The $\chi^2$ of our average with respect to the included measurements is $2.2$, corresponding to a p-value of 52\%. We do not include~\cite{BaBar:2006ztv}, as the analysis does not quote a partial branching fraction corrected for FSR radiation.
}
\label{tab:branc}
\end{table}

We conclude this section by stressing that new branching ratio measurements are imperative to clarify the mild tension between these two averages. In addition, new branching ratio measurements with (different) $q^2$ thresholds would be the natural input for the RPI \Vcb determination. Further measurements of $B\to X \ell \bar{\nu}_\ell$ would be highly desirable, such that they can be directly used in our analysis by implementing both the $b\to c$ and $b\to u$ in the local OPE \cite{Mannel:2021mwe}. This way uncertainties from subtracting the $b\to u$ contributions can be avoided, increasing the experimental precision. We plan to implement this strategy in a future version of our analysis. 

\subsubsection{\boldmath Belle and Belle~II $q_n$ measurements}\label{sec:bandb2}
\begin{table}[t!]
\vspace{1ex}\centering
\begin{tabular}{lc}
\hline
  & $q^2_\mathrm{cut} \; [{\rm{GeV}}^2]$   \\
 \hline
 Belle~\cite{Belle:2021idw} & $3,4,5,6,7,8$ 
 \\
 Belle~II~\cite{The:2022cbm} &1.5, 2.5, 3.5, 4.5, 5.5, 6.5, 7.5   \\
 \hline
\end{tabular}
\caption{Analyzed measured $q^2$ moments from Belle and Belle~II.}
\label{tab:subset}
\end{table}

Belle \cite{Belle:2021idw} and Belle~II \cite{The:2022cbm} recently presented first measurements of $q^2$ moments.
The Belle measurement presents separate moments for electrons and muons with a minimal $q^2$ threshold selection of \SI{3.0}{GeV^2} up to a maximum value of \SI{10.5}{GeV^2}. We average the Belle measured electron and muon $q^2$ moments, fully correlating identical systematic uncertainty sources. 
Belle~II provides measurements with a $q^2$ threshold starting at \SI{1.5}{GeV^2} up to \SI{8.5}{GeV^2}. Due to the high degree of correlations between the measured moments, we do not analyze the full set of moments of each experiment, but focus on the subset listed in Table~\ref{tab:subset}. We stress that the result of the analysis is fairly insensitive on which precise subset is chosen. We do not use moments with thresholds larger than \SI{8}{GeV^2}, to avoid the endpoint of the $q^2$ spectrum, whose contributions are dominated by exclusive states. We fully correlate the systematic uncertainties between both measurements related to the composition of the $X_c$ system and form factor uncertainties.


%% file: results.tex
\subsection{\boldmath First $V_{cb}$ determination from $q^2$ moments}\label{sec:default}

\begin{table}[t!]
    \centering
    \vspace{0.5em}
    \input{values_combined_central.tex}
        \caption{Results of our default fit using both Belle and Belle~II data for $\absVcb$, $m_b^{\rm kin}(1\;{\rm GeV})$, $\overline{m}_c(2\;{\rm GeV})$, the HQE parameters, and the correlation parameters $\rho_\mathrm{cut}$ and $\rho_\mathrm{mom}$. All parameters are expressed in GeV at the appropriate power. 
        }
         \label{tab:parameter_values_post_fit2}
\end{table}

In our default fit, we combine the first four central moments of the Belle and Belle~II measurements and the branching ratio Eq.~\eqref{eq:btoxc}. The parameters \absVcb, \rhoD, \rG, and  \rE are free parameters in the fit, while \mb, \mc, \muG, and \mupi are included with Gaussian constraints as discussed in Section~\ref{sec:const}. We set the parameters \sE, \sB, and \sqB, to which we have limited sensitivity, to zero (cf. discussion in Section~\ref{sec:m4considerations}), and we introduce the correlation parameters $\rhomom$ and $\rhocut$ as nuisance parameters in the fit. The result of this fit is reported in Table~\ref{tab:parameter_values_post_fit2}, while the correlation between the fitted parameters is given in Fig.~\ref{fig:correlation_matrix} in Appendix~\ref{app:addimat}. We find
\begin{align}\label{eq:ourresult}
    \absVcb &= (41.69 \pm 0.27|_{\mathcal{B}} \pm 0.31|_{\Gamma} \pm 0.18|_{\rm Exp.} \pm 0.17|_{\rm Theo.} \pm 0.34|_{\rm Constr.})  \cdot 10^{-3} \ , \nonumber \\
    &= (41.69 \pm 0.59)  \cdot 10^{-3} \ ,
\end{align}
where the uncertainties stem from the experimental branching ratio $\mathcal{B}$, the theoretical uncertainty on the total rate $\Gamma$, the experimental and theoretical uncertainties on the $q^2$ moments and the uncertainty from the external constraints. This first determination shows the potential of this method, allowing to determine $V_{cb}$ with percent level uncertainty. 

\begin{figure}[t!]
    \centering
    \includegraphics[width=0.4\textwidth]{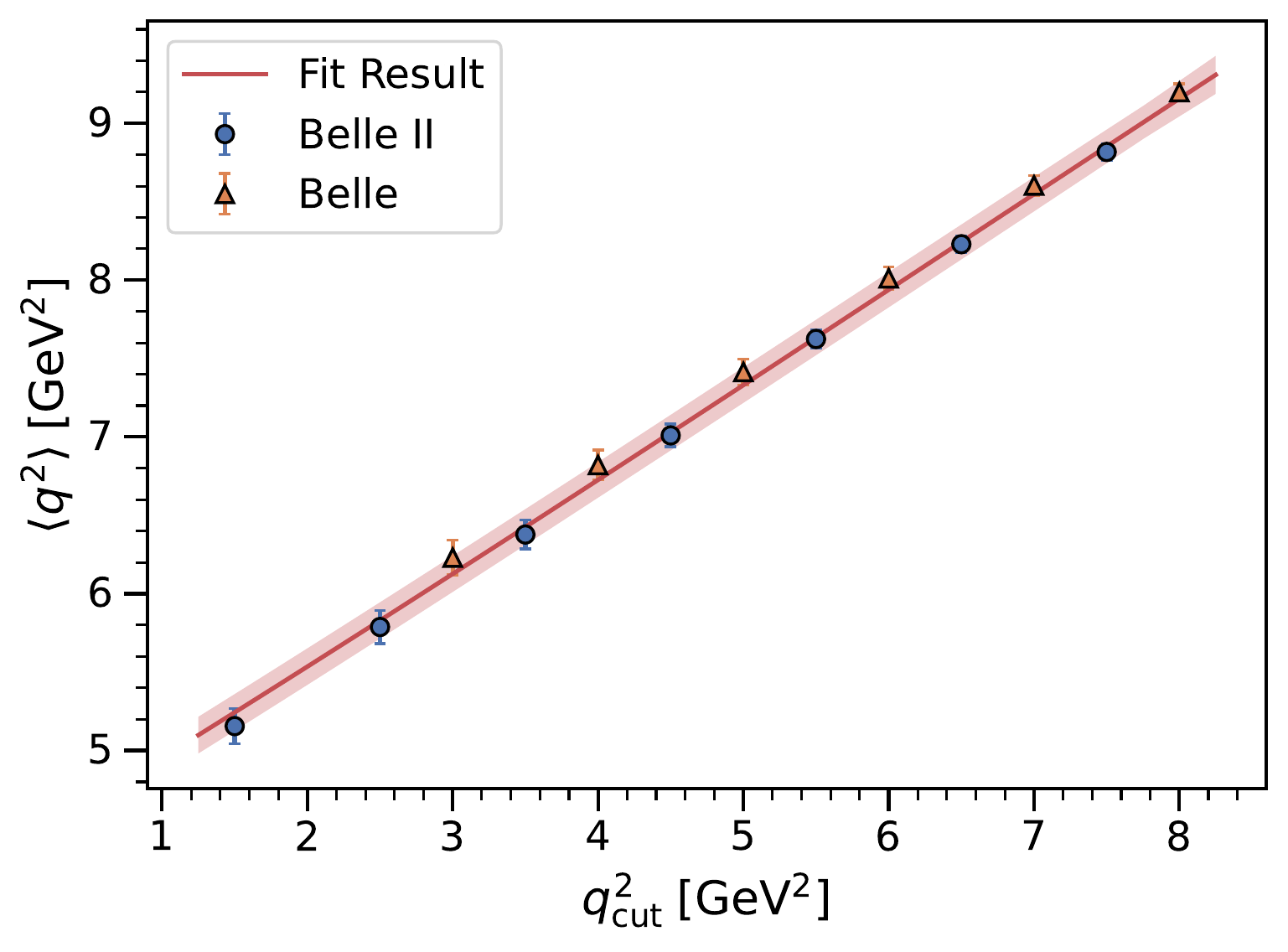}
    \includegraphics[width=0.4\textwidth]{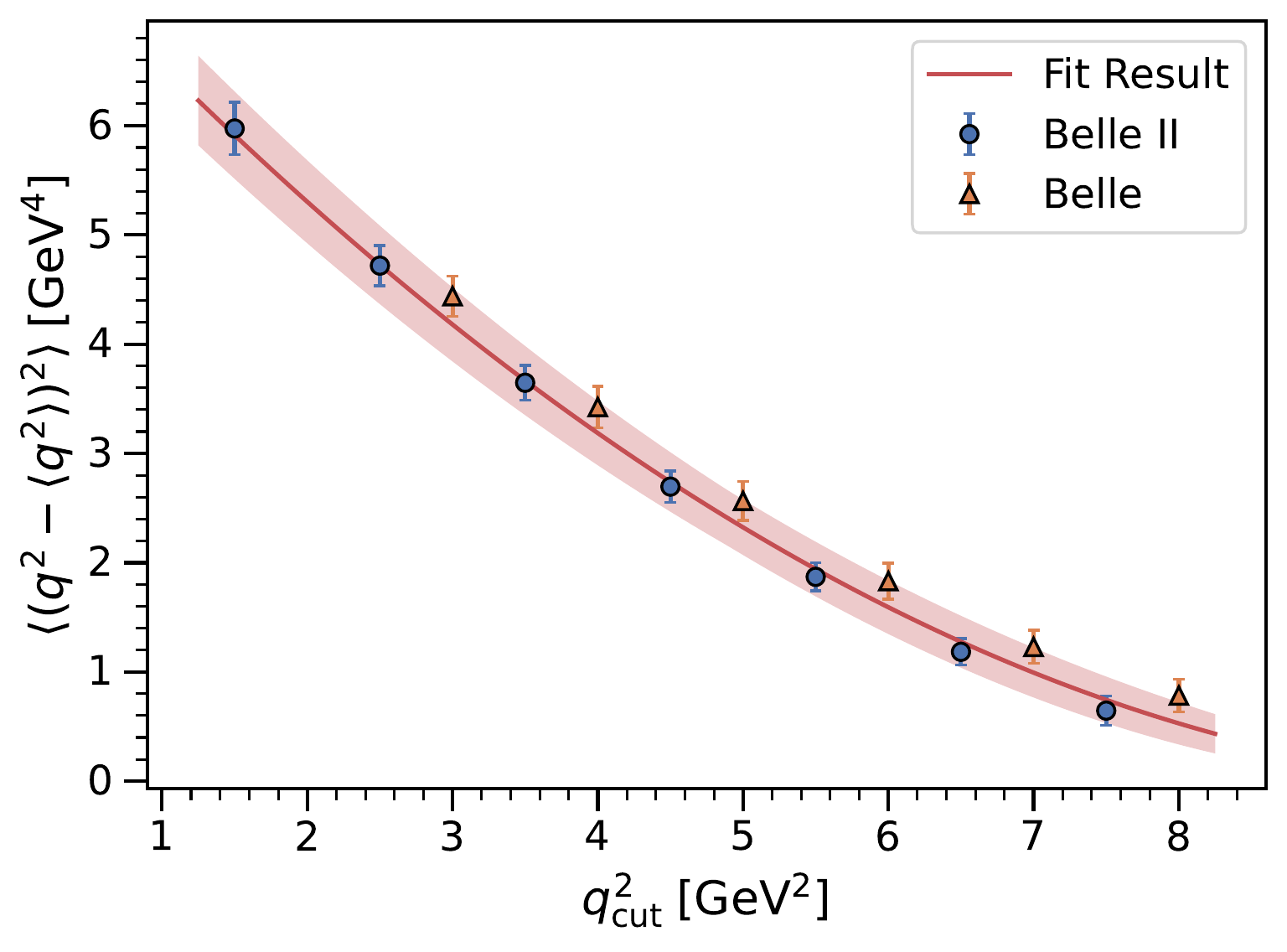}\\
    \includegraphics[width=0.4\textwidth]{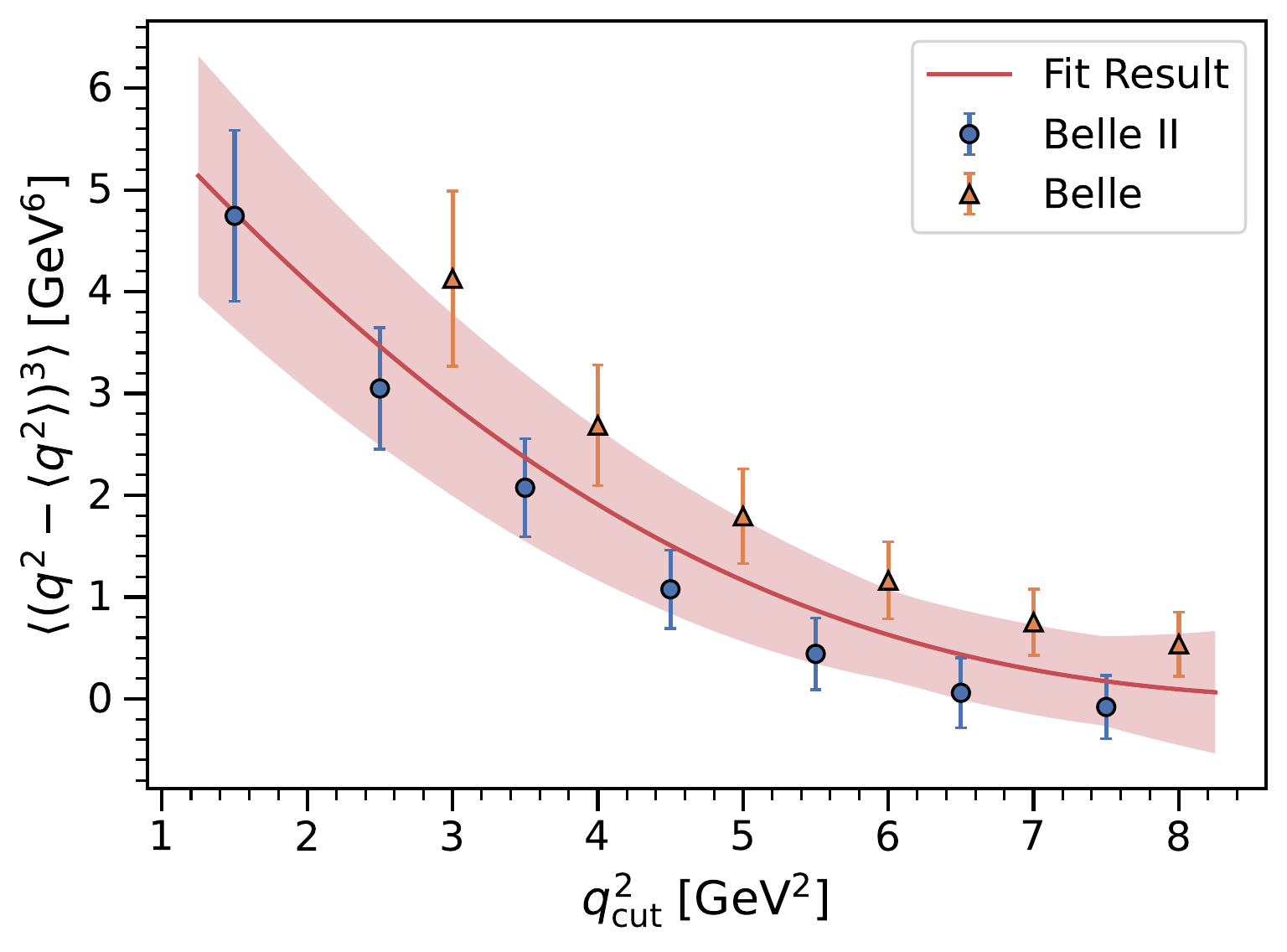}
    \includegraphics[width=0.4\textwidth]{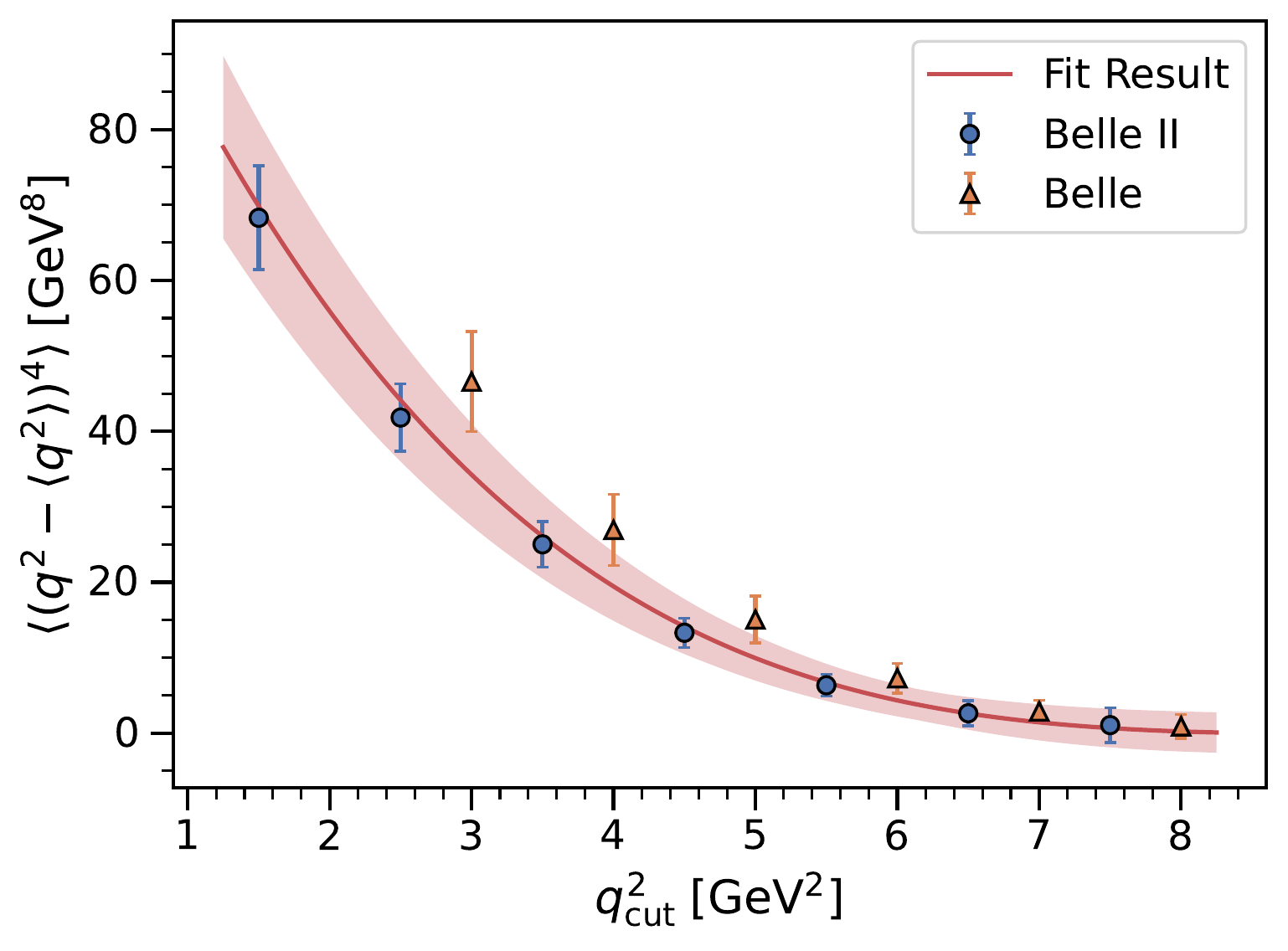}
    \caption{Fit projections for the central $q^2$ moments as a function of the $q^2$ threshold, combined with the measurement moments from both Belle and Belle~II.}
    \label{fig:central_moments_post_fit}
\end{figure}

\begin{figure}[t]
    \centering
    \includegraphics[width=0.4\textwidth]{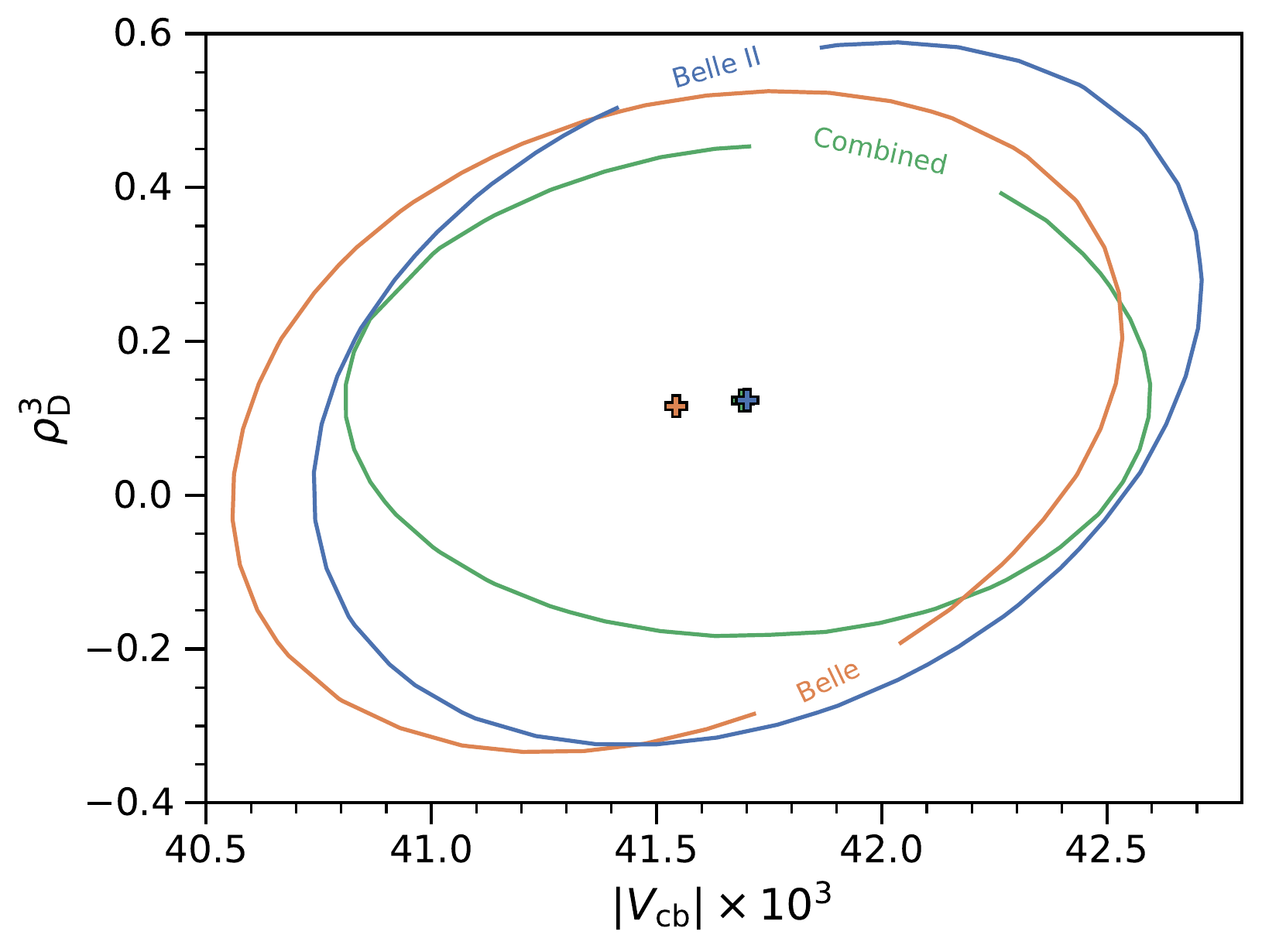}
    \caption{Comparison between Belle, Belle~II and the combined fit for the correlation between \absVcb and \rhoD. The crosses indicate the best-fit points.
    }
    \label{fig:likelihood_profiles_2d_comb_belle_belleII}
\end{figure}

In Fig.~\ref{fig:central_moments_post_fit}, we show the fit results for the four central $q^2$ moments as a function of the $q^2$ threshold, together with the analyzed Belle and Belle~II measurements. The fit has $\chisqmin=7.17$ with 49 degrees of freedom (dof), indicating an excellent fit. As Refs.~\cite{Bordone:2021oof,Alberti:2014yda,Gambino:2013rza} we recover a poor fit $\chi^2 / \mathrm{dof} = 5.02$ without the inclusion of a theory covariance into the fit. In appendix~\ref{sec:belleonly}(\ref{sec:belle2only}), we also present results using only Belle or Belle~II data. The obtained \absVcb and \rhoD values are compared in Fig.~\ref{fig:likelihood_profiles_2d_comb_belle_belleII}, indicating good agreement between the combined Belle and Belle~II and individual fits. 

Finally, we note that for a different input of the $B\to X_c\ell\bar\nu_\ell$ branching ratio, the corresponding $V_{cb}$ value can be easily obtained by a rescaling:
\begin{equation}
    |V_{cb}|  = \sqrt{\frac{\mathcal{B}(B\to X_c \ell\bar\nu)}{ (10.48 \pm 0.13)\% }} \times (41.69 \pm 0.59)  \cdot 10^{-3}\ .
\end{equation}
Using \eqref{eq:btoxc2}, we find
\begin{align}
    \absVcb &= (41.99 \pm 0.65)  \cdot 10^{-3}  \ .
\end{align}
This value is in excellent agreement with the results \cite{Bordone:2021oof}, especially when considering their fit results which include estimates for the power corrections, which decreases the value given in \eqref{eq:vcb1} by $0.25\%$. 

\subsection{\boldmath Theory correlations and HQE parameters versus $V_{cb}$}
In our default fit, we introduce the two correlation parameters $\rhomom$ and $\rhocut$. In Table~\ref{tab:parameter_values_post_fit2}, we report the best-fit point for these parameters. In Appendix~\ref{app:addimat}, the one dimensional $\chi^2$ profiles of these parameters are also given, showing a non-parabolic behaviour. To estimate the  confidence intervals quoted in Table~\ref{tab:parameter_values_post_fit2}, we use ensembles of pseudo-experiments. Of particular interest is also the effect of the chosen values of \rhomom and \rhocut on $V_{cb}$.
 
\begin{figure}[tb]
    \centering
    \includegraphics[width=0.4\textwidth]{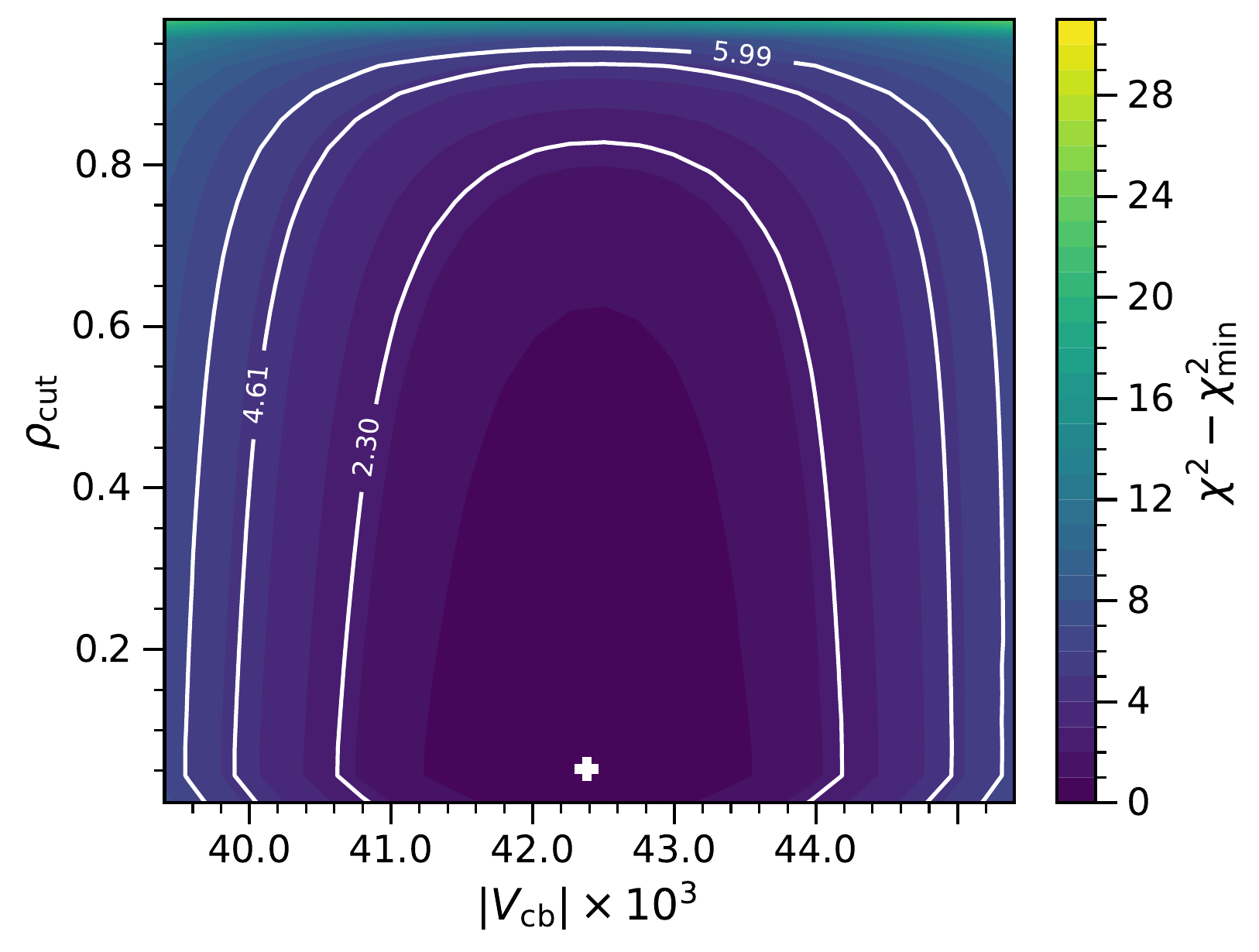}
    \includegraphics[width=0.4\textwidth]{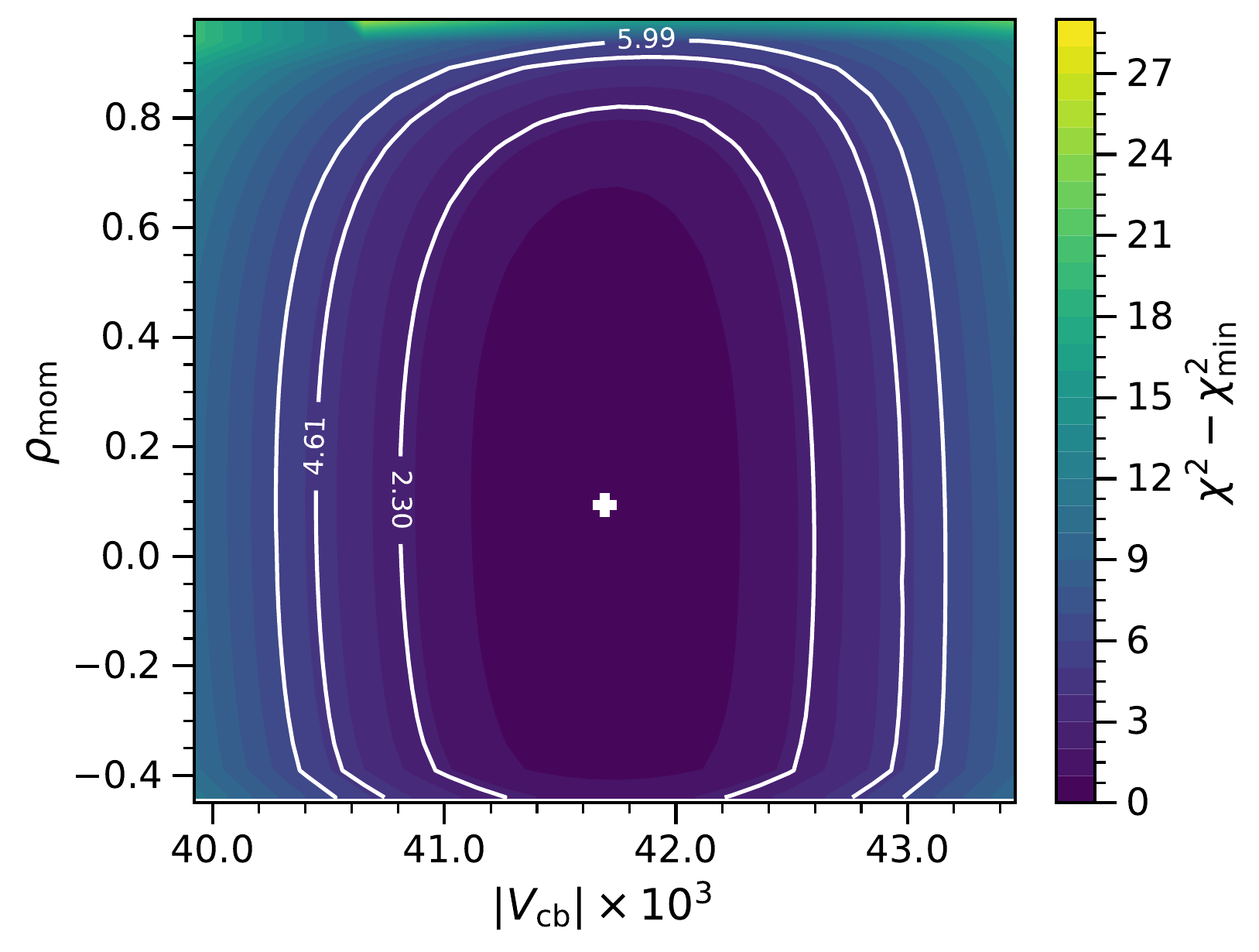}
    \caption{2D $\chi^2$ profile scans of \absVcb versus $\rhocut$ and $\rhomom$. The minimum $\chi^2_\mathrm{min}$ is subtracted from the $\chi^2$ function.}
    \label{fig:likelihood_profiles_2d_vcb_corr}
\end{figure}

To investigate this, we show in  Fig.~\ref{fig:likelihood_profiles_2d_vcb_corr} the 2D scans of the $\Delta \chi^2 = \chi^2 - \chi^2_{\mathrm{min}}$ contours for $V_{cb}$ versus \rhocut and \rhomom. Profiling over a large range of both \rhomom and \rhocut only has a small impact on the determined value of $V_{cb}$. Or phrased differently, our uncertainty on $V_{cb}$ includes a large range of possible correlation coefficients. We therefore conclude that $V_{cb}$ is very stable with respect to both of these nuisance parameters. As these parameters are a priori unknown this is an important finding. 

\begin{figure}[tb]
    \centering
    \includegraphics[width=0.3\textwidth]{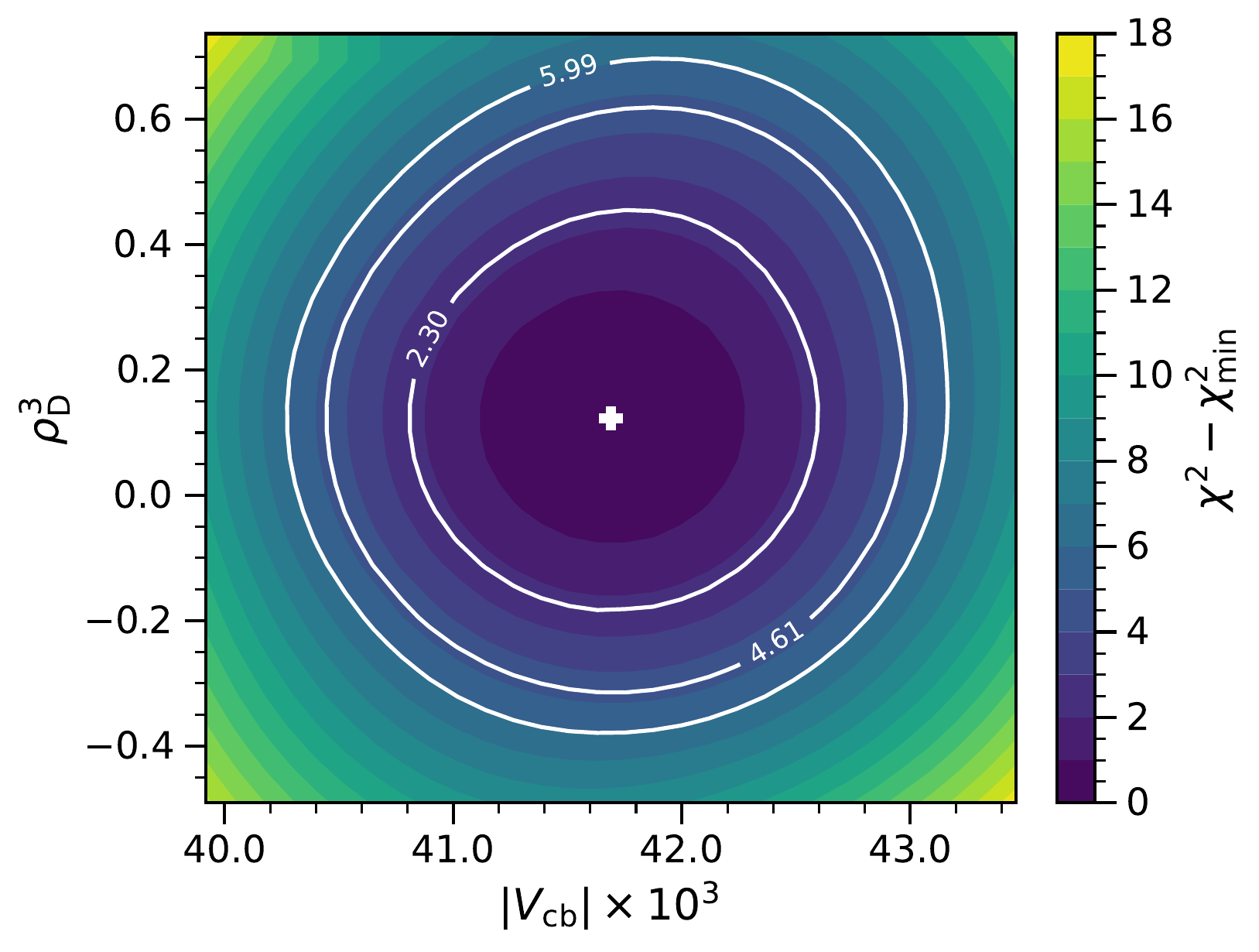}
    \includegraphics[width=0.3\textwidth]{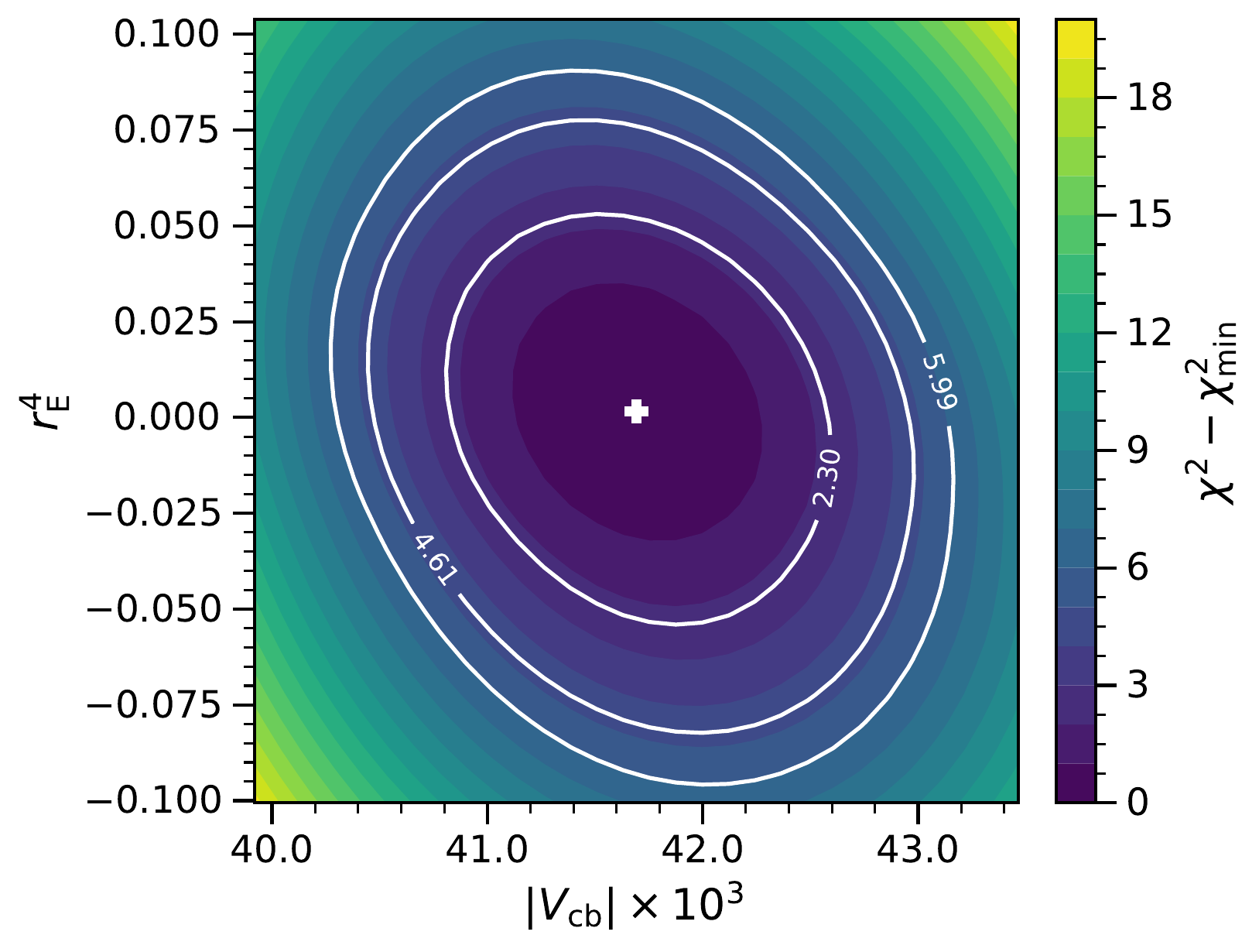}
    \includegraphics[width=0.3\textwidth]{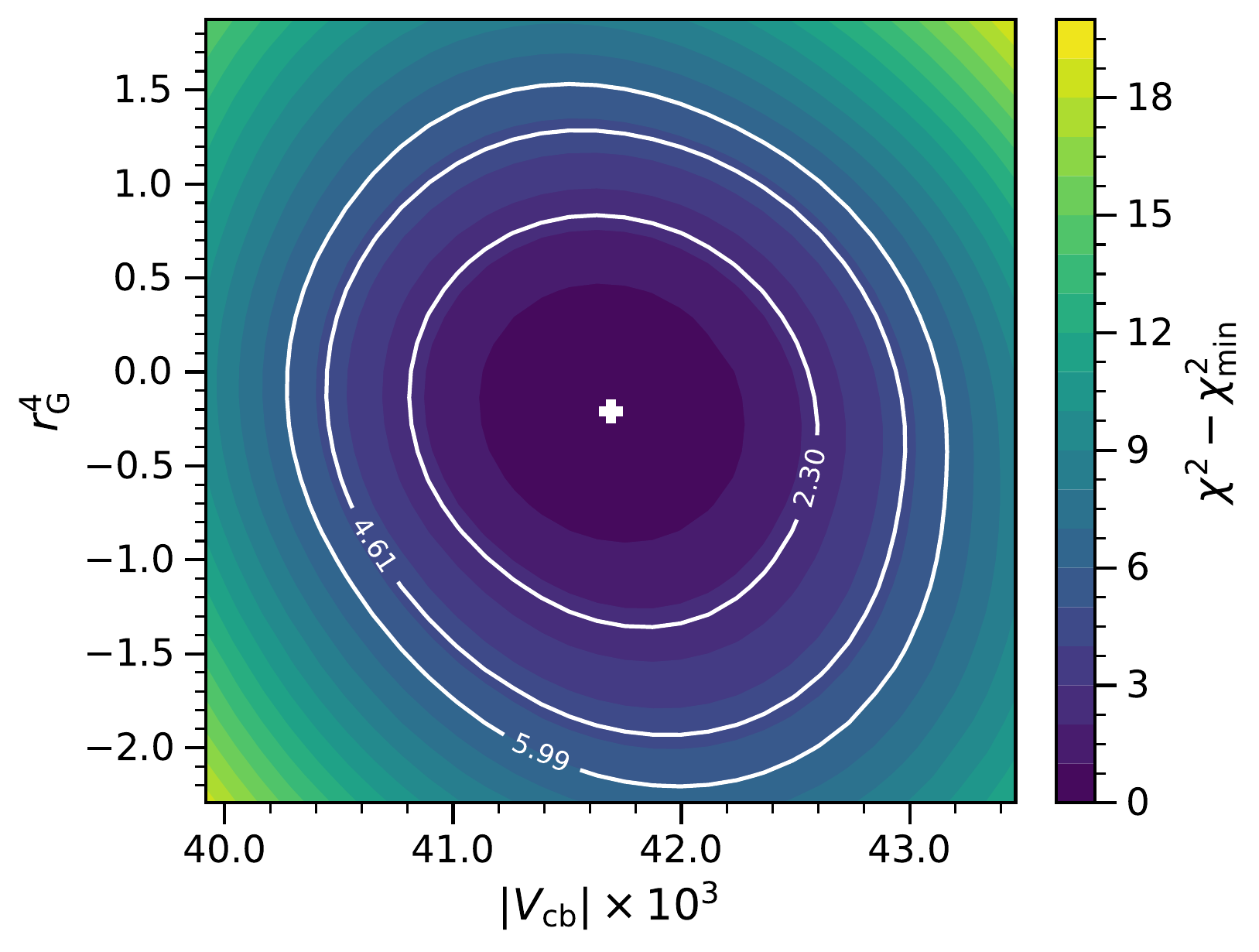}
    \caption{2D $\chi^2$ profile scans of \absVcb versus $\rho_D^3, r_E^4$ and $r_G^4$.  The minimum $\chi^2_\mathrm{min}$ is subtracted from the $\chi^2$ function.}
    \label{fig:likelihood_profiles_2d_vcb}
\end{figure}

In Fig.~\ref{fig:likelihood_profiles_2d_vcb}, we show the two-dimensional profile scan of $\Delta \chi^2$ between \rhoD, \rE and \rG versus $V_{cb}$. No sizeable correlations are observed.

For completeness, we also performed fits for fixed values of \rhomom and \rhocut. The fit results for $V_{cb}$, $\rhoD,\rE$ and \rG are given in Appendix~\ref{sec:corrscan}. These scans also show, that $V_{cb}$ is stable against variations of \rhomom and \rhocut. A similar conclusion was presented in \cite{Gambino:2013rza}.

\subsection{\boldmath Determination of the HQE parameters}\label{sec:det_hqe_pars}
Besides our determination of $V_{cb}$, also the determination of the HQE parameter is interesting. In our fit, both $\mu_\pi^2$ and $\mu_G^2$ are constrained by external inputs, and our fit results in Table~\ref{tab:parameter_values_post_fit2} show that we have limited sensitivity to constrain both parameters beyond their input values. On the other hand, in our default fit scenario, $\rhoD,\rE$ and \rG are free parameters that can be determined from the $q^2$ moments. We find
\begin{equation}
    \rhoD= (0.12 \pm 0.12|_{\rm Exp.}\pm 0.13|_{\rm Theo.}\pm 0.11|_{\rm Constr.})\;{\rm GeV}^4 = (0.12 \pm 0.20)\; {\rm GeV}^3  ,
\end{equation}
where the uncertainty stems from the experimental and theoretical uncertainty on the moments and on the external inputs, respectively. 
For the $1/m_b^4$, we are able to constrain \rE and \rG for the first time completely in a data-driven way. We find
\begin{align}
    \rE & = (0.02 \pm 0.21|_{\rm Exp.}\pm 0.27|_{\rm Theo.}\pm 0.00|_{\rm Constr.})\cdot 10^{-1}\;{\rm GeV}^4 \nonumber \\
    & = (0.02 \pm 0.34)\cdot 10^{-1}\;{\rm GeV}^4 \ , \\
     \rG &= (-0.21 \pm 0.42|_{\rm Exp.}\pm 0.49|_{\rm Theo.}\pm 0.25|_{\rm Constr.}) \;{\rm GeV}^4 \nonumber \\
    & = (-0.21 \pm 0.69)\;{\rm GeV}^4  \ .
\end{align}
Both values are small and compatible with zero within their uncertainties. We note that \rE $-$ the parameter we are most sensitive to $-$ is constrained to be very well below 1 GeV$^4$ or even $\Lambda_{\rm QCD}^4$, and also our results of \rG exclude spuriously (unexpected) large values for these parameters. This implies that the HQE seems well behaved and holds up to this order. 

\begin{figure}[tb]
    \centering
    \includegraphics[width=0.3\textwidth]{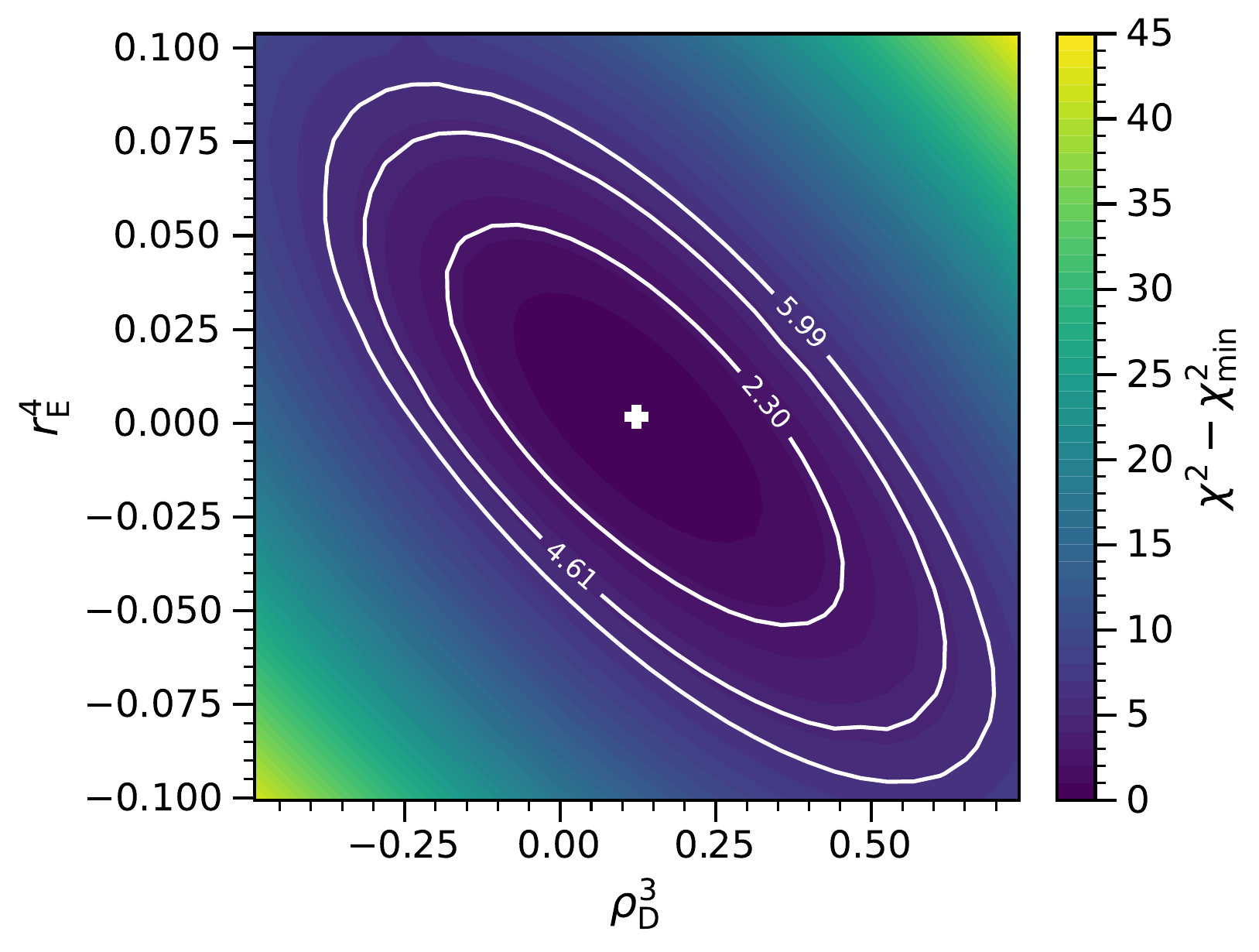}
    \includegraphics[width=0.3\textwidth]{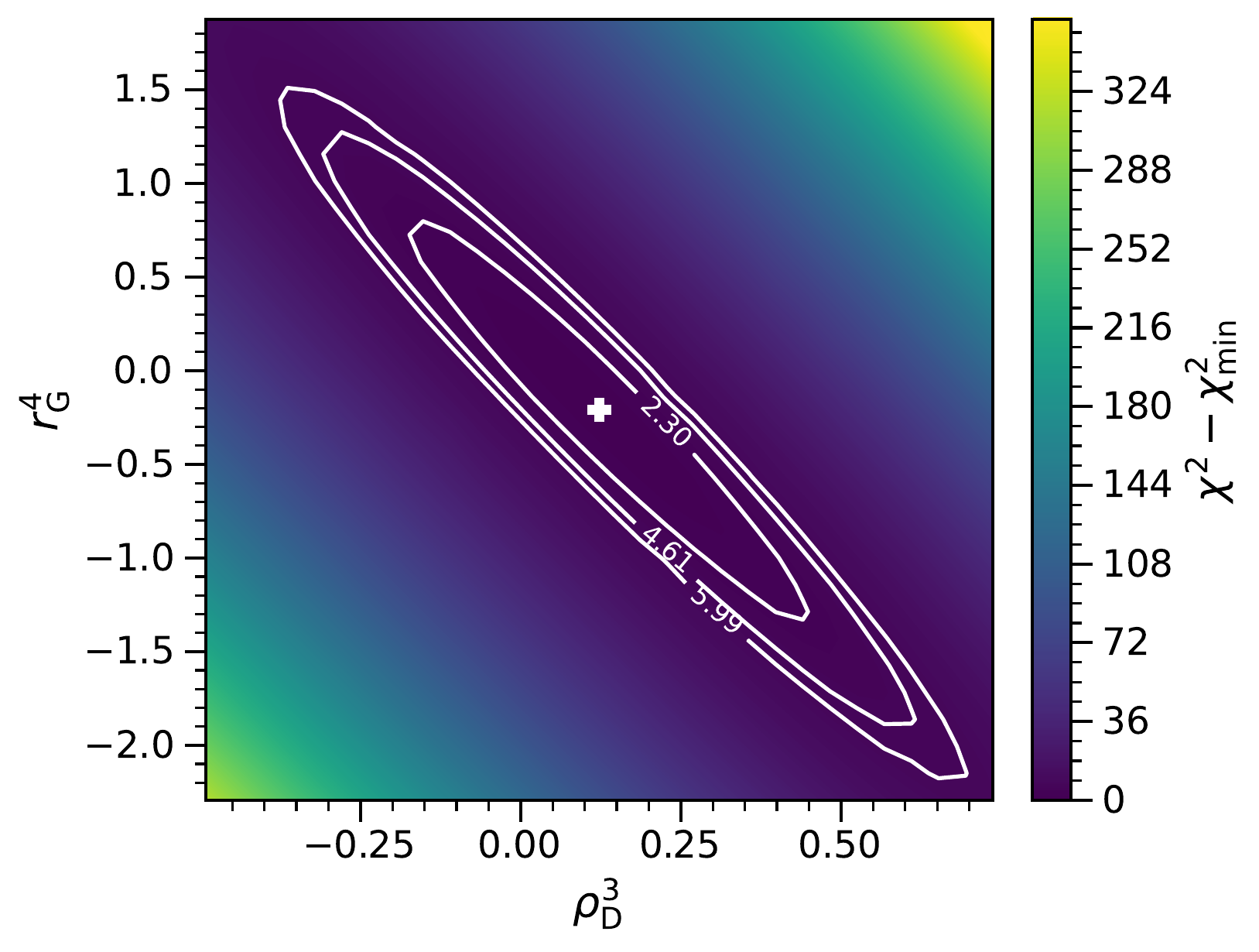}
       \includegraphics[width=0.3\textwidth]{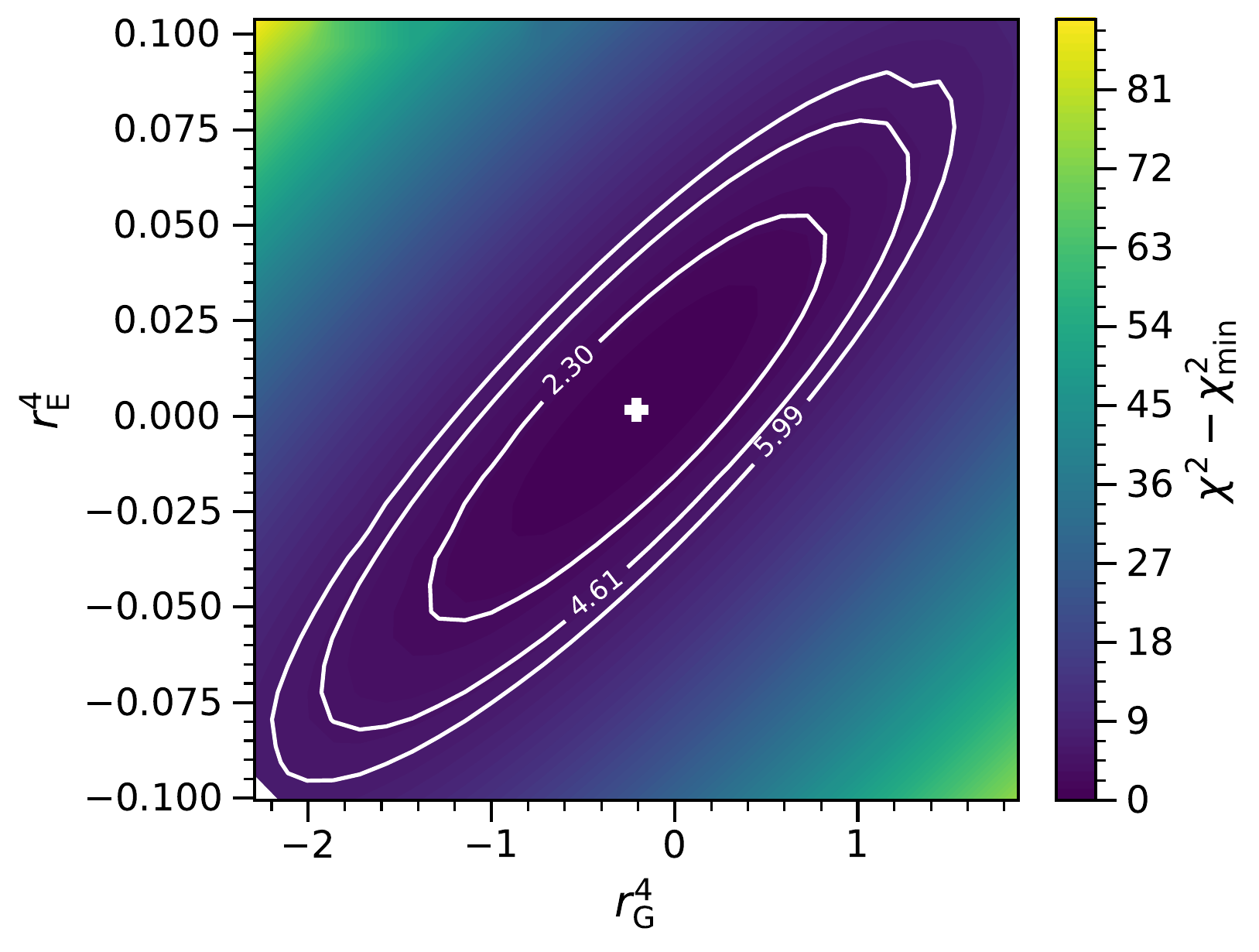}
    \caption{2D $\chi^2$ profile scans of \rhoD versus (left) \rE and (middle) \rG, and of \rE versus \rG (right). The minimum $\chi^2_\mathrm{min}$ is subtracted from the $\chi^2$ function.}
    \label{fig:likelihood_profiles_2d_rhod}
\end{figure}

Our determination of \rhoD is compatible with zero. From Appendix~\ref{sec:corrscan}, we observe that this holds true also for all fixed \rhomom and \rhocut choices. We note that our definition of \rhoD differs from that used in \cite{Bordone:2021oof}, because we include an additional $1/m_b$ contribution, its RPI completion  (see Appendix~\ref{app:sec:HQEpara} and \cite{Mannel:2018mqv}). Therefore, their result, which reads $\rhoD =(0.185\pm0.031) \rm{GeV}^3$ \cite{Bordone:2021oof}, cannot be directly compared to ours. However, it is worth exploring the apparent difference in sensitivity for \rhoD between the $q^2$ method and the moment fit of Ref.~\cite{Bordone:2021oof}. To understand this differences, we investigated the two-dimensional contours in $\Delta \chi^2$ of $\rhoD : \rE$ and \rG and of $\rE: \rG$. These scans are given in the left, middle and right panel of Fig.~\ref{fig:likelihood_profiles_2d_rhod}, respectively. The contours reveal a large (anti)correlation between these three parameters. In fact, the shape changes of the moments as function of $q^2_{\rm cut}$ are similar for these three parameters, indicating that they can compensate each other. Especially the shape changes induced by different value of \rhoD and \rG are very similar. This seems to indicate that at least in this analysis, we are only sensitive to a linear combination of these parameters. This combined with rather conservative theoretical uncertainties may explain our limited sensitivity to \rhoD. Although all these parameters are independent in the HQE, this may be a hint that a further reduction of parameters is possible, for example when expression everything in full QCD states (see discussion in \cite{Mannel:2018mqv}). This interesting observation deserves further study, which we leave for future work. 

On the other hand, for a more direct comparison with \cite{Bordone:2021oof}, we may also consider a fit with all $1/m_b^4$ corrections set to zero. We find the results listed in Table~\ref{tab:parameter_values_mb4_zero_post_fit}, specifically $\rhoD = 0.03\pm 0.02$. This determination differs substantially from that obtained by \cite{Bordone:2021oof} and may indicate that the $q^2$ moments actually add additional information on the HQE parameters.

\subsection{\boldmath Including all $1/m_b^4$ terms}

\begin{table}[t]
    \centering
    \vspace{0.5em}
    \resizebox{\textwidth}{!}{%
        \input{values_combined_central_mb4_constrained.tex}
    }
        \caption{Fit result including all $1/m_b^4$ parameters with a standard normal Gaussian constraint. All parameters are expressed in GeV at the appropriate power. }
            \label{tab:parameter_values_post_fit_mb4_constrained}
\end{table}

In our default fit, we only consider the $1/m_b^4$ terms \rE and \rG. To study the effect of the neglected parameters, we also perform a fit including \sB, \sE and \sqB. In this case, we have to include a constraint on the parameters to obtain a well-behaved fit. We include standard normal Gaussian constraints (i.e. mean of zero, standard deviation one) for all $1/m_b^4$ terms. The result is given in Table~\ref{tab:parameter_values_post_fit_mb4_constrained} and
we observe no significant deviations from the default fit results. As expected, this fit shows that the most sensitive $\mathcal{O}(1/\mb^4)$ HQE parameters are \rG and \rE, since the post-fit parameter uncertainties can be reduced from unity.
For the remaining $\mathcal{O}(1/\mb^4)$ HQE parameters, no significant uncertainty reduction is seen. Most importantly, we obtain exactly the same $V_{cb}$ value as from our default fit:
\begin{equation}
    |V_{cb}| = (41.69 \pm 0.59) \cdot 10^{-3} \ .
\end{equation}

To account for missing higher-order corrections from neglecting \sE, \sB and \sqB, we add an additional uncertainty to our default fit in \eqref{eq:ourresult} by including variations of these parameters by $\pm 1$ GeV$^4$. 
In total, we find a variation of $V_{cb}$ of $0.23\cdot 10^{-3}$, dominated by the contribution of \sE. Our final default result is therefore
\begin{equation}
    |V_{cb}| = (41.69 \pm 0.59|_{\rm fit} \pm 0.23|_{\rm h.o.}) \cdot 10^{-3} = (41.69 \pm 0.63) \cdot 10^{-3} \ ,
\end{equation}
where we added both uncertainties in quadrature.

%% file: values_combined_central.tex
\begin{tabular}{lrrrrrrrrrr}
\toprule
{} &  $|V_{cb}| \times 10^3$ &  $m_b^{\rm kin} $ &  $\overline{m}_c $ &  $\mu_G^2 $ &  $\mu_\pi^2 $ &  $\rhoD $ &  $\rG $ &  $ \rE \times 10$ &  $\rho_\mathrm{cut}$ &  $\rho_\mathrm{mom}$ \\
\midrule
Value       &                          41.69 &             4.56 &             1.09 &                 0.37 &          0.43 &                  0.12 &              -0.21 &                        0.02 &                 0.05 &                 0.09 \\
Uncertainty &                           0.59 &             0.02 &             0.01 &                 0.07 &          0.24 &                  0.20 &               0.69 &                        0.34 &                $^{+0.03}_{-0.01}$
&                 $^{+0.10}_{-0.10}$ \\
\bottomrule
\end{tabular}

%% file: values_combined_central_mb4_constrained.tex
\begin{tabular}{lrrrrrrrrrrrrr}
\toprule
{} &  $|V_{cb}| \times 10^3$ &  $m_b^{\rm kin} $ &  $\overline{m}_c $ &  $\mu_G^2 $ &  $\mu_\pi^2 $ &  $\rhoD $ &  $\rG $ &  $\rE \times 10$ &  $s_E^4 $ &  $s_{qB}^4 $ &  $s_B^4 $ &  $\rho_\mathrm{cut}$ &  $\rho_\mathrm{mom}$ \\
\midrule
Value       &                          41.69 &             4.56 &             1.09 &                 0.37 &          0.43 &                  0.10 &              -0.12 &                        0.04 &              -0.04 &               -0.02 &               0.04 &                 0.05 &                 0.10 \\
Uncertainty &                           0.59 &             0.02 &             0.01 &                 0.07 &          0.24 &                  0.18 &               0.68 &                        0.31 &               0.95 &                0.99 &               0.95 &                  $^{+0.03}_{-0.01}$ &                  $^{+0.10}_{-0.10}$ \\
\bottomrule
\end{tabular}

%% file: app.tex
\section{HQE parameters}\label{app:sec:HQEpara}
In this appendix, we list the HQE parameters up to order $1/m_b^4$ as given in \cite{Mannel:2018mqv}:
\begin{subequations}\label{eq:MEs}
\begin{eqnarray}
&& \langle \bar{b}_v b_v \rangle = 2 m_B \mu_3 \vphantom{\frac{1}{1} } \ ,\\
&& \langle \bar{b}_v  (i D_\alpha) (i D_\beta) (-i \sigma^{\alpha \beta} ) b_v \rangle =  2m_B \mu_G^2  \vphantom{\frac{1}{1} } \ , \\ 
&& \frac{1}{2}\langle \bar{b}_v \left[ (iD_\mu) \, , \,   \left[ \left( i vD + \frac{1}{2m_b} (iD)^2 \right)  \, , \, (i D^\mu) \right] \right]   b_v \rangle 
= 2 m_B \rho_D^3 \ ,\\ 
&& \langle \bar{b}_v \left[ (iD_\mu) \, , \,  (iD_\nu) \right]  \left[ (iD^\mu)  \, , \, (i D^\nu) \right]   b_v \rangle 
= 2 m_B r_{G}^4  \vphantom{\frac{1}{1} }\ ,  \\ 
\label{eq:rE4}&& \langle \bar{b}_v \left[ (ivD ) \, , \,  (iD_\mu) \right]  \left[ (ivD)  \, , \, (i D^\mu) \right]   b_v \rangle 
= 2 m_B r_{E}^4  \vphantom{\frac{1}{1} } \ , \\
&& \langle \bar{b}_v \left[ (iD_\mu) \, , \,  (iD_\alpha) \right]  \left[ (iD^\mu)  \, , \, (i D_\beta) \right]   (-i \sigma^{\alpha \beta})  b_v \rangle 
= 2 m_B s_{B}^4  \vphantom{\frac{1}{1} }\ ,  \\ 
&& \langle \bar{b}_v \left[ (ivD) \, , \,  (iD_\alpha) \right]  \left[ (ivD)  \, , \, (i D_\beta) \right]   (-i \sigma^{\alpha \beta})  b_v \rangle 
= 2 m_B s_{E}^4  \vphantom{\frac{1}{1} }\ ,  \\ 
&& \langle \bar{b}_v  \left[ iD_\mu \, , \, \left[ iD^\mu \, , \,  \left[ iD_\alpha \, , \, iD_\beta \right] \right] \right]  (-i \sigma^{\alpha \beta}) b_v 
\rangle = 2 m_B s_{qB}^4  \vphantom{\frac{1}{1} } \ ,
\label{eqn:sqB}
\end{eqnarray} 
\end{subequations}
 Here we have redefined $\rho_D^3$ to include its RPI completion as discussed in Ref.~\cite{
Mannel:2018mqv}, and omitted the tilde introduced in \cite{Mannel:2018mqv}. 

We emphasize that here the HQE parameters are defined using full covariant derivatives, as this is more beneficial to see the RPI connections. These definitions, therefore, differ from those often used in literature (e.g. \cite{Bordone:2021oof,Gambino:2016jkc}) where the 
covariant derivative is split into a spatial and a time derivative via 
\begin{equation}\label{eq:perpbasis}
i D_\mu  = v_\mu ivD + D_\mu^\perp \ .
\end{equation}
We refer to Appendix A in \cite{Fael:2018vsp} for the conversion between the different bases. 

\section{Additional Material for Default Fit}\label{app:addimat}
\begin{figure}[tb]
    \centering
    \includegraphics[width=0.5\textwidth]{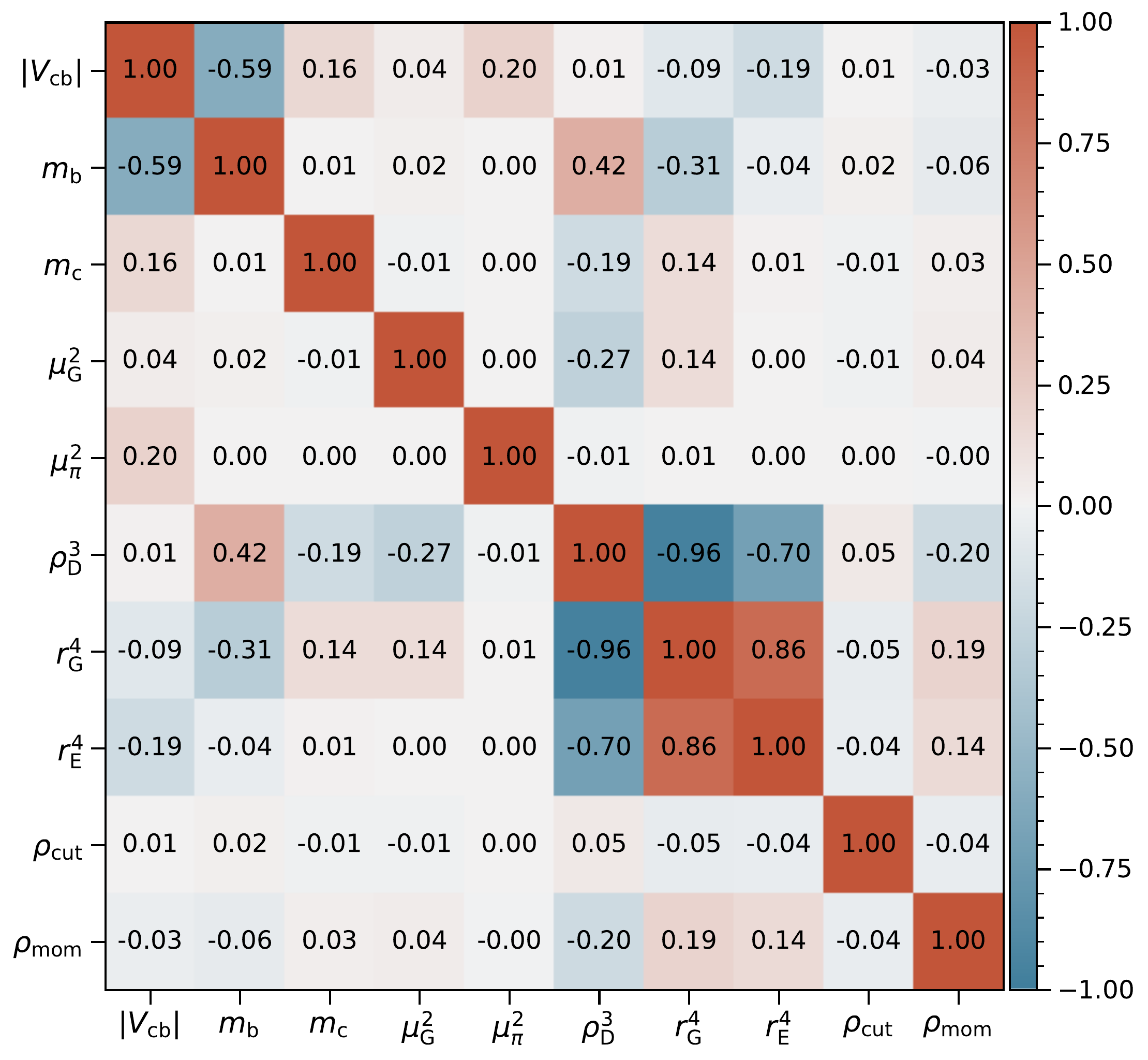}
    \caption{Correlation matrix for $\absVcb$, the HQE parameters, and the correlation parameters $\rhocut$ and $\rhomom$.}
    \label{fig:correlation_matrix}
\end{figure}

\begin{figure}[tb]
    \centering
    \includegraphics[width=0.4\textwidth]{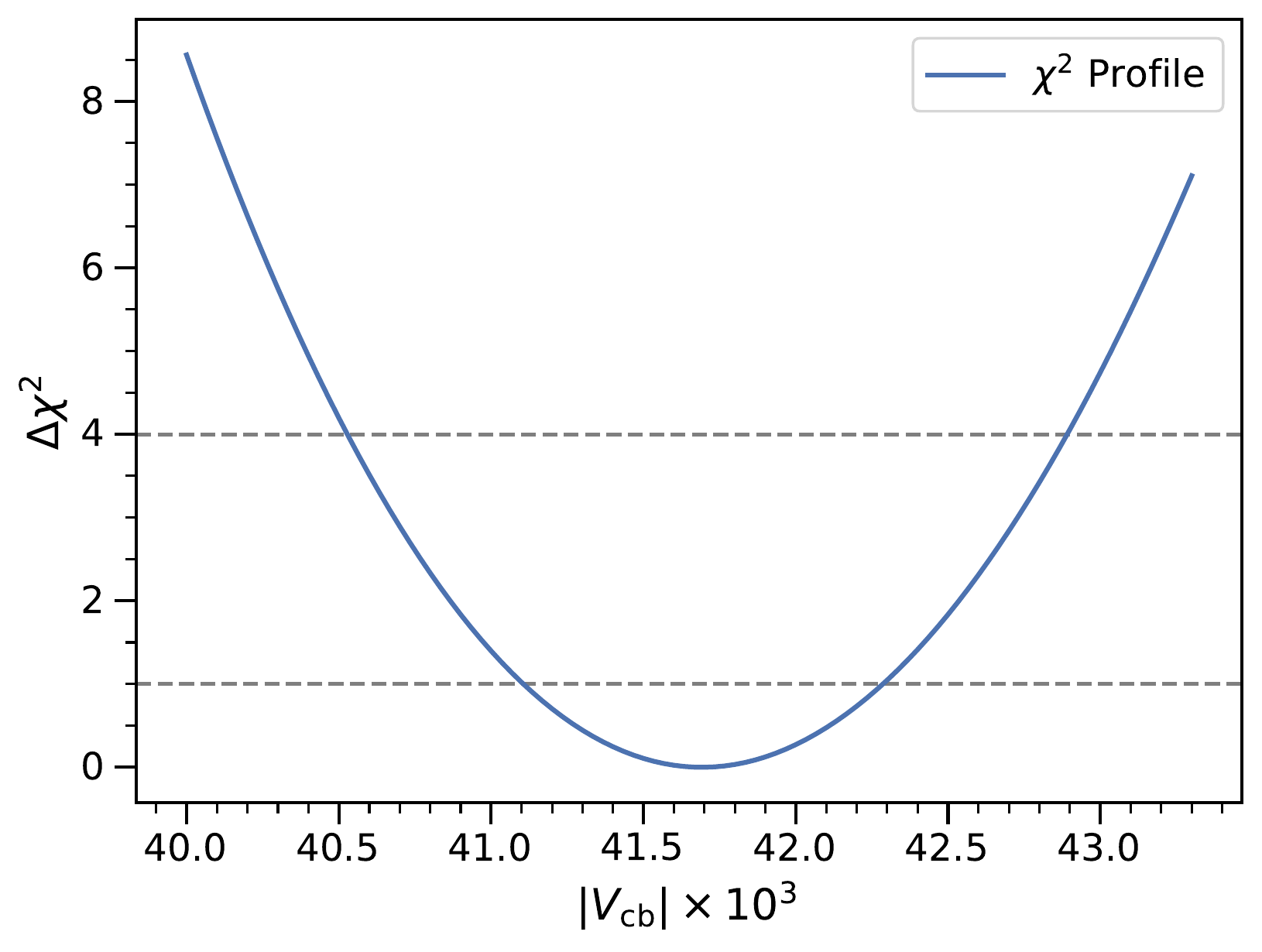}
    \includegraphics[width=0.4\textwidth]{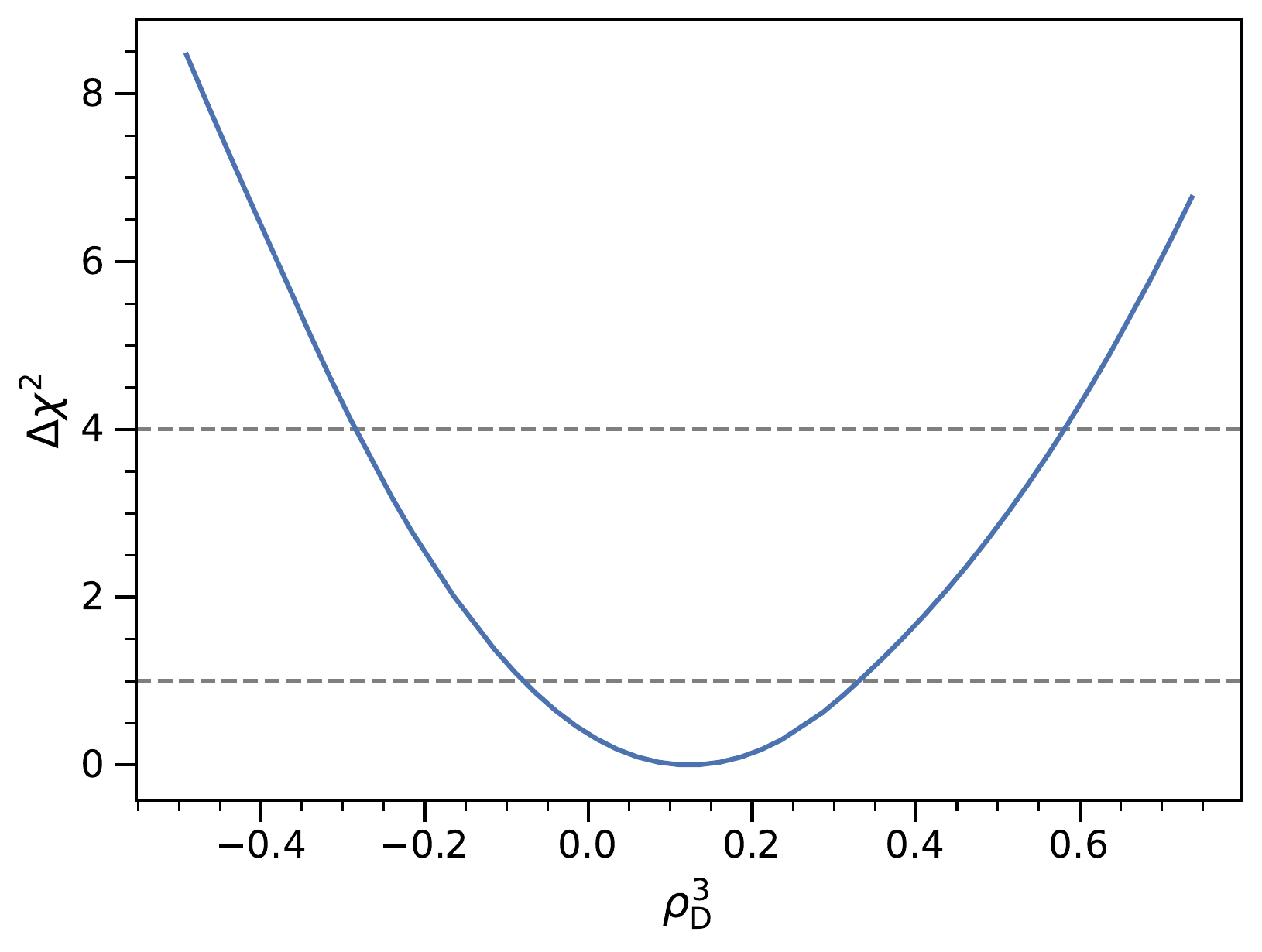}\\
    \includegraphics[width=0.4\textwidth]{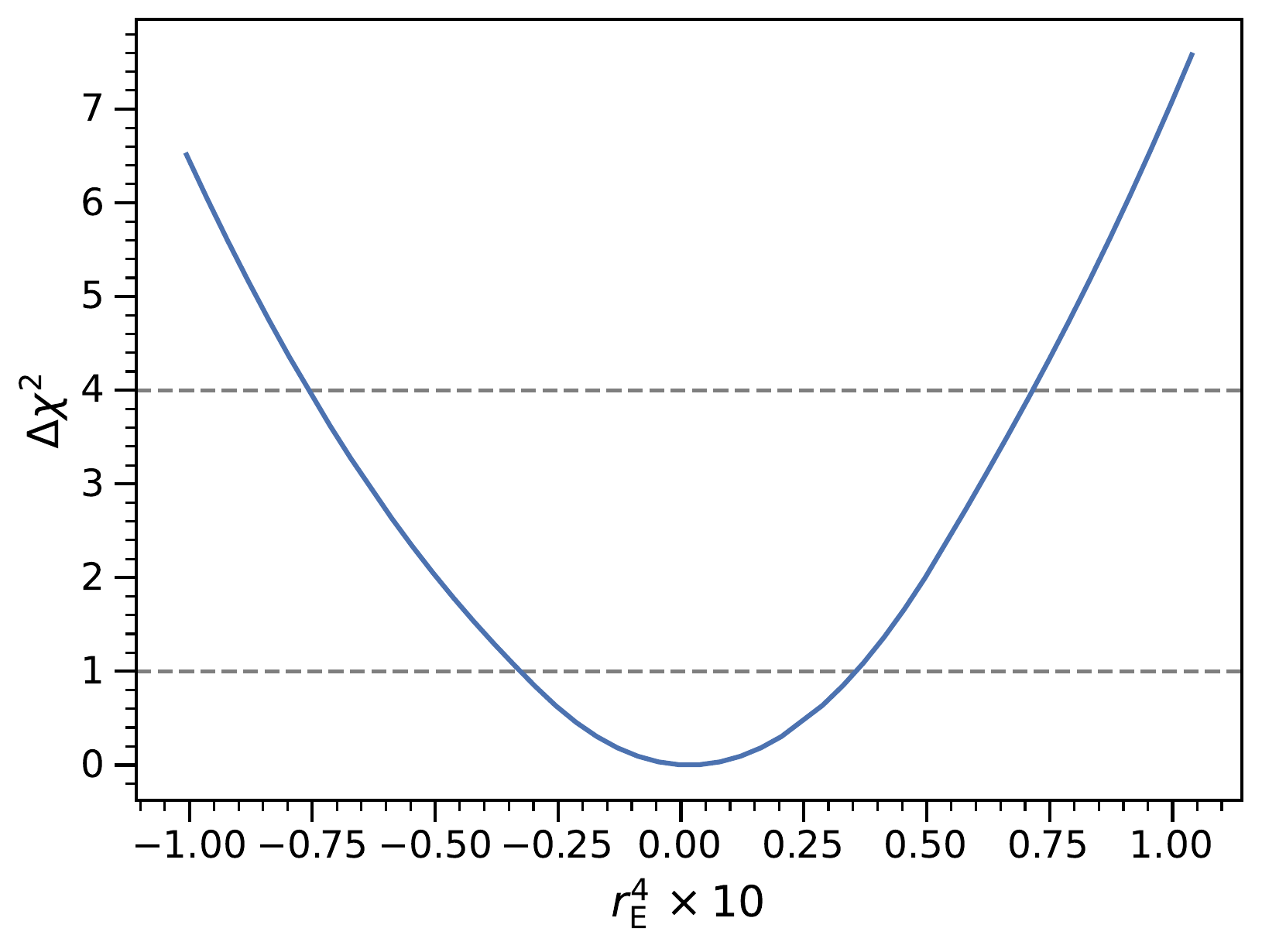}
    \includegraphics[width=0.4\textwidth]{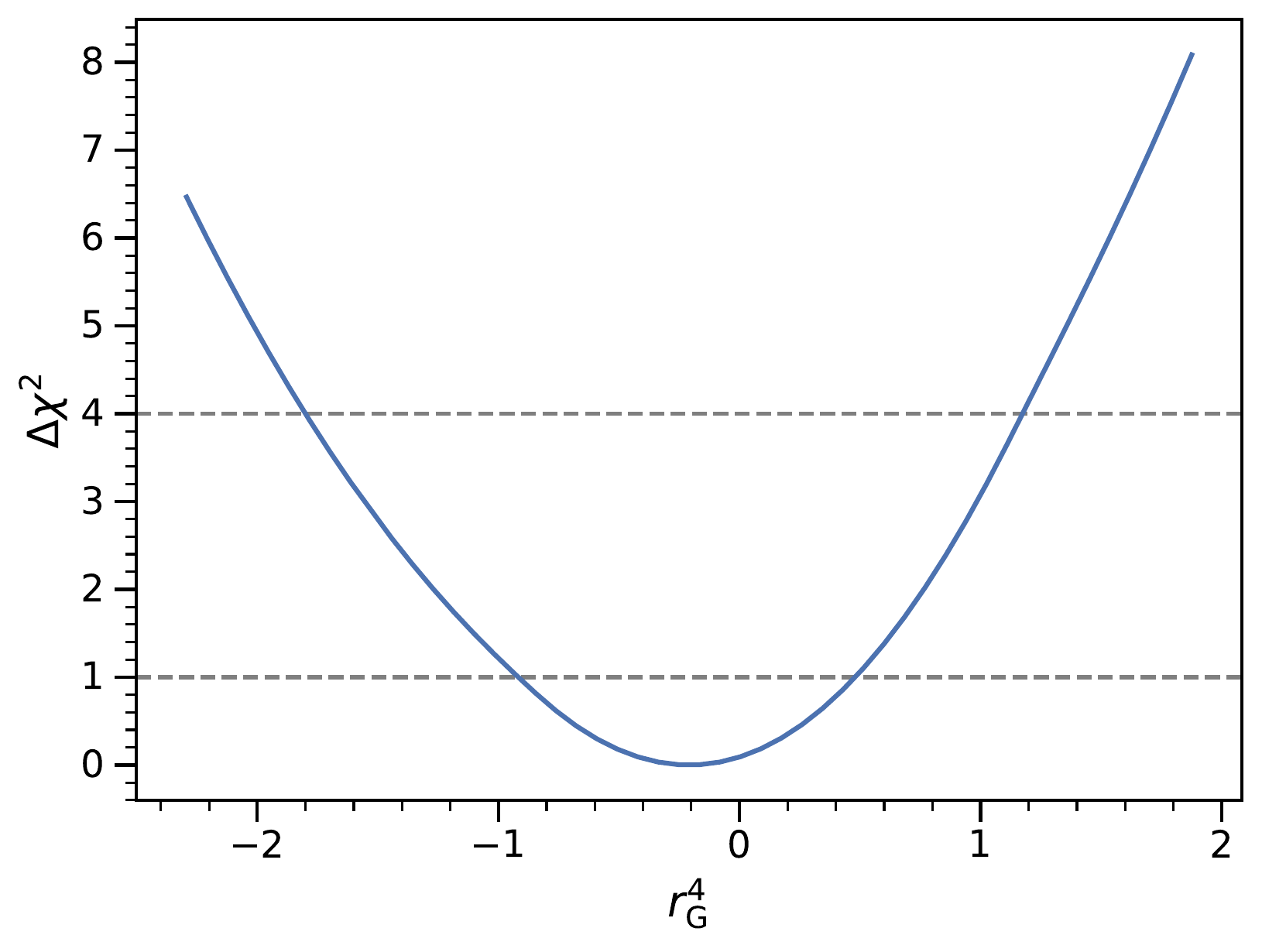}
    \\
    \includegraphics[width=0.4\textwidth]{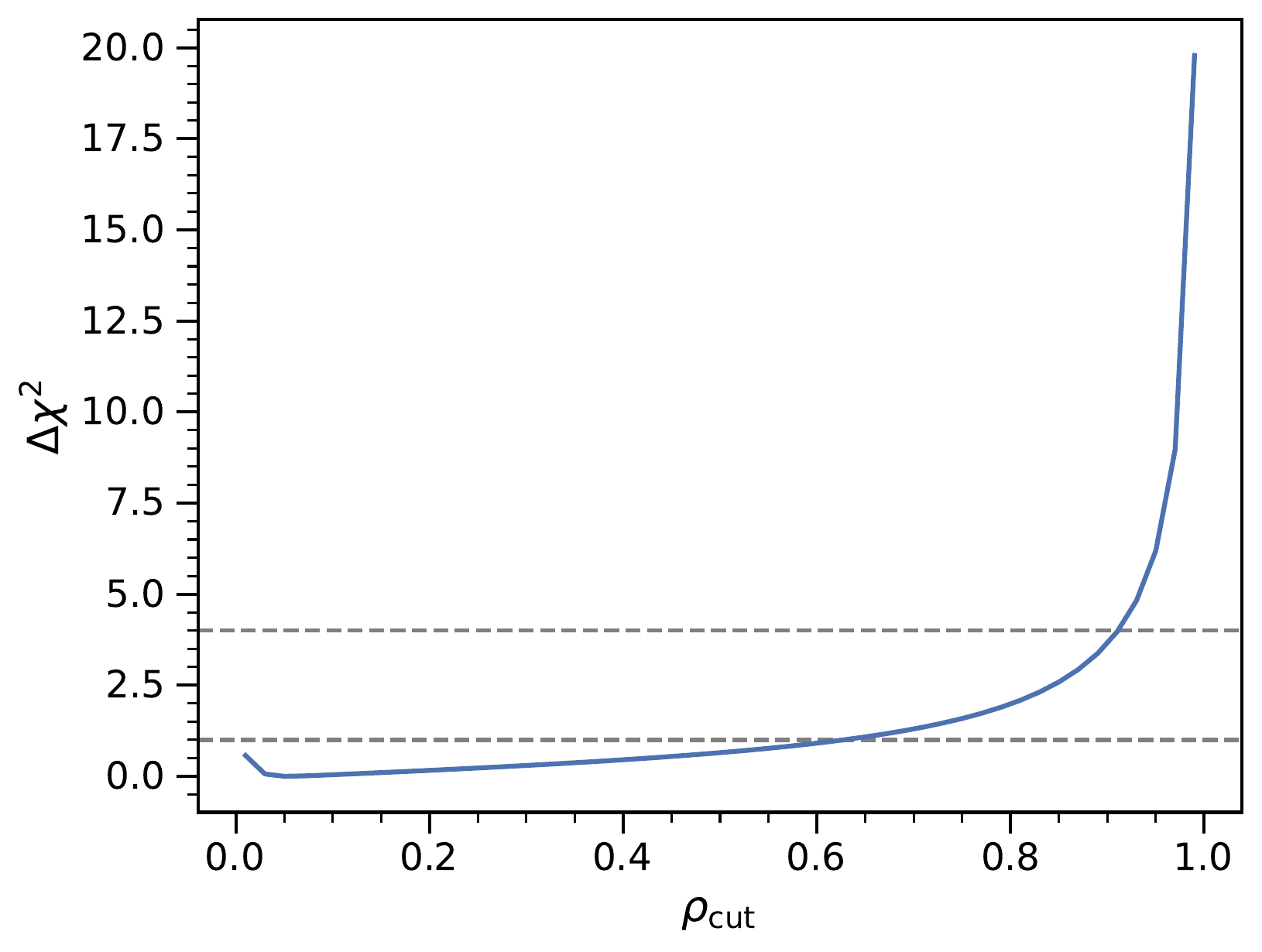}
    \includegraphics[width=0.4\textwidth]{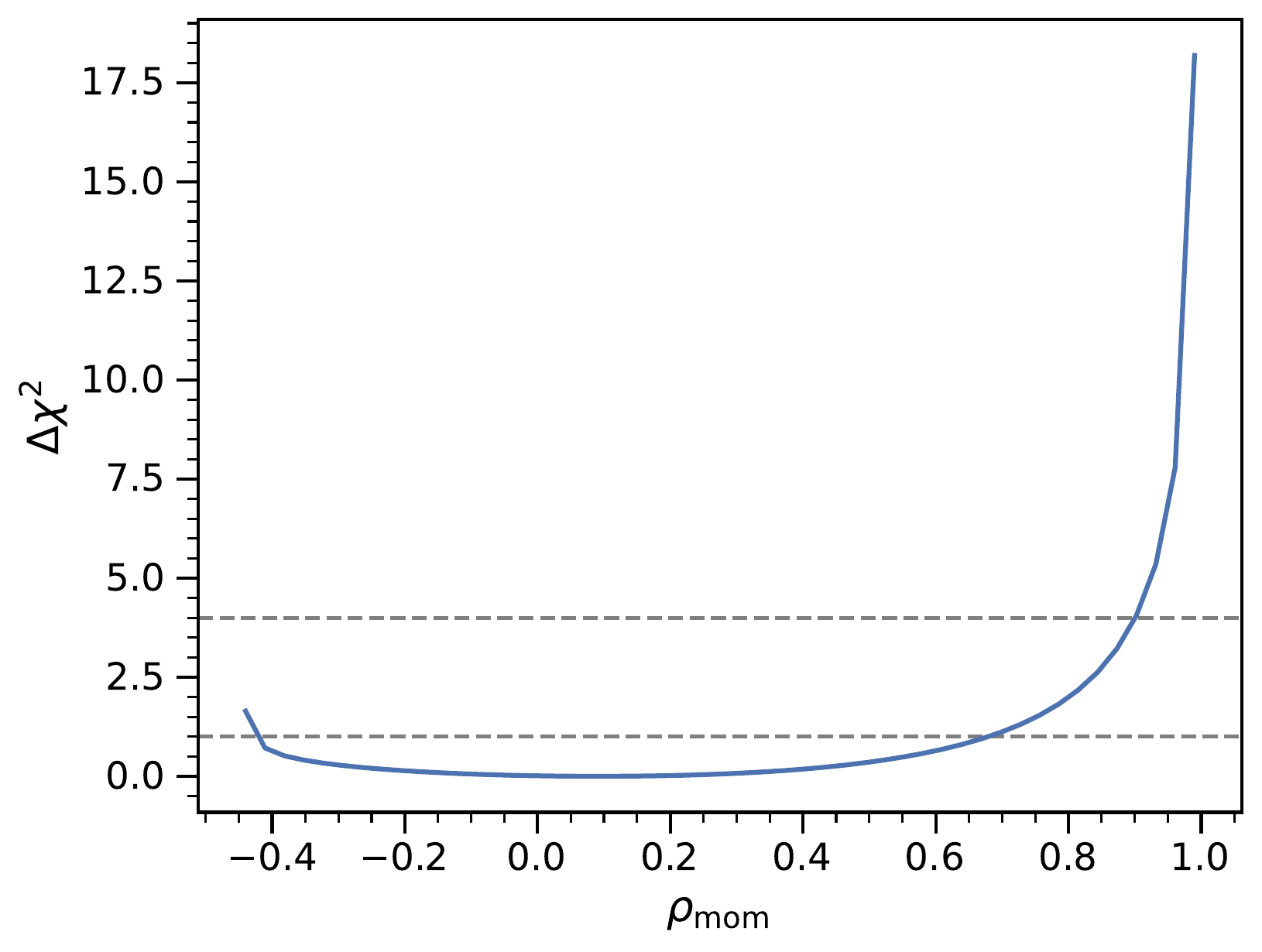}
    \caption{1D $\chi^2$ profile scans for $\absVcb\times10^3$, \rhoD, $\rE\times10$, and \rG. The minimum $\chi^2_\mathrm{min}$ is subtracted from the $\chi^2$ function.}
    \label{fig:likelihood_profiles_1d}
\end{figure}

The correlation matrix of our default fit, Table~\ref{tab:parameter_values_post_fit2}, is shown in Figure~\ref{fig:correlation_matrix}. We note that \absVcb has a small correlation to \mc and the HQE parameters, while a larger negative correlation of $-0.59$ with \mb is observed due to the \mb dependence of the total rate. 
In Table~\ref{tab:parameter_values_post_fit2}, we present the uncertainties as calculated from the Hessian matrix. We also carry out $\Delta\chi^2$ contour scans to validate the Hessian uncertainties. One-dimensional scans are shown in Figure~\ref{fig:likelihood_profiles_1d} and two-dimensional scans are shown in Figures~\ref{fig:likelihood_profiles_2d_vcb} and \ref{fig:likelihood_profiles_2d_rhod}. We validated the coverage and unbiased extraction of all parameters using pseudo-experiments / toy MC. Ensembles of toy data sets are created using a multivariate Gaussian distribution with mean values set to the prediction of the $q^2$ moments as evaluated at the best-fit points and the full covariance used in the construction of the best-fit \chisq function and fitted. The mean values of the ancillary constraints on \mb, \mc, \muG and \mupi are varied around the best-fit value using the corresponding prior uncertainties. For all parameters, with the exception of \rhomom and \rhocut, we find pull distributions compatible with standard normal distributions. We use the pull distributions of \rhomom and \rhocut to construct approximate 68\% confidence levels of both parameters. 

Using a a similar approach, we also estimate the contributions of the experimental, theoretical and constraint uncertainties to the fit parameter uncertainties. We separate the uncertainties stemming from the experimental branching ratio $\mathcal{B}$, the theoretical uncertainty on the total rate $\Gamma$, the theoretical and experimental uncertainties on the $q^2$ moments and the uncertainty from the external constraints. 

In Table~\ref{tab:parameter_values_mb4_zero_post_fit} the results for our default fit with all $1/m_b^4$ parameters set to zero is given. The extracted value of $|V_{cb}|$ only differs marginally from our default fit. 

\begin{table}[t]
    \centering
    \vspace{0.5em}
    \input{values_combined_central_mb4_zero.tex}
        \caption{Results of our default fit using both Belle and Belle~II data for $\absVcb$, $m_b^{\rm kin}(1\;{\rm GeV})$, $\overline{m}_c(2\;{\rm GeV})$, \rhoD, and the correlation parameters $\rho_\mathrm{cut}$ and $\rho_\mathrm{mom}$. The $\mathcal{O}(1/\mb^4)$ terms are set to zero. All parameters are expressed in GeV at the appropriate power.}
    \label{tab:parameter_values_mb4_zero_post_fit}
\end{table}

\section{Scan over the correlation parameters}\label{sec:corrscan}
In this appendix, we perform  our default fit but for fixed combinations of $\rhocut$ and \rhomom. We present fits using Belle and Belle~II data separately, and combined. 

\begin{figure}[htb]
    \centering
    \includegraphics[width=.8\textwidth]{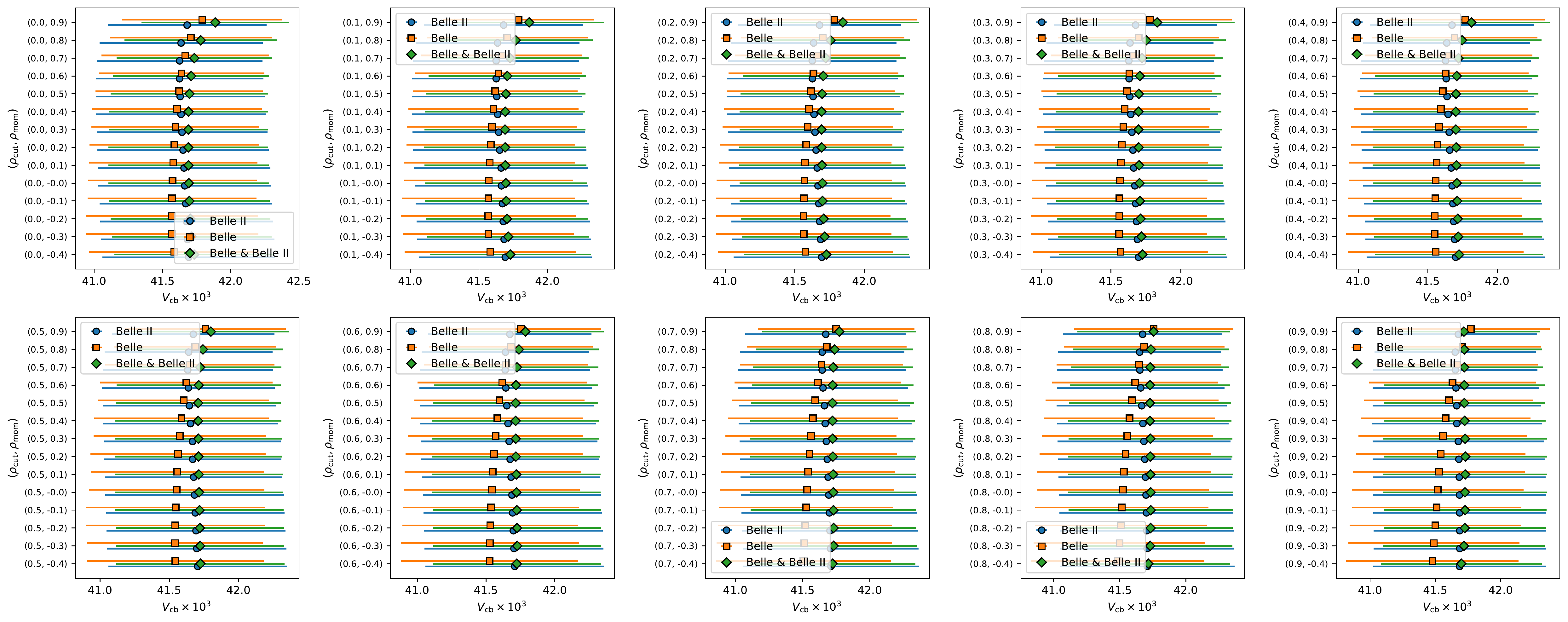}
    \caption{Fit results for \absVcb for different combinations of correlation parameter values.}
    \label{correlation_scan_vcb}
\end{figure}

\begin{figure}[htb]
    \centering
    \includegraphics[width=.8\textwidth]{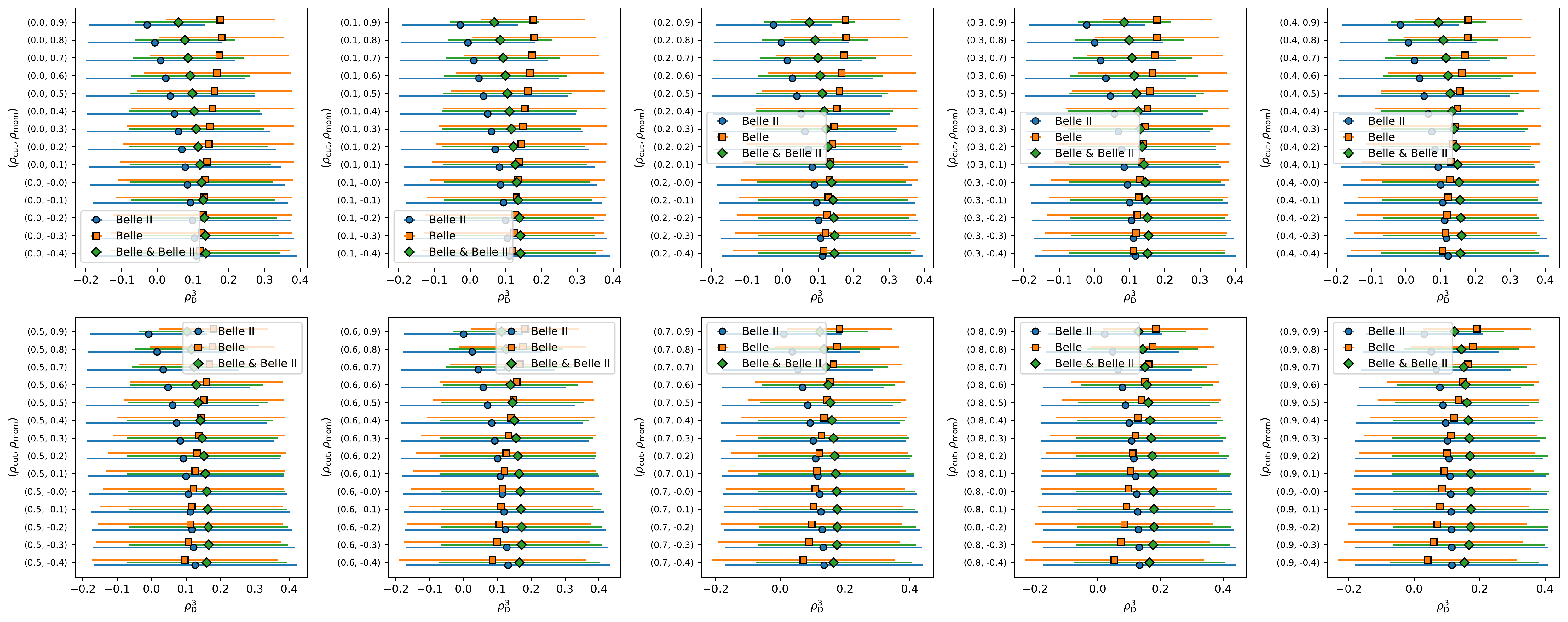}
    \caption{Fit results for \rhoD for different combinations of correlation parameter values.}
    \label{correlation_scan_rhod}
\end{figure}

\begin{figure}[htb]
    \centering
    \includegraphics[width=.8\textwidth]{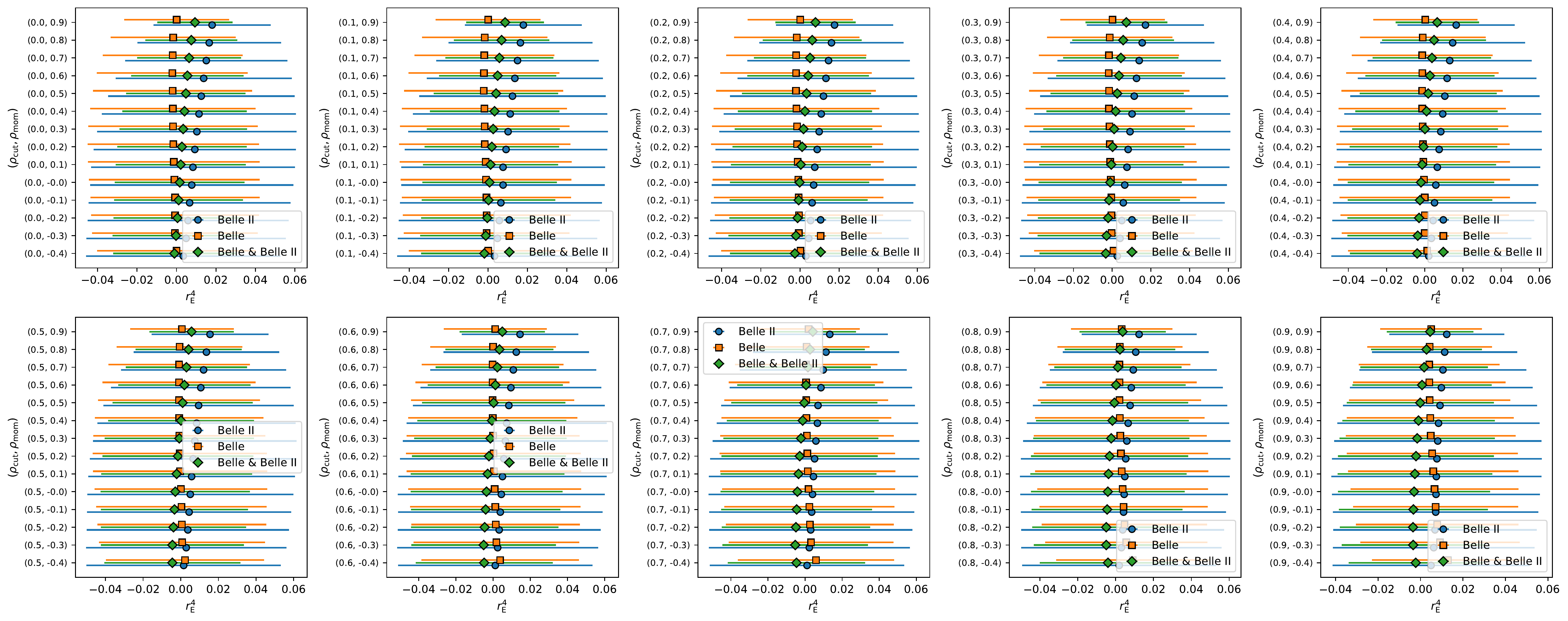}
    \caption{Fit results for \rE for different combinations of correlation parameter values.}
    \label{correlation_scan_re}
\end{figure}

\begin{figure}[htb]
    \centering
    \includegraphics[width=0.8\textwidth]{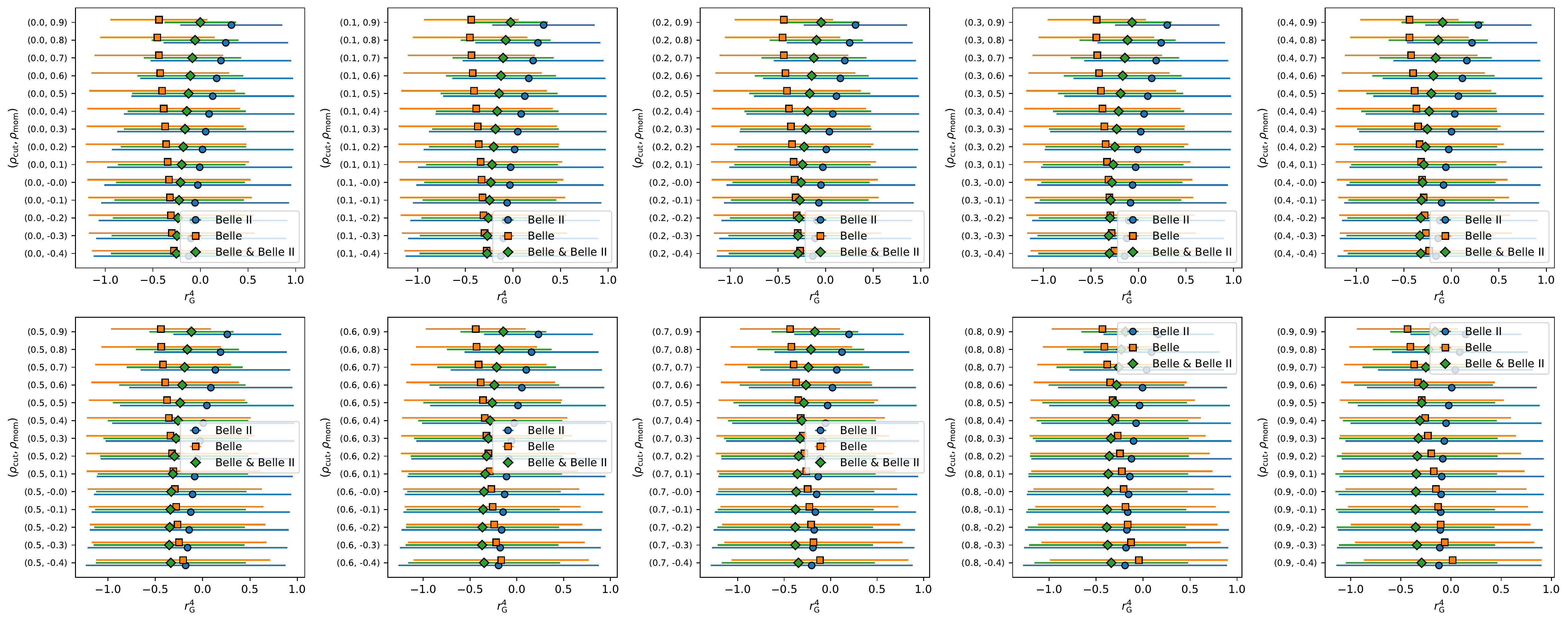}
    \caption{Fit results for \rG for different combinations of correlation parameter values.}
    \label{correlation_scan_rg}
\end{figure}

\section{Belle Only Results}\label{sec:belleonly}
In this appendix, we present the results using Belle \cite{Belle:2021idw} results only. 
\begin{table}[b]
    \centering
    \vspace{0.5em}
    \label{tab:parameter_values_post_fit_belle}
    \input{values_belle_central.tex}
        \caption{Results of our default fit using only Belle data for $\absVcb$, $m_b^{\rm kin}(1\;{\rm GeV})$, $\bar{m}_c(2\;{\rm GeV})$, the HQE parameters, and the correlation parameters $\rhocut$ and $\rhomom$. All parameters are expressed in GeV at the appropriate power.}
    \end{table}

\begin{figure}[tb]
    \centering
    \includegraphics[width=0.4\textwidth]{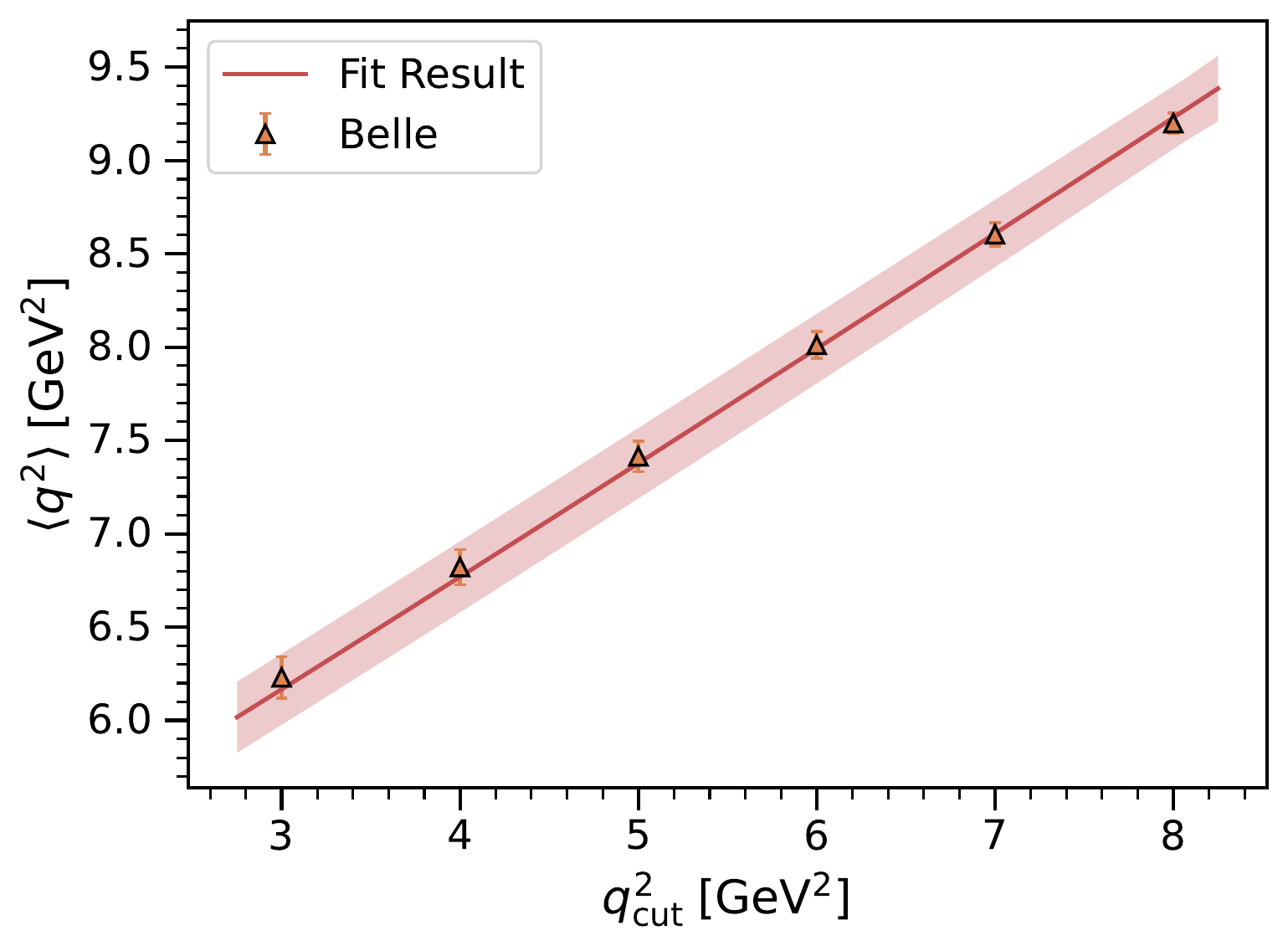}
    \includegraphics[width=0.4\textwidth]{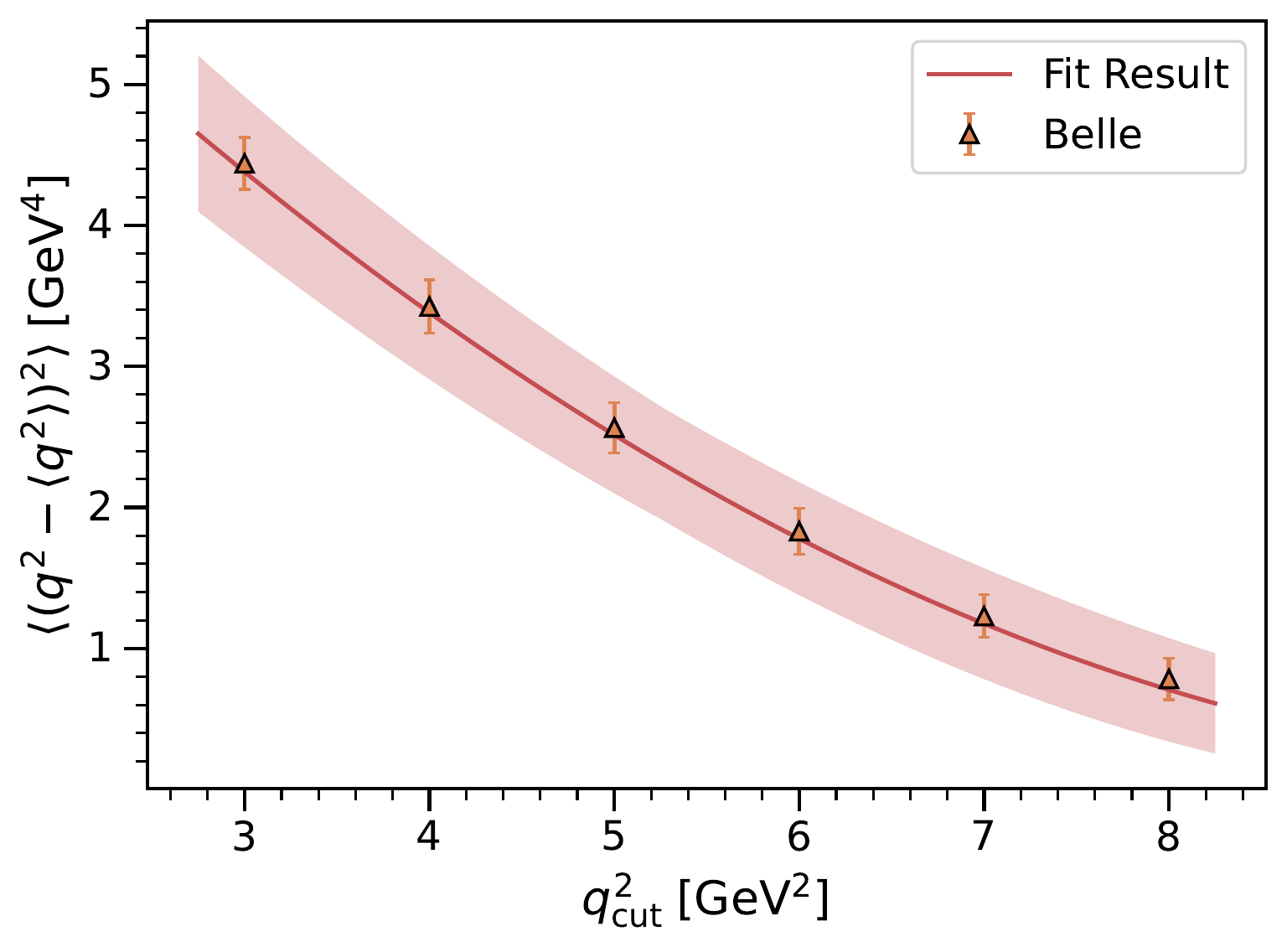}\\
    \includegraphics[width=0.4\textwidth]{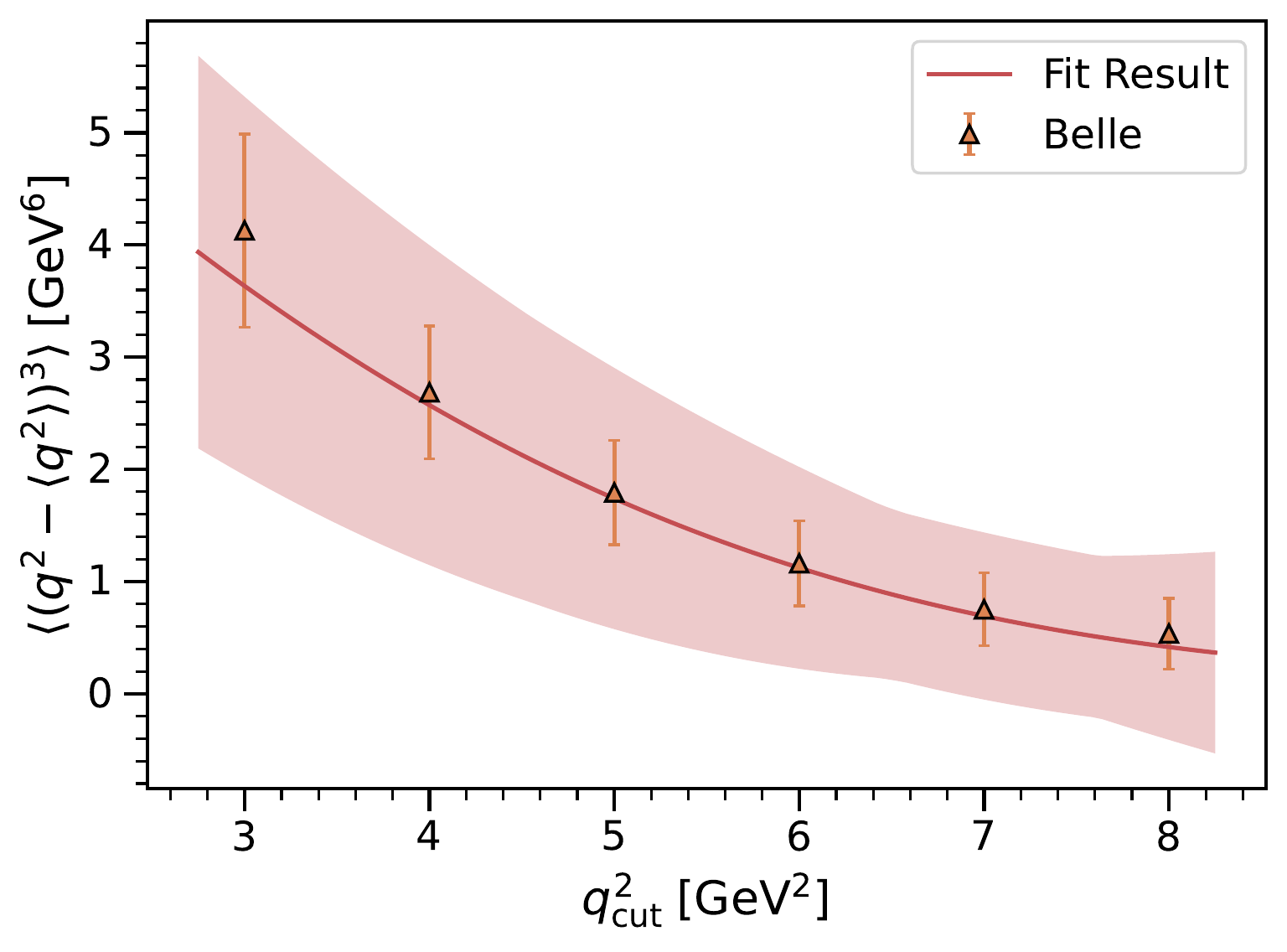}
    \includegraphics[width=0.4\textwidth]{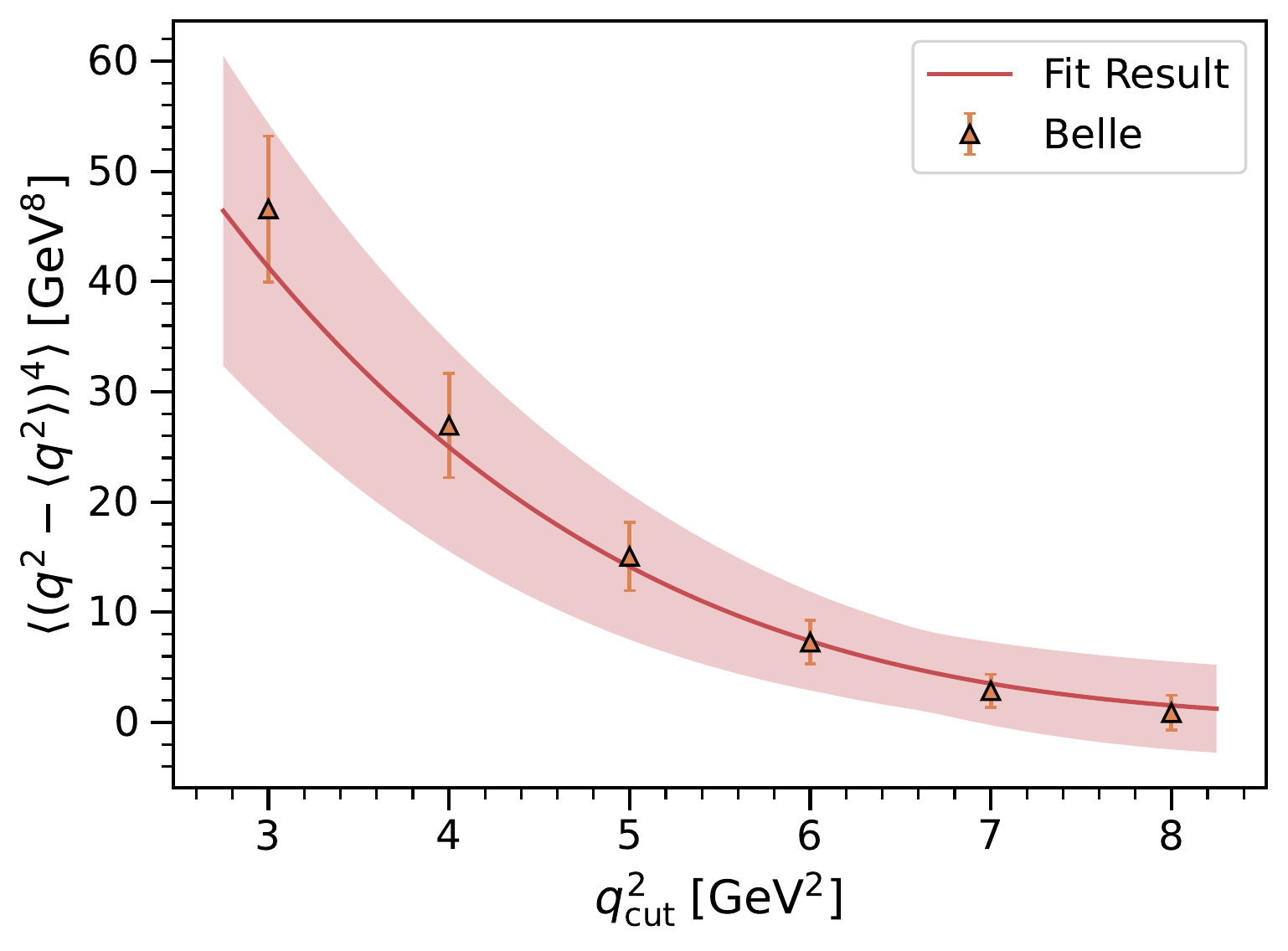}
    \caption{Fit projections for $q^2$ moments using Belle data only. }
    \label{fig:central_moments_post_fit_belle}
\end{figure}

\begin{figure}[tb]
    \centering
    \includegraphics[width=0.5\textwidth]{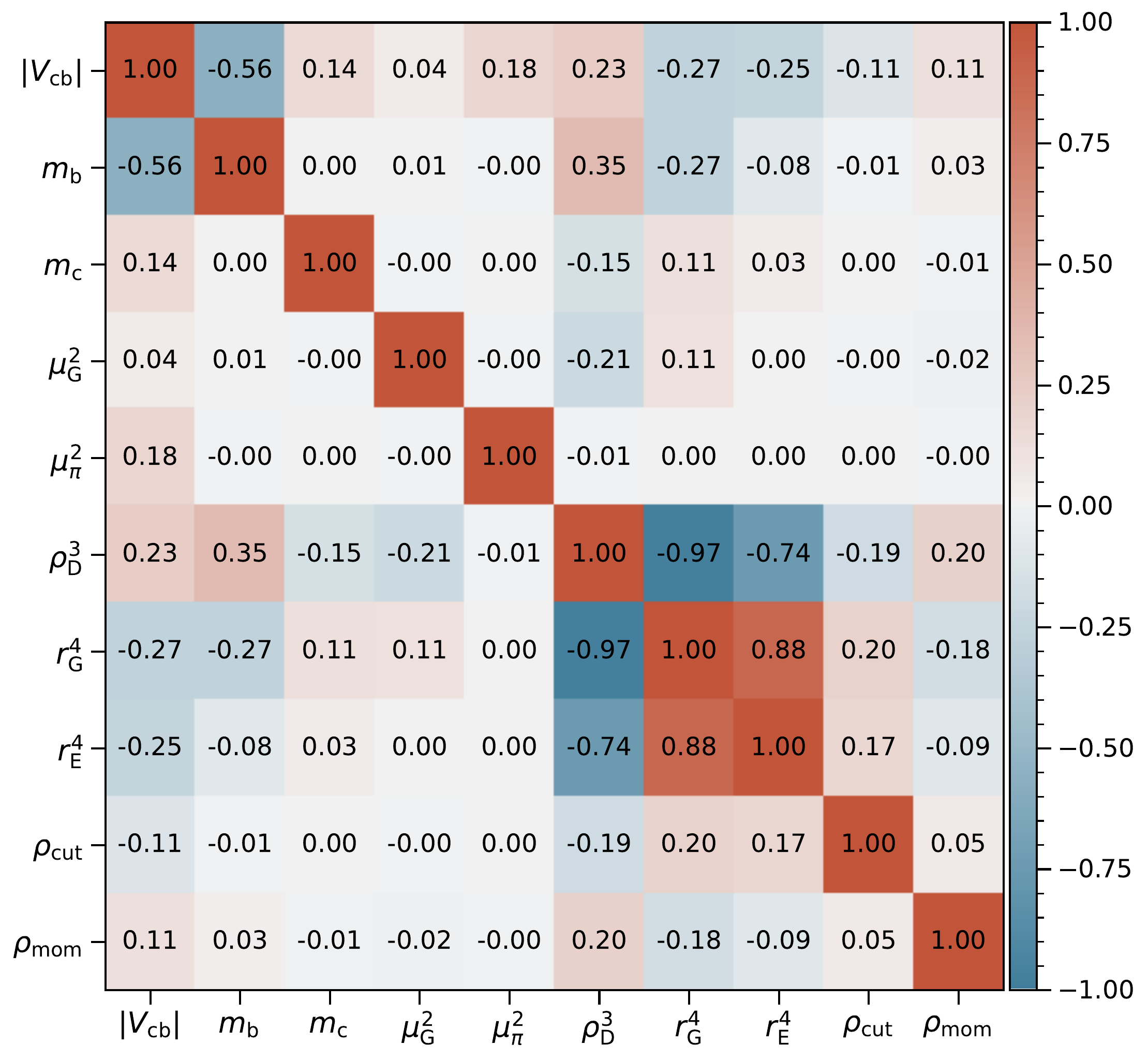}
    \caption{Correlation matrix for $\absVcb$, the HQE parameters, and the correlation parameters $\rhomom$ and $\rhocut$ using Belle data only.}
    \label{fig:correlation_matrix_belle}
\end{figure}

\section{Belle~II Only Results}\label{sec:belle2only}
In this appendix, we present the results using Belle II results only.

\begin{table}[h]
    \centering
    \vspace{0.5em}
    \label{tab:parameter_values_post_fit_belle2}
    \input{values_belle2_central.tex}
        \caption{Results of our default fit using only Belle~II data for $\absVcb$, $m_b^{\rm kin}(1\;{\rm GeV})$, $\bar{m}_c(2\;{\rm GeV})$, the HQE parameters, and the correlation parameters $\rhocut$ and $\rhomom$. All parameters are expressed in GeV at the appropriate power. }
    \end{table}

\begin{figure}[h]
    \centering
    \includegraphics[width=0.4\textwidth]{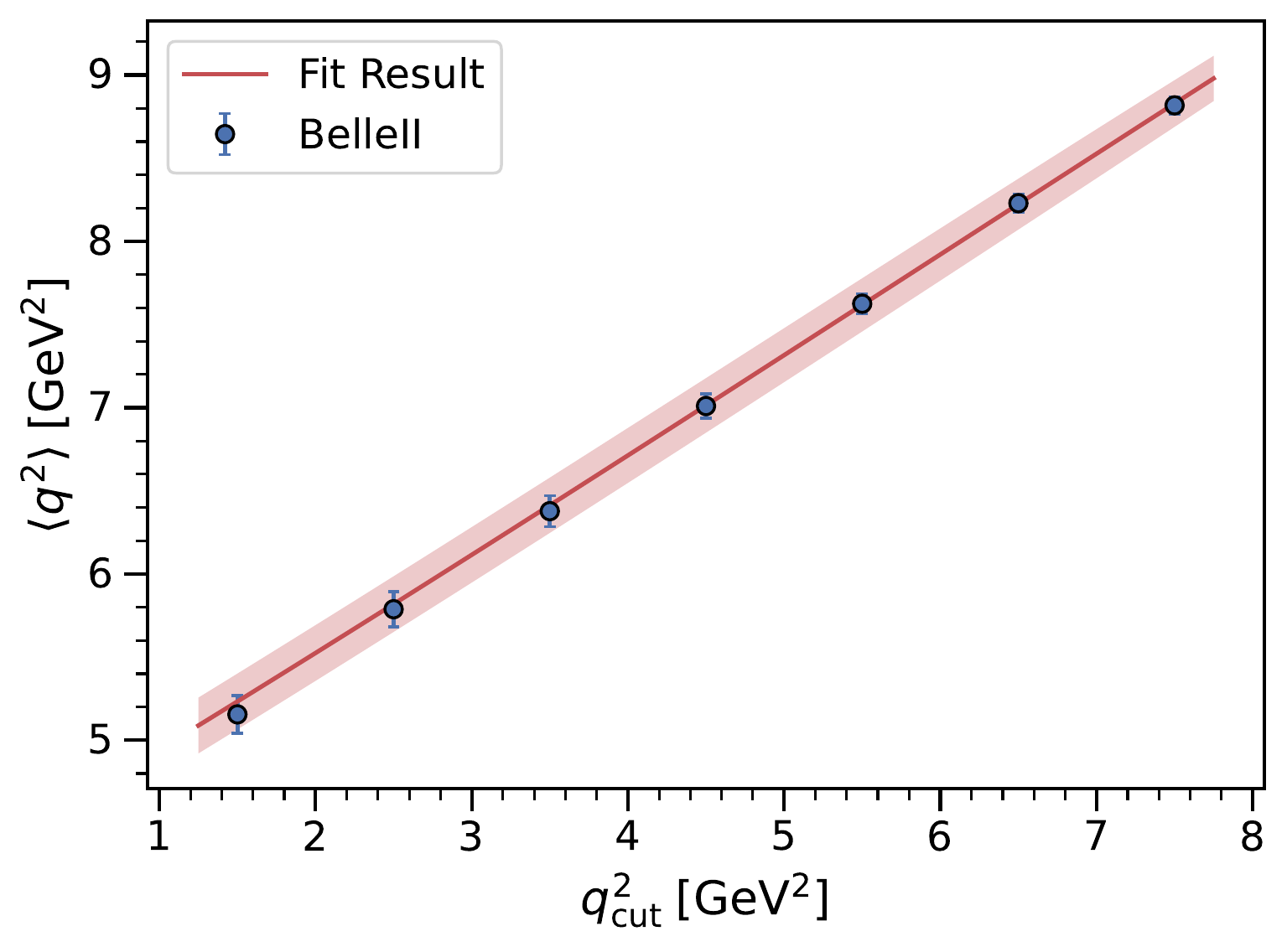}
    \includegraphics[width=0.4\textwidth]{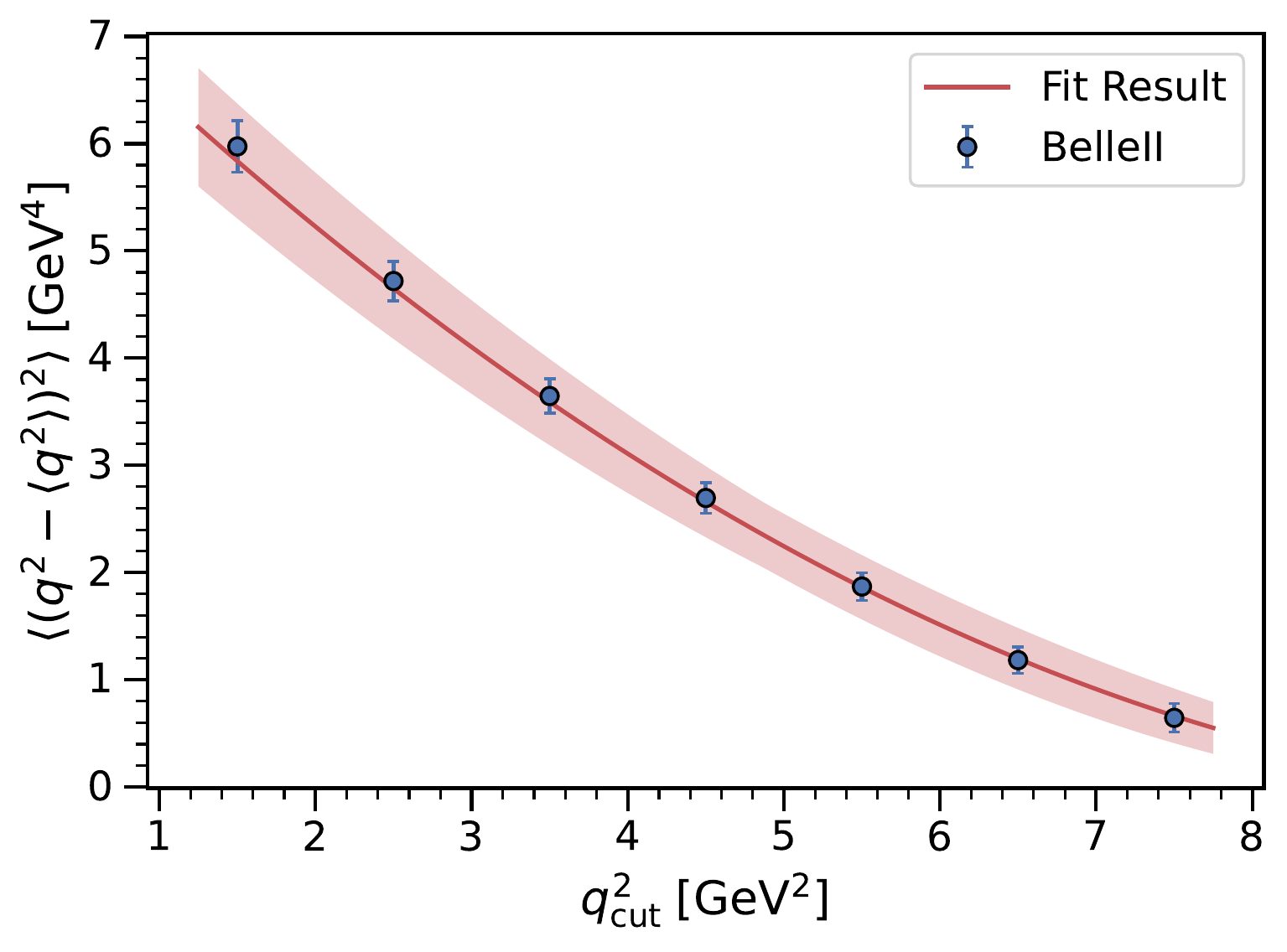}\\
    \includegraphics[width=0.4\textwidth]{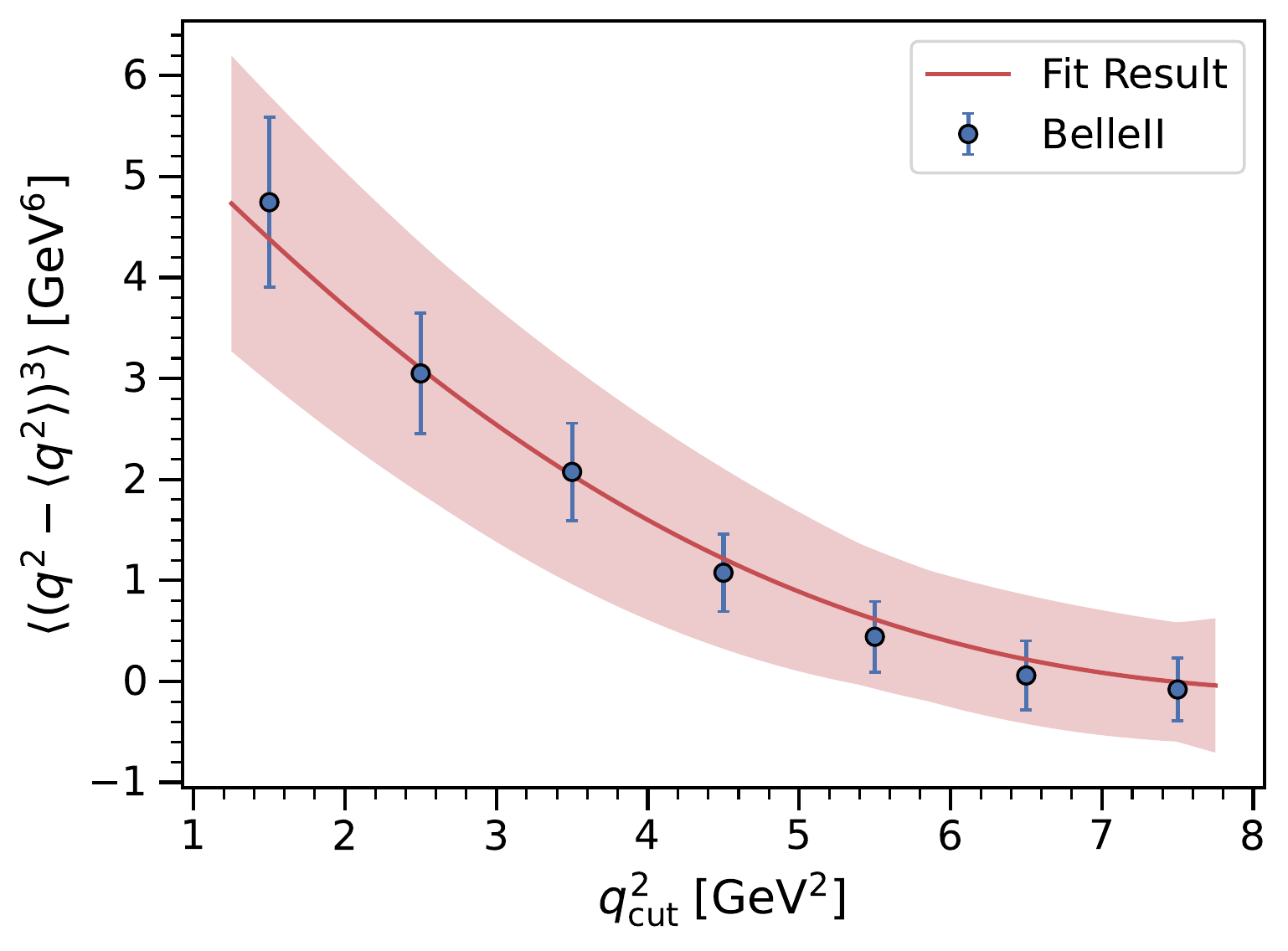}
    \includegraphics[width=0.4\textwidth]{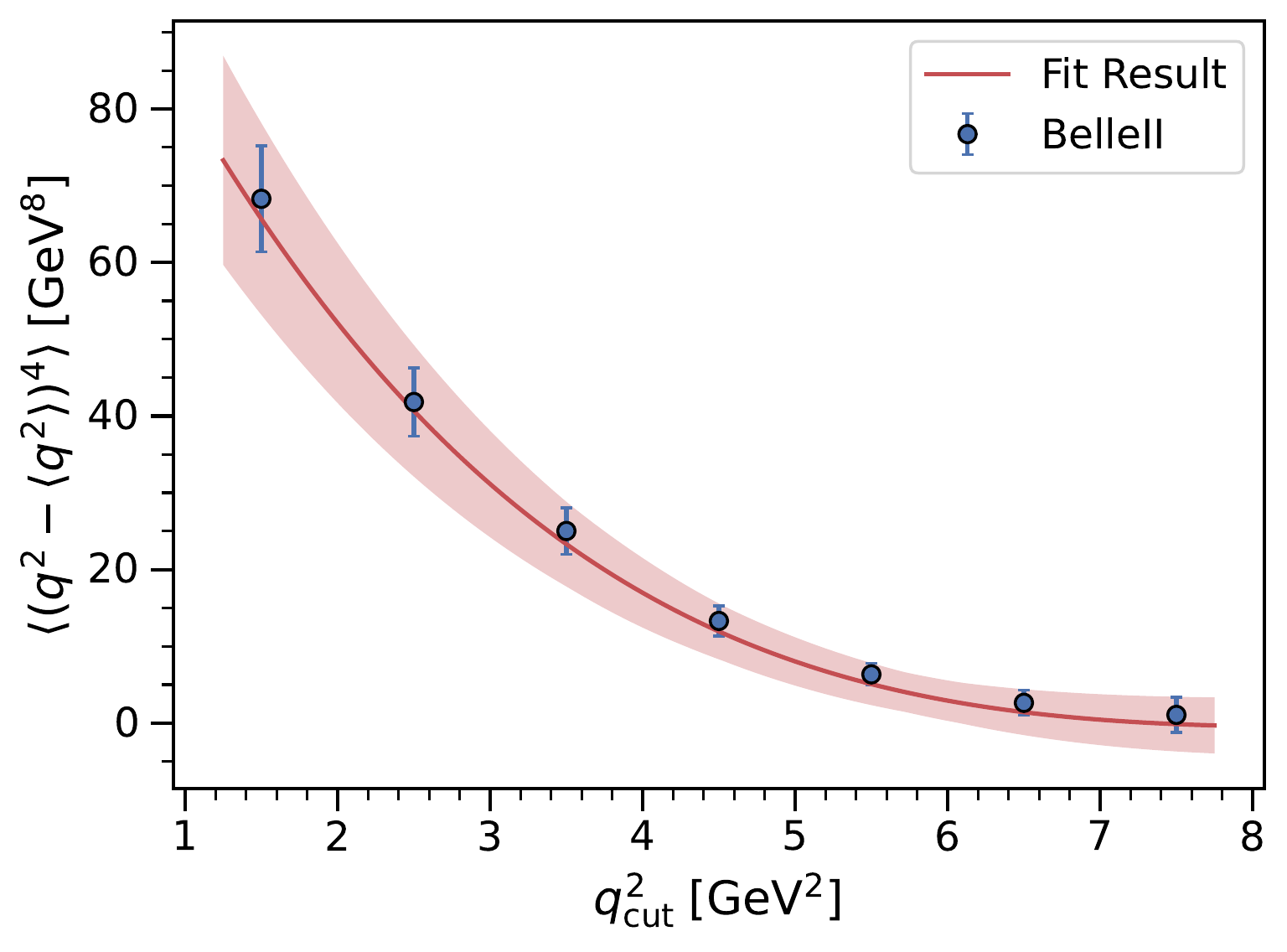}
    \caption{Fit projections for $q^2$ moments using Belle~II data only. }
    \label{fig:central_moments_post_fit_belle2}
\end{figure} 

\begin{figure}[bt]
    \centering
    \includegraphics[width=0.5\textwidth]{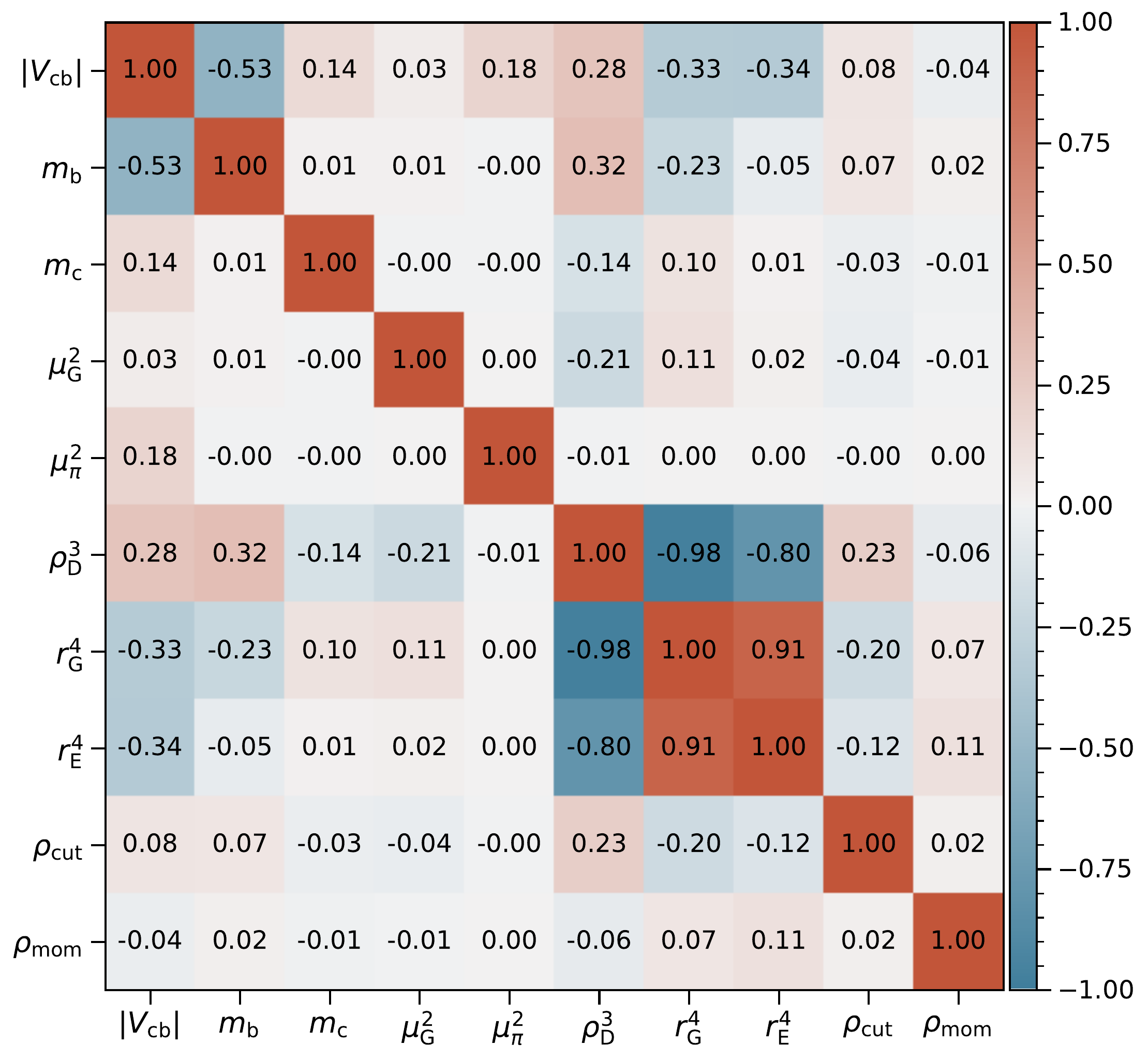}
    \caption{Correlation matrix for $\absVcb$, the HQE parameters, and the correlation parameters $\rhomom$ and $\rhocut$ using Belle~II data only.}
    \label{fig:correlation_matrix_belle2}
\end{figure}

%% file: values_combined_central_mb4_zero.tex
\begin{tabular}{lrrrrrrrr}
\toprule
{} &  $|V_\mathrm{cb}| \times 10^3$ &  $m_b^{\rm kin} $ &  $\overline{m}_{c} $ &  $\mu_{G}^2 $ &  $\mu_\pi^2 $ &  $\rhoD$ &  $\rho_\mathrm{cut}$ &  $\rho_\mathrm{mom}$ \\
\midrule
Value       &                          41.76 &             4.56 &             1.10 &                 0.38 &          0.43 &                  0.03 &                 0.05 &                 0.14 \\
Uncertainty &                           0.57 &             0.02 &             0.01 &                 0.07 &          0.24 &                  0.02 &                 - &                 - \\
\bottomrule
\end{tabular}

%% file: values_belle_central.tex
\begin{tabular}{lrrrrrrrrrr}
\toprule
{} &  $|V_{cb}| \times 10^3$ &  $m_{b} $ &  $m_{c} $ &  $\mu_{G}^2 $ &  $\mu_\pi^2 $ &  $\rho_{D}^3 $ &  $r_{G}^4 $ &  $r_{E}^4 \times 10$ &  $\rho_\mathrm{cut}$ &  $\rho_\mathrm{mom}$ \\
\midrule
Value       &                          41.54 &             4.56 &             1.09 &                 0.36 &          0.43 &                  0.12 &              -0.27 &                        0.01 &                 0.62 &                 0.02 \\
Uncertainty &                           0.66 &             0.02 &             0.01 &                 0.07 &          0.24 &                  0.28 &               0.99 &                        0.48 &                 - &                 - \\
\bottomrule
\end{tabular}

%% file: values_belle2_central.tex
\begin{tabular}{lrrrrrrrrrr}
\toprule
{} &  $|V_{cb}| \times 10^3$ &  $m_{b}^{\rm kin} $ &  $\overline{m}_{c} $ &  $\mu_{G}^2 $ &  $\mu_\pi^2 $ &  $\rho_{D}^3 $ &  $r_{G}^4 $ &  $r_{E}^4 \times 10$ &  $\rho_\mathrm{cut}$ &  $\rho_\mathrm{mom}$ \\
\midrule
Value       &                          41.70 &             4.56 &             1.09 &                 0.36 &          0.43 &                  0.12 &              -0.17 &                        0.02 &                 0.47 &                -0.38 \\
Uncertainty &                           0.65 &             0.02 &             0.01 &                 0.07 &          0.24 &                  0.30 &               1.08 &                        0.53 &                 - &                 - \\
\bottomrule
\end{tabular}